%% file: 160107_document_2.tex
\documentclass[a4paper,titlepage,10pt,DIV=13]{ scrartcl}
\usepackage[all]{xy}

\usepackage{amsfonts}
\usepackage{amsmath}
\usepackage{amssymb}
\usepackage{amsthm}
\usepackage{tabularx}
\usepackage{graphicx}
\usepackage{eurosym}
\usepackage{multirow}
\usepackage{pdfpages}
\usepackage{float}
\usepackage{todonotes}
\usepackage{hyperref}

\usepackage{xcolor}
\usepackage{mdframed}
\newmdenv [linecolor=black, frametitle=Conclusion,skipabove=0.5cm,skipbelow=0.5cm,leftmargin=0.5cm,rightmargin=0.5cm,backgroundcolor=lightgray]{infobox}

\usepackage[latin1]{inputenc}
\usepackage[T1]{fontenc}

\bibstyle{alpha}
\usepackage{cite}

\usepackage[manualmark,
  headsepline,
  plainheadsepline,
  footsepline,
  plainfootsepline]{scrpage2}
\usepackage{helvet}
\newcolumntype {Z}{ >{\centering\arraybackslash }X <{}}
\newcolumntype {Y}{ >{\centering\arraybackslash }p{0.20\textwidth} <{}}

\newcommand{\percent}{\%}

\begin{document}
%{\today\\ \svnInfoRevision,\ \svnInfoDate\ \svnInfoTime}

%\renewcommand\refname{ }

%\extratitle{\svnInfoRevision,\ \svnInfoDate\ \svnInfoTime}
\titlehead{\centering \LARGE Final report}
\title{Riblet Sensor - Light Scattering on Micro Structured Surface Coatings} 
\author{\textbf{Prof. Dr. Mirco Imlau, Hauke Bruening, Kay-Michael Voit,}\\ \textbf{Juliane Tschentscher and Volker Dieckmann} \\ Universit\"at Osnabr\"uck}

\publishers{\Large{Name of the coordinating person: Prof. Dr. Mirco Imlau}
 \vspace{2cm}{ 
\centering
  \vspace{2cm}
	\includegraphics[]{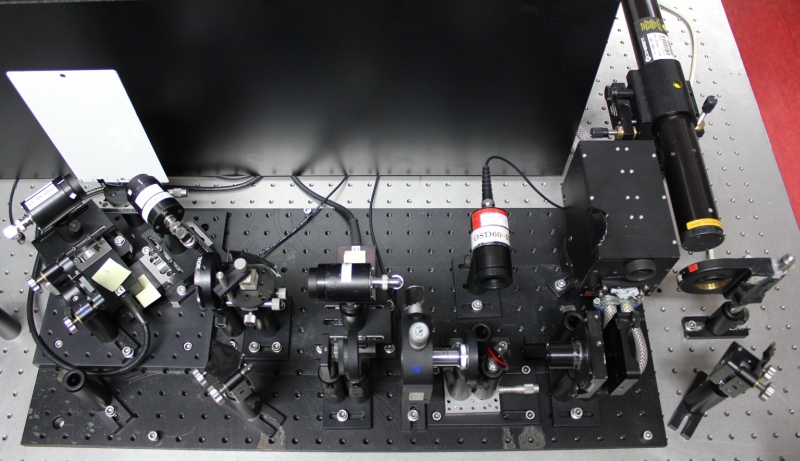}
} \newline
\textnormal{Call identifier:}\\
SP1-JTI-CS-2011-02
}
%\date{\today\\ \svnInfoRevision,\ \svnInfoDate\ \svnInfoTime}

%\subject{\textnormal{Call identifier:}\\
%SP1-JTI-CS-2011-02}

\date{ }
\maketitle

%\listoftodos
\tableofcontents

\section*{Abstract}

With the application of appropriate surface structuring on aircrafts, up to 8\% fuel may be saved in regular air traffic.
This not only decreases costs, but especially reduces exhaust of greenhouse gases significantly.

Before these techniques can be introduced into productive environments, a controlling method for the quality of surface structuring had to be established to be used during fabrication and service, ensuring persistent quality of the structured coatings and a justified decision for surface renewal. 
In this project, these important requirements for achieving the improvements defined above are fulfilled.
We have shown that fast sampling is possible using noncontacting laser probing, and we have presented a working preliminary configuration for the sensor. 

In the theoretical part, a model for the interaction between a probing laser beam and the surface is developed and the resulting wavefront is derived. This is done using a combination of Huygens-Fresnel diffraction theory and geometrical optics. The model is then used to counsel the design of the experimental setup, to interpret the emerging data and to develop characteristic quantities for the sample, their derivation from the data and their signal-to-noise ratio.

In the experimental part, the interaction of laser light with the structured riblet surface is studied. For this purpose an optical setup was installed to perform measurements of undegraded and degraded surfaces depending on a variety of experimental parameter like probe wavelength or angle of incidence.
The results of these measurements in the form of intensity distributions as a function of angles are constantly compared and checked with the theoretical calculations. A preliminary configuration of an optical setup with an optimized laser system is now available for further studies. 
It allows for very sensitive measurements of even slight degradations of the surfaces. 
Here, it is regardless if the damage to the riblets is symmetrical or asymmetrical due to mechanical loss of material or if it is deriving from changes of reflectivity due to chemical processes of the riblet material itself.

A fast implementation in commercial products should be possible on the basis of this report.

The research leading to these results has received funding from the European Union Seventh Framework Programme (FP7\textbackslash 2007-2013) for the Clean Sky Joint Technology Initiative under grant agreement number CSJU-GAM-SFWA-2008-001.

\section{Introduction}

The interaction of light with micro- and nanostructured dielectrics strongly attracts researcher's interest in the field of (nonlinear) optics and photonics in recent years. This interest is particularly driven by the rapid success and the tremendous technical improvements in the fabrication of micro- and nanostructured materials using for instance EUV-photolithography, optical micropatterning with fs-laserpulses and e-beam writing (cf. \cite{Bratton2006,Kuroiwa2004,Tseng2003}). It enabled the fabrication of structures with spatial variation of the susceptibility on a scale far below the light wavelength, i.e. with a spatial precision on the nanometer scale. At the same time, the fabrication of three-dimensional, spatially structured dielectrics was introduced (cf. \cite{Lenzner1998,Krol2008}). Applying these technologies for photonic industries, the targeted design of optical components with given features has become possible.

The impact of the scientific research in light-matter interaction and the outcome of the various research activities becomes apparent regarding the variety of novel technologies that were particularly developed for the field of photonics and that are already established on the market: photonic crystals and photonic crystal fibers, next-generation integrated optical components (lasers and filters in silica chips), Bragg-filters for small-bandwidth laser systems, DWDMs for telecommunication purposes or dielectrics with periodically-poled nonlinearities for frequency conversion (cf. \cite{Joannopoulos1997,Russell2003,Lumeau2010,Yu2010,Lu2000}).

We note that in any of these systems, periodic and aperiodic structures as well as structures with defects or phase-jumps are at the origin of the functionality. For instance, chirped mirrors for the optimization of fs-laser cavities, are built from intelligently structured dielectric layers and allow for a reflectivity in a broad wavelength regime with flat-top profile and wavelength-dependent phase shift (cf. \cite{Szipocs1994}).

\begin{figure}[ht]
  \centering
    \includegraphics[width=0.6\textwidth,trim=0in 5in 0in 0in]{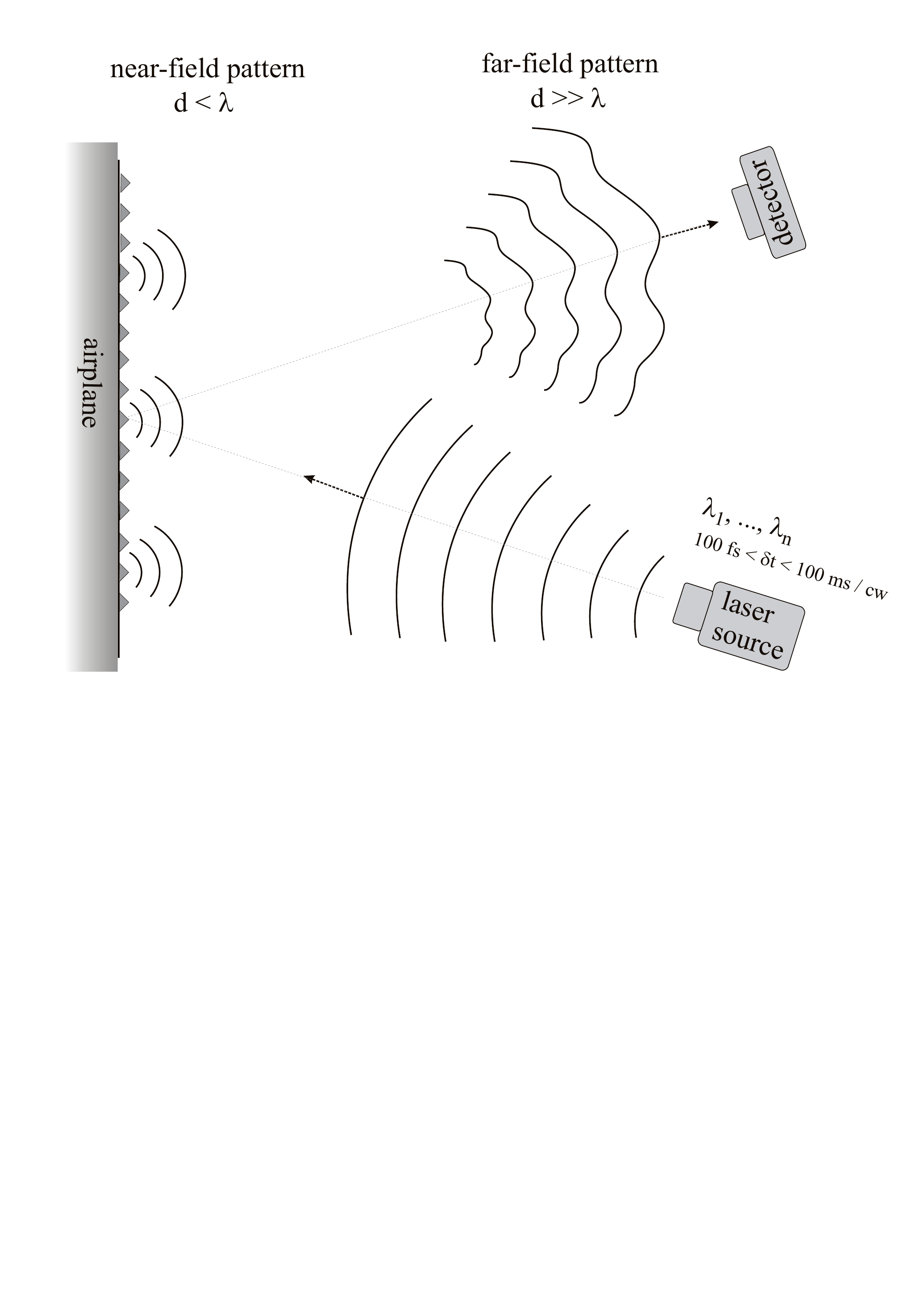}
  \caption{Details of sensor development: principle scheme of wave-coupling, the holographic principle, differences between far- and near-field studies and geometric arrangements of laser source and detector.}
 \label{figure1}
\end{figure}

It is this most-recent state-of the art knowledge on light-matter interactions with micro- and nanostructured dielectrics that has been the starting point and base for our outmost innovative project approach. 
According to the overall task, microstructured surfaces that particularly show aging effects of their structure on the sub-$\mu$m scale were inspected by a simple optical system. 
We have solved this task by combining the analysis of diffraction features and properties based on geometrical optics emerging from the interaction of the structured surface under study with plain electromagnetic waves as sketched in figure \ref{figure1}. 
%We have solved this task by using a controlled electric-field coupling of counterpropagating electromagnetic waves that interact with the structured surface under study. The principle scheme underlying the sensor development is sketched in figure \ref{figure1}.

The resulting setup can easily be transferred to a low-cost, low-weight, hand-held or robot-driven optical system with low power consumption but high precision using state-of-the-art semiconductor laser systems in the near-infrared spectral range.
In comparison to other surface sensitive methods, such as low-coherence microscopy or reflectometry, our approach further allows a rapid scan of the surface, i.e. it allows for short-time structural inspection on the ms-time scale that is advantageous in view of scanning large surface areas and/or repetitive measurements of surface regimes in order to increase the signal-to-noise ratio. The large parameter set enables further amilioration of the signal-to-noise ratio. Further, we gain a more detailed insight to riblet degradation that is originating from particle deposition such as dust inclusions.

% \section{Project assignment}
% 
% Problemstellung aus dem Antrag

\input{theory.tex}

\clearpage

\input{experiment.tex}

\clearpage

\section{Comparison between theoretical and experimental scattering features}

\input{comparison.tex}

\clearpage

\section{Definition of characteristic quantities}

\input{snr.tex}

\clearpage

\section{Device development}

%Device catalogue aus Report I

\input{device.tex}

\clearpage

\input{socio.tex}

\clearpage

\section{Acknowledgment}

The research leading to these results has received funding from the European Union Seventh Framework Programme (FP7\textbackslash 2007-2013) for the Clean Sky Joint Technology Initiative under grant agreement number CSJU-GAM-SFWA-2008-001.

\clearpage
\section{References}
\bibliographystyle{unsrt}
\bibliography{library}

\newpage
\section{Appendix}

\input{appendix.tex}

\clearpage
%\subsection{Presentations}
%\clearpage
%
%%\input{second_report_gender}
%
%%%%%%%%%%%% Die Seitenzahl muss durch pages=- ersetzt werden, dann werden alle seiten kompiliert
%
%\includepdf[pages=-,nup= 1x2, delta=5mm 5mm,templatesize={10cm}{10cm}]{appendix/vortrag_0_monate.pdf}
%
%\includepdf[pages=-,nup= 1x2, delta=5mm 5mm,templatesize={10cm}{10cm}]{appendix/vortrag_6_monate.pdf}
%
%\includepdf[pages=-,nup= 1x2, delta=5mm 5mm,templatesize={10cm}{10cm}]{appendix/vortrag_12_monate.pdf}
%
%\includepdf[pages=-,nup= 1x2, delta=5mm 5mm,templatesize={10cm}{10cm}]{appendix/final_presentation.pdf}

%\includepdf[pages=-,nup= 2x2,frame= true, delta=3mm 3mm]{appendix/vortrag_0_monate.pdf}

%\clearpage
%\input{chapters/5-appendix}
 
% \clearpage
% \section{Presentation information}
% \subsection{date}
% 2013-01-25
% \subsection{presentation}
% The presentation will be presented within an open discussion at IFAM. All responsible project members from UOS that contributed to the results of WP1-WP4 participate at the meeting. The presentation will be distributed as paper-print on the meeting as well as as pdf-file on a CD-ROM.
\end{document}

%% file: theory.tex
\section{Theoretical simulations and considerations}

% \subsection{Tasks of related WPs}
% 
% The task of WP1 for the first six months of the projects was to theoretically
% study the interaction of coherent light waves, this means laser light, with the structured, and especially non-degraded Riblet surfaces.
% Structural changes in the Riblet design and new expertise since the first development of the work packages entered the theoretical considerations.
% An appropriate extension of the proposed model approach described in the proposal was realized.
% Particularly, a combination of wave optics based on the Huygens-Fresnel principle, the wave superposition principle including interference phenomena as well as ray tracing have been taken as a basis for the development of a set of analytical and numerical tools to study the interaction.
% 
% The task of WP1 for the second was to theoretically study the interaction
% of coherent (laser) light waves with degraded Riblet surfaces. Analytical and numerical tools for the description of scattering from non-degraded Riblets are extended by consideration of developments in the experimental/optical sensor setup. The numerical techniques are optimized regarding resolution and step-width. The main aspect of Riblet degradation is implemented by modeling spherical shapes of the apex according to definitions of WP 4.1 (six month project presentation). Predictions for degraded real structures are deduced from the numerical results.

\subsection{Principles of the simulations}

For a simulation of the intensity-apex angle plot, the Riblet parameters have been assumed to be  $50\,\mu$m as Riblet height, $100\,\mu$m as period length and $\alpha = 45^\circ$ as opening angle, as agreed on in the Kickoff meeting (cf. attached protocol) and according to the samples provided by the IFAM.
The desired wavelength of the probe beam lies in the optical and near IR range.
Based on these boundary conditions, the structure size of the diffraction image can be estimated via a set of equations for light-matter interaction with spatially periodic optical lattices.
In the present case of a lattice with spatial frequencies in the regime of $\approx 6\cdot 10^4\,$m$^{-1}$ (\textit{microstructures}),
this results in an expectation value of the diffraction structure size in the order of magnitude of around $0.1^\circ-1^\circ$.
While this is absolutely resolvable in common optical setups (and indeed is resolved in WP2), it opens possibilities for highly accurate solutions exceeding the approaches originally proposed in the WP description.

\subsubsection{Geometrical optics}

For a comprehensive analysis of the geometrical beam path, an appropriate analytical tool has been developed in \textit{Wolfram Mathematica 8.0}.
The script including results is attached to the report.
It has been used to trace the path for the case of perpendicular incidence (described so far) as well as to examine the influence of slanted incidence.

\begin{figure}[ht]
	\centering
	\includegraphics[height=4cm]{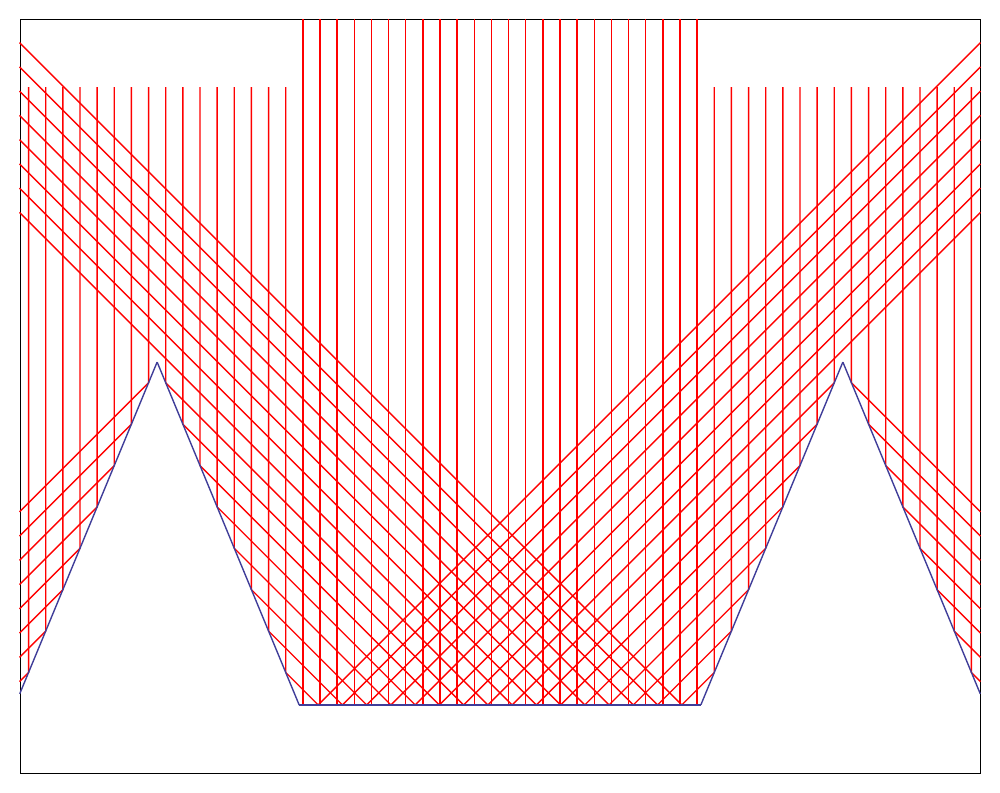}
	\includegraphics[height=4cm]{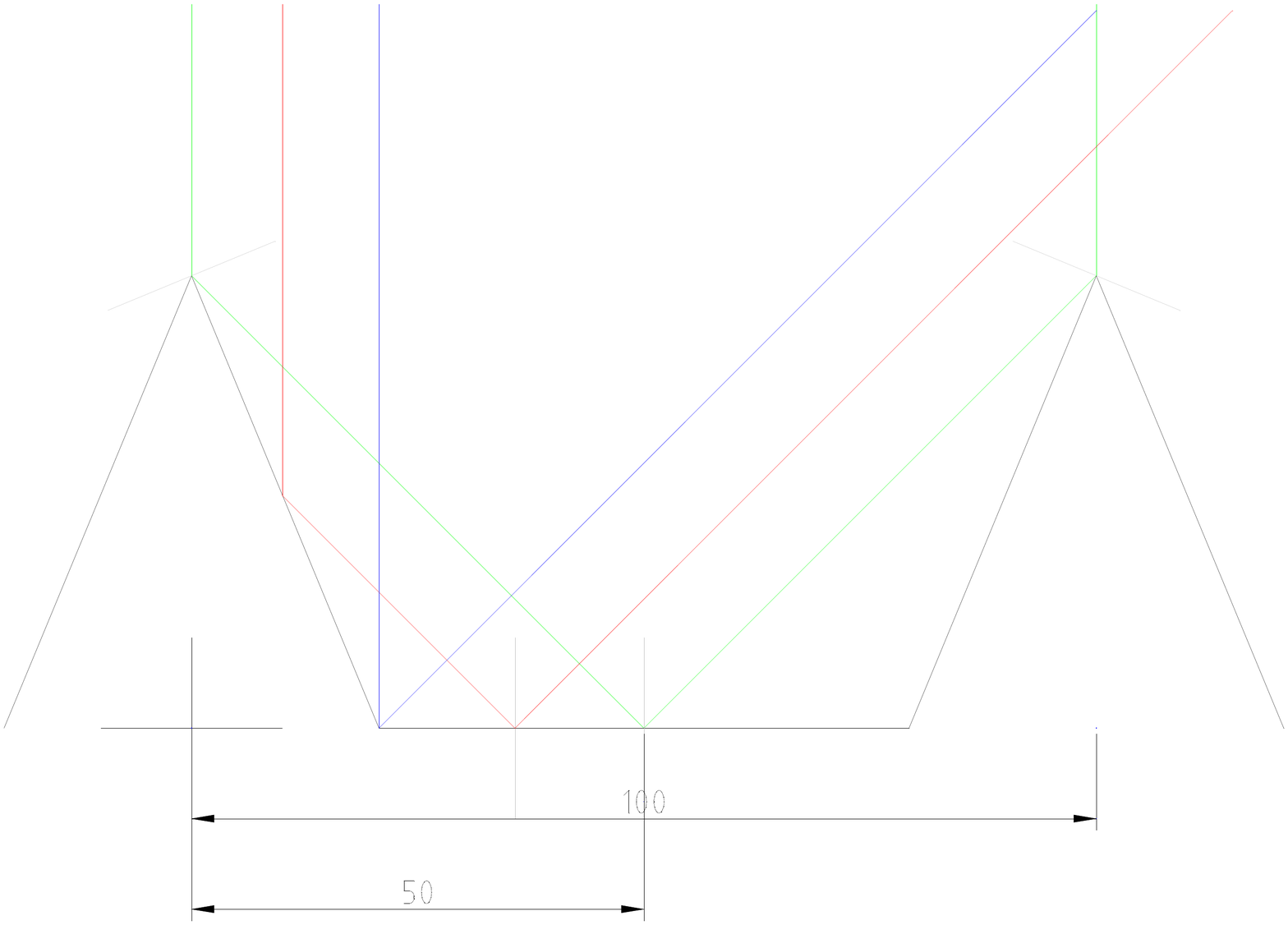}
	\caption{Beam path at $0^\circ$ (perpendicular) incidence, lengths in $\mu$m.}
	\label{fig:beampath0d}
\end{figure}

\begin{figure}[ht]
	\centering
	\includegraphics[height=4cm]{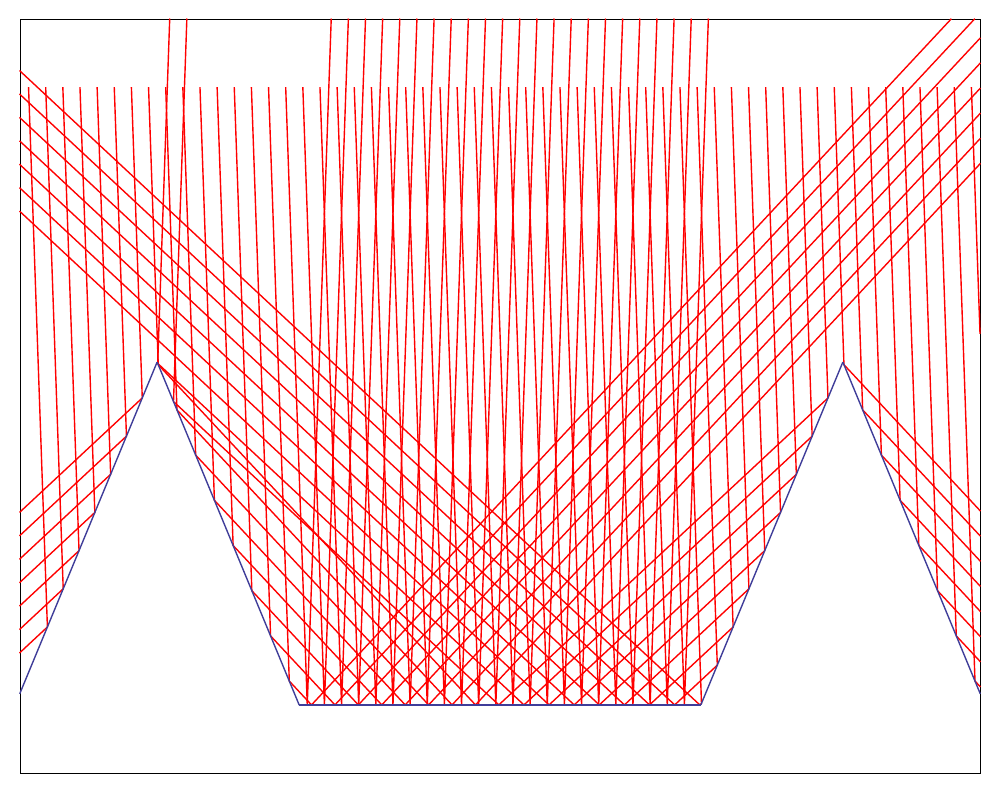}
	\includegraphics[height=4cm]{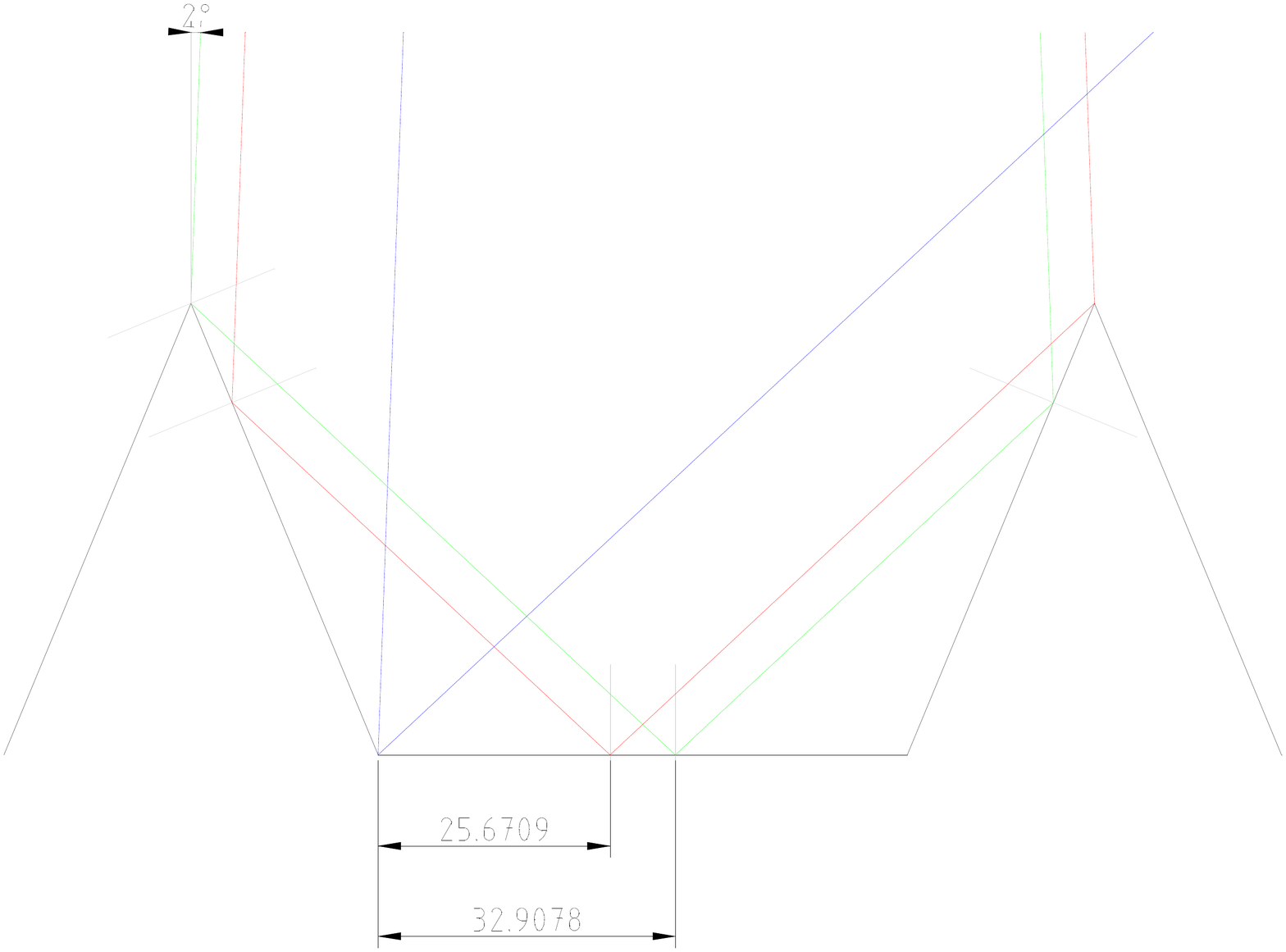}
	\caption{Beam path at $2^\circ$ incidence, lengths in $\mu$m.}
	\label{fig:beampath2d}
\end{figure}

For perpendicular incidence, Fig.~\ref{fig:beampath0d} shows that the complete intensity hitting the flank is in two steps reflected into directions of $\beta_1$.
Thus the spot at $\beta_0$ solely carries information on the plains between the Riblets, while the one at $\beta_1$ solely carries information on the Riblet flanks.
However, even slight deviations from perpendicular incidence lead to considerable shadowing effects.
Fig.~\ref{fig:beampath2d} shows an intensity loss of 22\,\percent in $\beta_1$-directions at a tilted incidence by 2$^\circ$ due to reflection in additional signal spots.

As a result of the beam path analysis, a carefully adjusted perpendicular incidence is strongly recommended in the view of a pronounced signal-to-noise ratio.
However, signals from tilted incidence may provide a more detailed insight into the degraded structure (e.g. the type of degradation may be deduced) and, thus, will be reconsidered within follow-up projects.

\subsubsection{Fresnel-Huygens diffraction theory}

To create a reference image a semi-analytical tool has been developed using \textit{Wolfram Mathematica 8.0} for fast just-in-time simulation of diffraction patterns starting from the Huygens-Fresnel principle by numerically summing over a finite number of spherical waves with distances smaller than the wavelength.
The images presented here have been calculated with 200 waves per $\lambda$, i.e. $\approx 3\cdot 10^8\,$m$^{-1}$ on a \textit{Dell Precision T7400} workstation.
The script including results is attached to the report.

For the simulations, a setup has been modeled in which a spatially restricted ($d=50\text{ periods}=5$mm), isotropic plain wave front with a wave vector perpendicular to the surface hits the sample.
In accordance with the experimental work package, the wavelength has been chosen in the near-infrared spectral range, exemplarily shown for $633\,$nm in the following.
The scattering pattern then consists of a set of spherical waves with origins on the surface, amplitudes determined by the local reflectivity and a phase offset according to the length difference of the geometrical beam path. Wave superposition and interference are considered for the calculation of the field amplitudes as a function of a coordinate parallel to the sample in a constant distance $a$. This geometry allows for considerable speed improvements of the calculations, due to the fact that only the diffraction pattern of one riblet has to be calculated. The full pattern may then be derived via shifted superposition of this pattern.
To apply these results to experimental data measured using different geometries, maps are introduced in the software. The files attached to this report include maps to circular geometries (constant distance to illuminated riblet spot, as used in original setup) and linear screens tiltet to the sample (as used in final setup).
  
This method results in three spots centered at $\beta_0=0^\circ$, $\beta_{+\bar{1}}=135^\circ$, and $\beta_{-\bar{1}}=225^\circ$. Different spatial extends of the beam have been simulated ($1\,$mm $ < d < 10\,$mm) and found to have no influence on the overall structure.

The three angles can be related to the angles of the Riblet surface planes involved ($0^\circ, \pm 67.5^\circ$) and constitute the conclusion from the geometrical considerations that reflection of the incoming probe beam has to be taken into account in addition to beam diffraction. In more detail, the probe beam is reflected both from the planes between the Riblets into the direction $\beta_0$ and from the flanks of the Riblets into $\beta_{\pm\bar{1}}$ direction.
Notice, however, that the beams in $\beta_{\pm\bar{1}}$-directions obey one or more additional interactions with the surface in the further beam path (cf. ray tracing discussed above) and the scattering spots will appear at rather different angles, experimentally (cf. Fig.~\ref{fig:results_overview}).

\begin{figure}[h]
	\centering
	\includegraphics[width=0.4\textwidth]{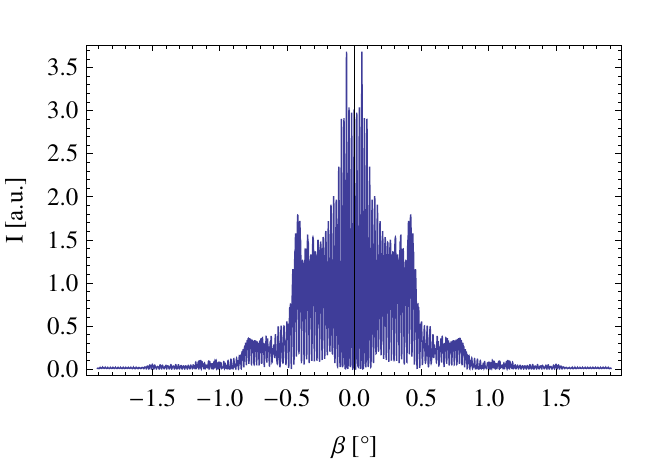}
	\includegraphics[width=0.4\textwidth]{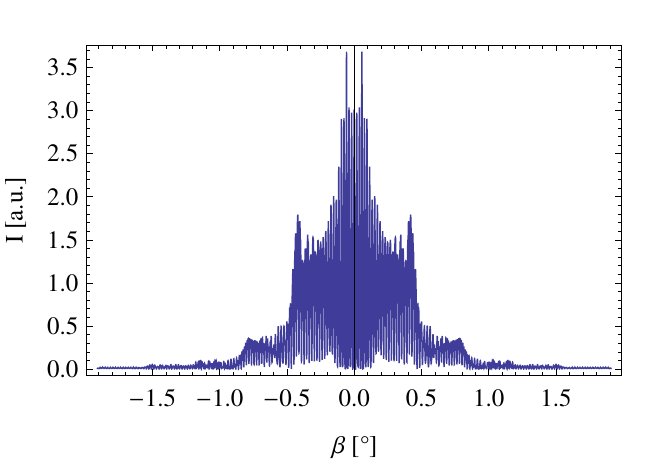}
	\caption{Intensity distribution around $\beta_0=0^\circ$ simulated on the basis of the Huygens-Fresnel principle. $x$-axis apex angle in $^\circ$, $y$-axis intensity in arbitrary units. Left side: intensity distribution from waves solely originating from the planar plains between the Riblets. Right side: result of wave superposition from the entire surface structure. The two images are found to be identical.}
	\label{fig:0orderHF}
\end{figure}
In a second step, reflection processes are considered in addition. For this purpose, the diffraction image related to (a) the planes between the Riblets and (b) to the flanks of the Riblets are treated in separate calculations. Results for (a) are depicted in Fig.~\ref{fig:0orderHF} (left), i.e. light is transmitting the structure in the areas of the Riblets. For comparison the right plot shows the calculation for the case that the entire surface structure is considered. Remarkably, no difference is found between the two plots, thus supporting the model approach of distinct treatment of the scattering spots.

\begin{figure}[h]
	\centering
	\includegraphics[width=0.9\textwidth]{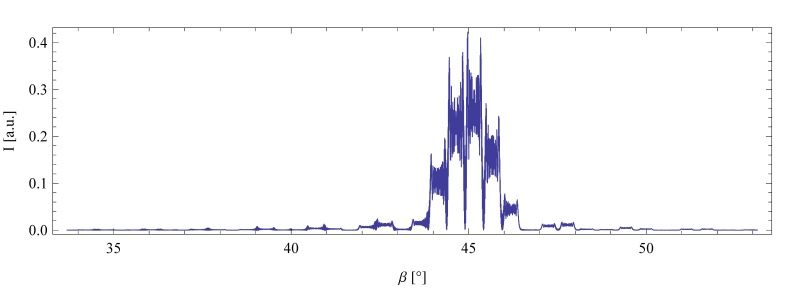}
	\caption{Intensity distribution around $\beta_1$, simulated on the basis of the Huygens-Fresnel principle. $x$-axis apex angle in $^\circ$, $y$-axis intensity in arbitrary units.}
	\label{fig:1orderPlainHF}
\end{figure}

In the next step, calculations are performed for case (b). For this purpose, a second interaction of the beams $\beta_{\bar{\pm 1}}$-directions needs to be considered (cf. ray tracing discussed below). It leads to an intensity distribution with two clearly distinguishable spots at  $\beta_1=\pm 45^\circ$, besides $\beta_0=0^\circ$. The intensity distribution with additional consideration of scattering around $\beta_1$ is shown in Fig. \ref{fig:1orderPlainHF}.

\begin{figure}[ht]
	\centering
	\includegraphics[width=0.9\textwidth]{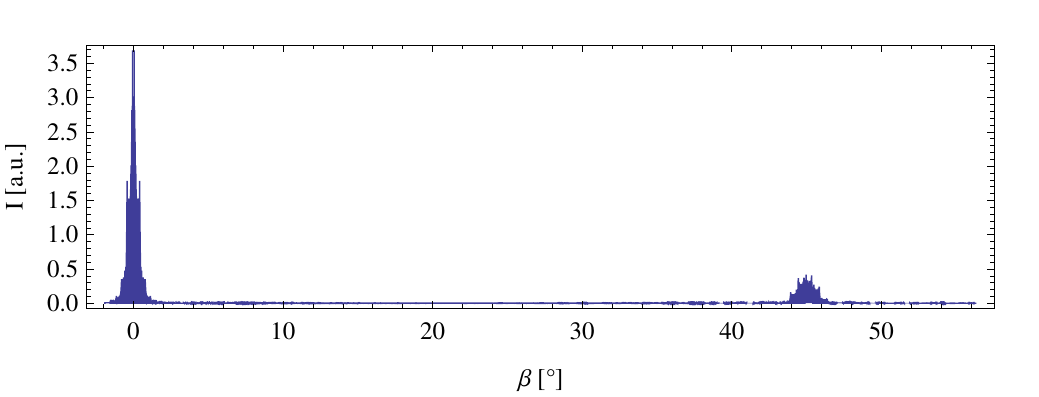}
	\caption{Full intensity distribution simulated on the basis of the Huygens-Fresnel principle. 1. axis apex angle in $^\circ$, 2. axis intensity in arbitrary units.}
	\label{fig:fullPatternHF}
\end{figure}

Combining the wave package development from the structure from Huygens-Fresnel principle, the wave superposition including interference and ray tracing for the full Riblet structure, we end up in the final angular intensity distribution that is depicted in Fig.~\ref{fig:fullPatternHF}.
Because intensities and scattering angles are deduced from the calculations, the plot can be directly compared with the appropriate data of the experimental study (cf. Fig.~\ref{fig:results_overview}).

\subsection{Theoretical implementation of riblet degradation}

\begin{figure}[ht]
	\centering
	\includegraphics[width=8cm]{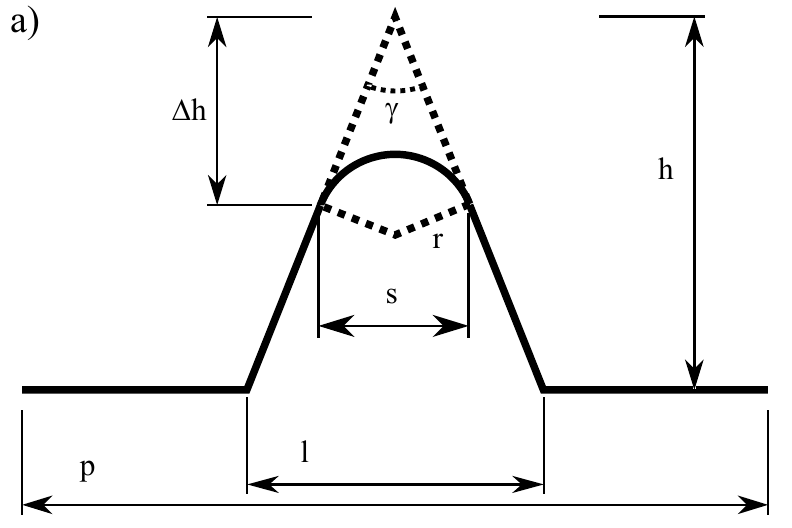}
	\includegraphics[width=8cm]{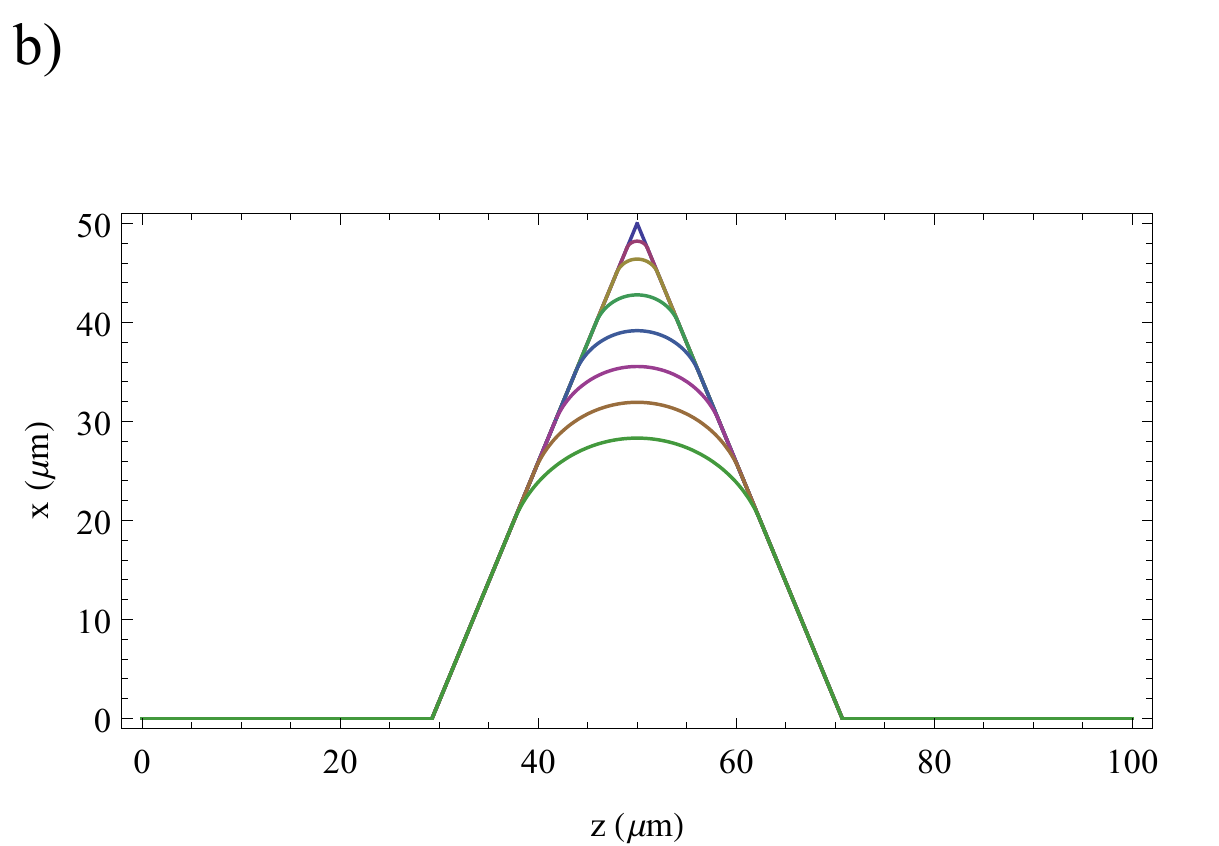}
	\caption{Schematical sketch of the model approach for the implementation of Riblet degradation in the numerical simulations. (a) Principle shape of the Riblet cut and its characteristic measures. The dotted and straight lines mark the non-degraded and degraded Riblet shapes. (b) Right figure shows structures for $d=.05$, $.1$, $.2$, $.3$, $.4$, $.5$, $.6$.}
	\label{fig:riblet-deg}
\end{figure}

Degradation of Riblet structures is implemented by modeling a spherical shape of the Riblet apex while keeping the slope of the lateral faces as shown in the sketch of a Riblet profile in Fig.~\ref{fig:riblet-deg}a). Model parameters are the radius of the sphere $r$ and the riblet width $s$ at the height ($h-\Delta$h). Here, h is the height of the non-degraded Riblet, and $\Delta$h denotes its decrease by degradation. This yields a measure for the degree of degradation $d$ that is given by the quotients $s/l=2r\cos(22.5^{\circ})/l$ and $\Delta h/h$ with $l$ the base width of the Riblet and $0\leq d\leq 1$ .

The non-degraded Riblet is characterized by $p=96\,\mu$m, $h=48\,\mu$m and a tip angle of $\gamma=45^\circ$.
Numerical calculations are performed with discrete values of the degree of degradation $d=0.05$, $0.1$, $0.2$, $0.3$, $0.4$, $0.5$, $0.6$, i.e. between 5$\percent$ and 60$\percent$.
The corresponding shapes are depicted schematically by means of Riblet cuts in Fig.~\ref{fig:riblet-deg}b)

\begin{infobox}[]
The interaction between light and periodical structure is modeled using a combination of spherical waves -- following the Huygens--Fresnel principle -- and geometrical optics. The optical path length to the final position of interaction is calculated via ray-tracing and subsequently introduced to the phase of the emitted spherical wave. 
\end{infobox}

\subsection{Results of simulation}

\subsubsection{Calculation on experimental parameters}

According to the geometry changes in the experimental setup and for reasons of a consistency check, calculations on the spatial resolution that allow for an appropriate detection of the scattering intensity profiles were performed.
The required spatial resolution depends strongly on the diameter of the laser beam used:
For a beam diameter of $1\,$mm, as used in the experimental setup, the simulation yields a spatial distance between two intensity peaks in the scattering fine structure of $\approx10\,\mu$m. This distance decreases as a function of increasing beam diameter, e.g. with a laser beam of $5\,$mm diameter we get peak-distances in the scattering fine structure of $\approx1\,\mu$m.

At the same time, the fine structure of the scattering pattern does not include relevant features (from the current point of view) and, hence, does not have to be resolved completely by the detection system.
Pinholes that are mounted on the Si-PIN detector diodes can be chosen with a diameter larger than the peak-distances in the scattering fine structure without loosing significant information on the Riblet shape and its degree of degradation.

In the simulations, the overall shape of the scattering features is best captured using a maximal pinhole size of $10\,\mu$m for $5\,$mm beam diameter and $50\,\mu$m for $1\,$mm.

\subsubsection{Influence of degradation on scattering pattern}

The $0^\circ$ peak is caused by light reflected from the plains between the Riblets, while the $\pm45^\circ$ peaks emerge from a two-step reflection from the flanks of the Riblets (step 1) and subsequent reflection from the plains (step 2).
Consequently, from the theoretical point of view, no major changes of the scattering pattern around $0^\circ$ are expected if Riblet degradation is considered, i.e. $0<d<1$.
Reflection from the spherical tips of the degraded Riblets will cause a noise background that superimposes the scattering intensity pattern in the entire sphere.

The contributions of scattered light into directions of $0$ and $45^{\circ}$ can be analyzed in very detail from the simulation.
Table~\ref{table1} summarizes the intensity fractions of the incoming light that are scattered from different areas of the degraded Riblet into different directions:
(i) denotes the fraction of the incoming intensity that impinges onto the spherical tips and not onto the plains between the Riblets or their flanks.
This intensity will be distributed into a wide angle range of about -135$^\circ$ to 135$^\circ$ and can be sub-divided into three parts:
\begin{enumerate}
\item The angular range between -55$^\circ$ to 55$^\circ$. Here, scattered waves are captured by the Si-PIN detectors without a second step of scattering or reflection. Its integral intensity related to the incoming intensity is given in (ii).
  \item The angular range $\pm$90$^\circ$ to $\pm$135$^\circ$, that originate from waves scattered within a two- or three-step process. E.g., the incoming wave is first scattered in a direction of a neighboring Riblet, then into the sensor via a second and sometimes third reflection process.
  The corresponding integral intensity fraction related to the incoming intensity is given in (iii). Note that this intensity is further reduced significantly in case of lower reflectivities.
  \item The angular range between $\pm$55$^\circ$ to $\pm$90$^\circ$, that is not contributing to the scattering background.
\end{enumerate}

\begin{table}[h]
\begin{center}
    \begin{tabular}{ | p{5cm} || c | c | c | c | c | c |}
    \hline
     & $5\percent$ & $10\percent$ & $20\percent$ & $30\percent$ & $40\percent$ & $50\percent$ \\ \hline\hline
     (i) Fraction of incoming intensity hitting the spherical tips & $2.1\percent$ & $4.1\percent$ & $8.3\percent$ & $12.4\percent$ & $16.6\percent$ & $20.7\percent$ \\ \hline
     (ii) Integral scattering intensity between -55$^\circ$ to 55$^\circ$& $1.0\percent$ & $2.1\percent$ & $4.1\percent$ & $6.2\percent$ & $8.3\percent$ & $10.4\percent$ \\ \hline
     (iii) Integral scattering intensity between $\pm90^\circ$-to $\pm135^\circ$ & $0.5\percent$ & $1.0\percent$ & $1.9\percent$ & $2.9\percent$ & $3.9\percent$ & $4.9\percent$ \\ \hline
    \end{tabular}
    \caption{Intensity fractions of incoming and scattered waves related to the spehrical tip and different angular regimes. Detailed explanation is given in the text.}
    \label{table1}
\end{center}
\end{table}

\begin{figure}[ht]
	\centering
	\includegraphics[width=8cm]{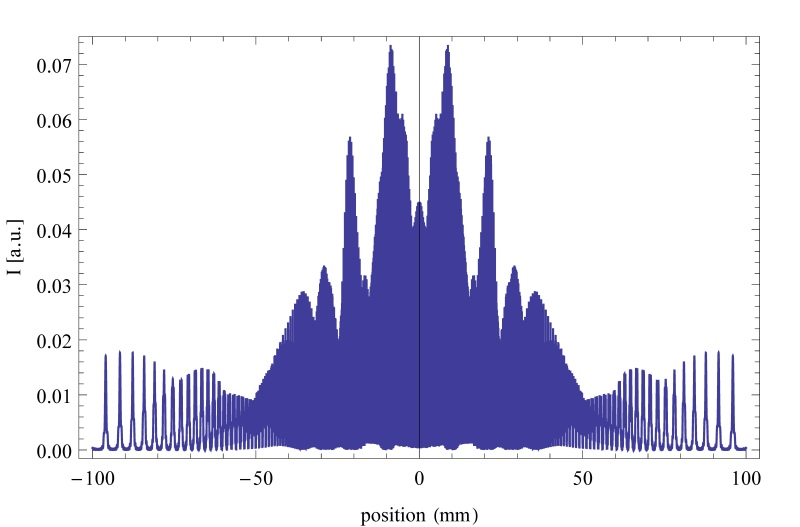}
	\includegraphics[width=8cm]{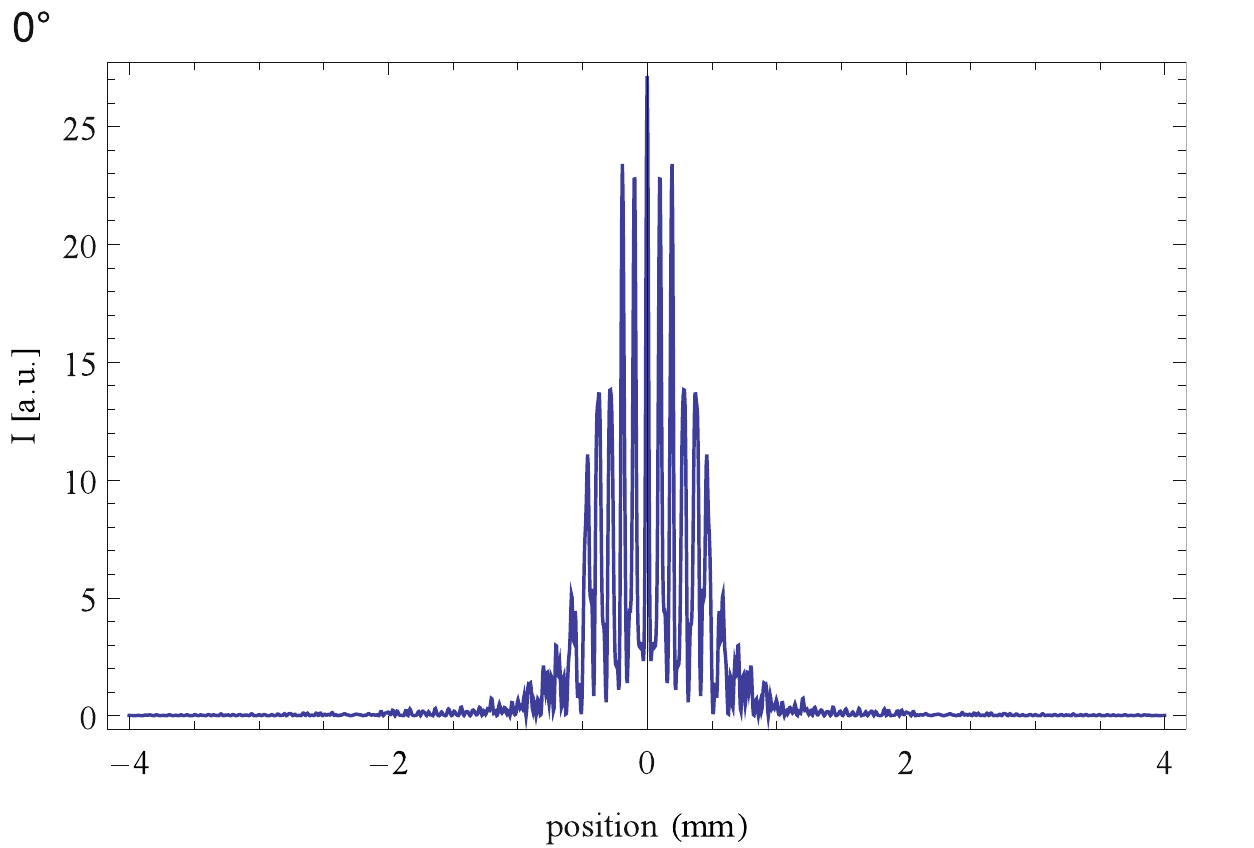}
	\caption{Left figure: Scattering pattern simulated solely for scattering from the spherical tips of a degraded Riblet structure with $d=50\percent$ in a plane-screen distance of $4\,$cm. Right figure shows full scattering intensity around $0^\circ$ for comparison of magnitudes.
	%The intensity axis is comparable to Fig.\,\ref{fig:pattern-undeg}.
	}
	\label{fig:pattern-tops-d50}
\end{figure}

A simulation of the scattering intensity distribution originating solely from the spherical Riblet tips according to (ii) and, exemplarily, for a degree of degradation of $d=0.5$ is depicted in Fig.\,\ref{fig:pattern-tops-d50}.
The spatial coordinate corresponds with an angular regime of about $\pm70^{\circ}$.
The scattering intensity is determined to be more than two orders of magnitude smaller compared with the scattering intensities obtained for a non-degraded pattern.
Thus, the contribution of the rounded tips to the overall scattered intensity can be neglected, but determines the background noise.
The same can be concluded for the scattered waves from the spherical tips that propagate to the next-neighboring Riblet and are scattered subsequently within a two- or three-step process.

\begin{figure}[ht]
	\centering
	\includegraphics[width=8cm]{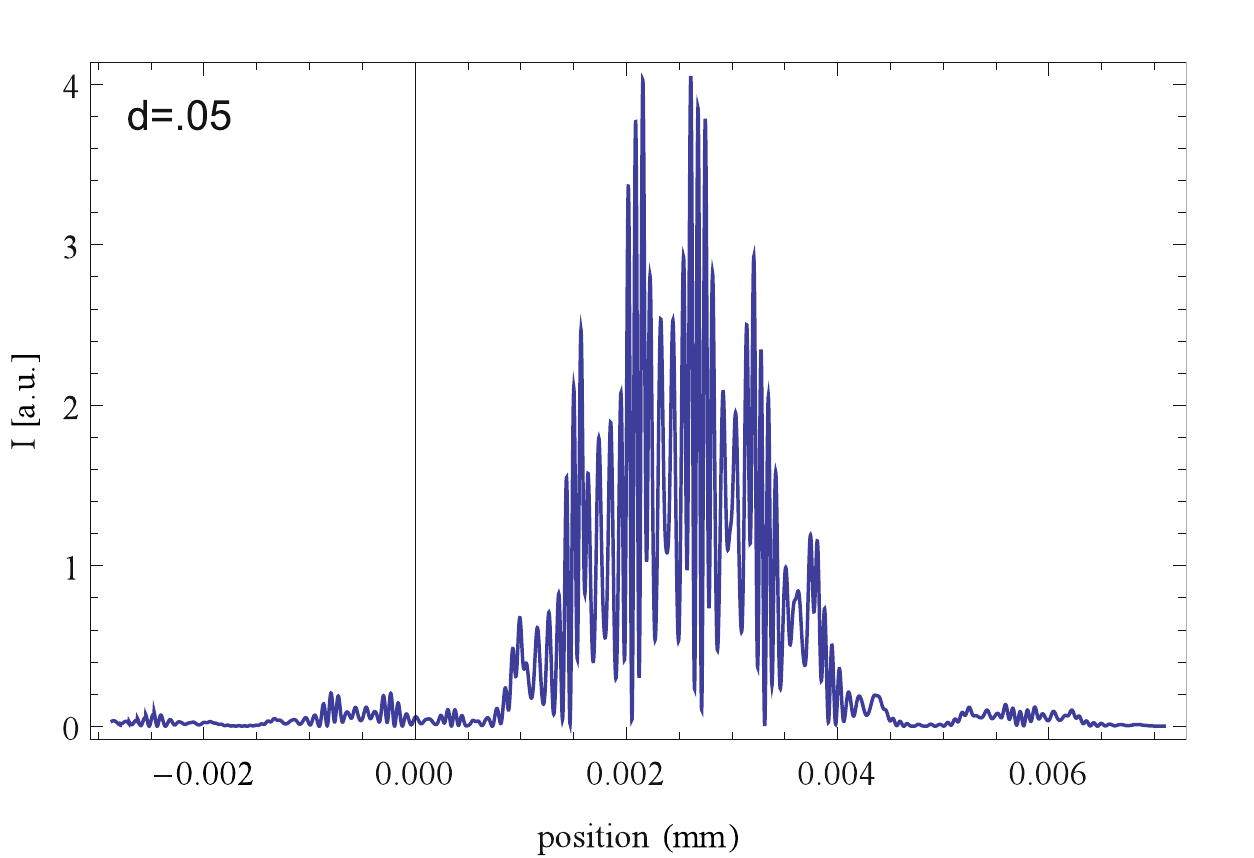}
	\includegraphics[width=8cm]{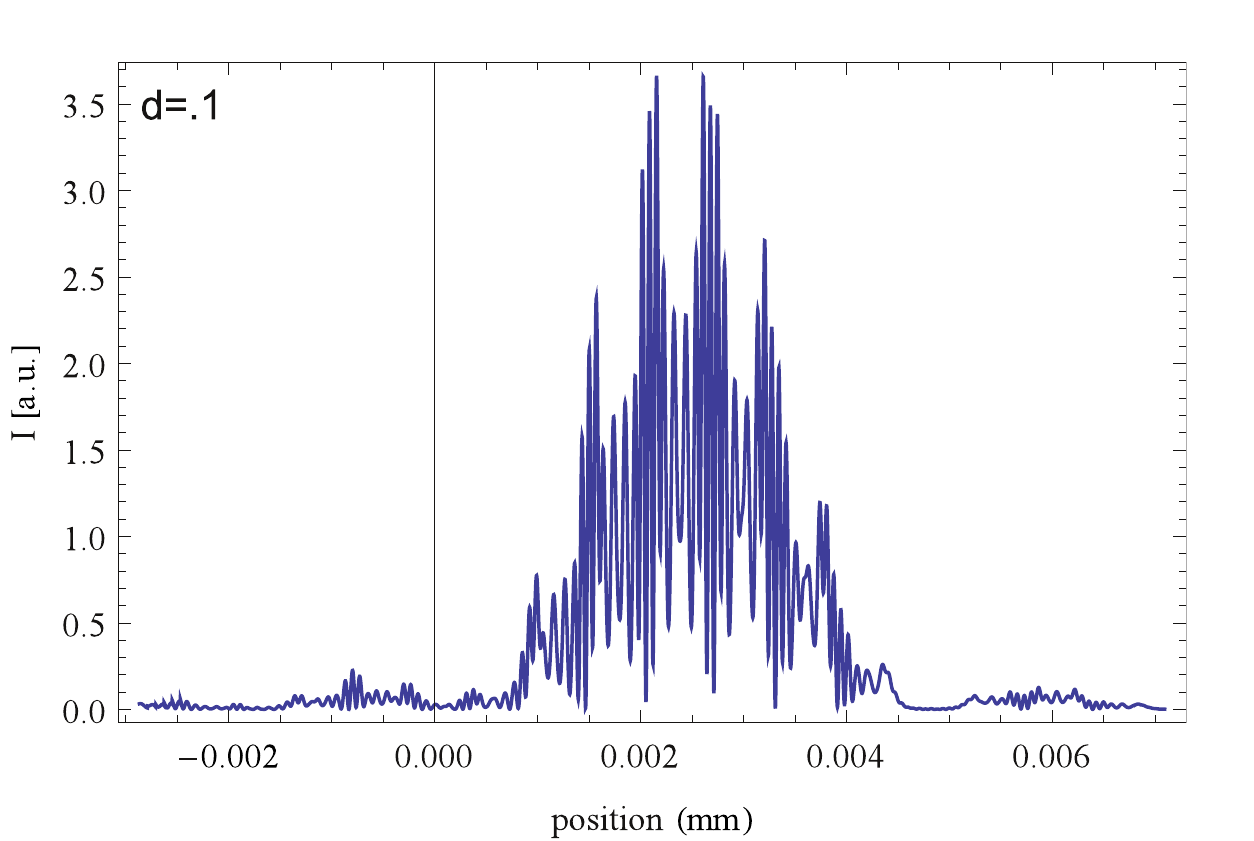}
	\includegraphics[width=8cm]{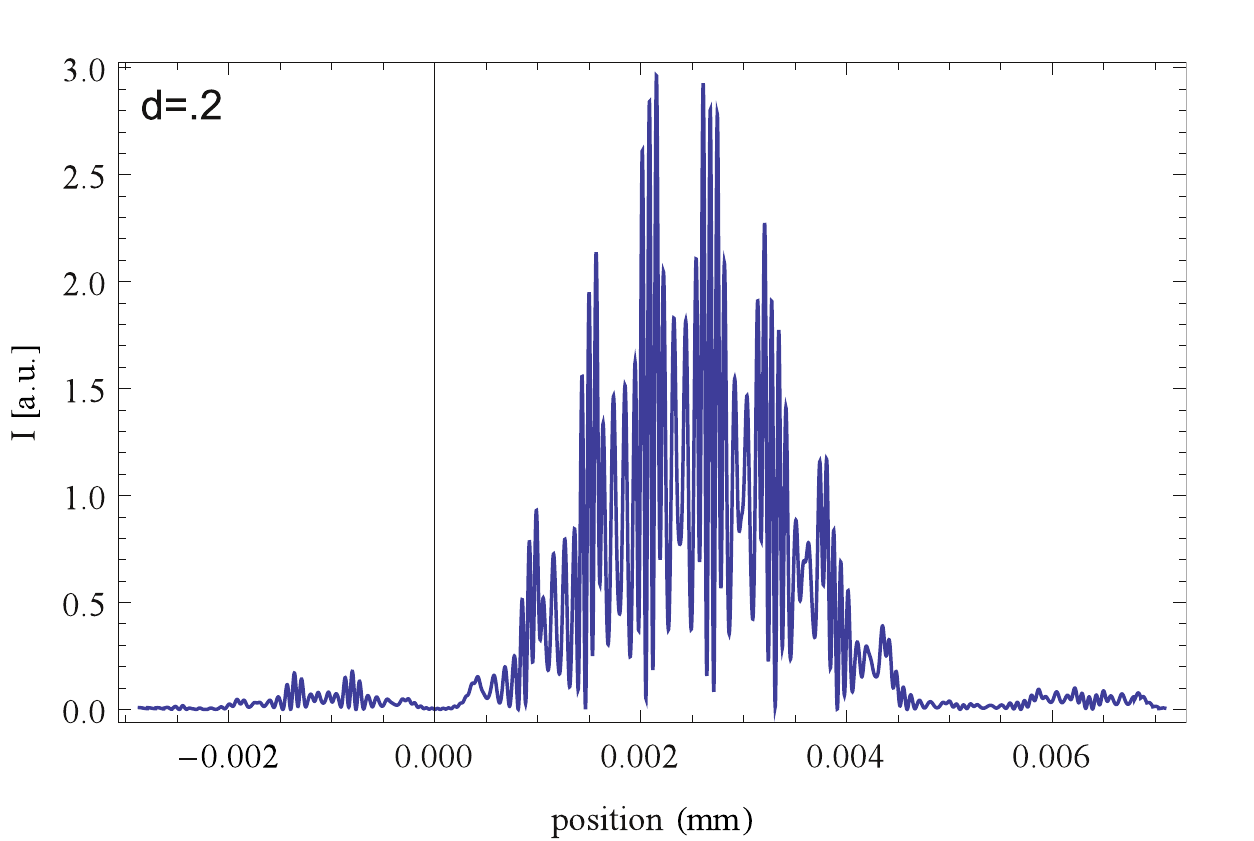}
	\includegraphics[width=8cm]{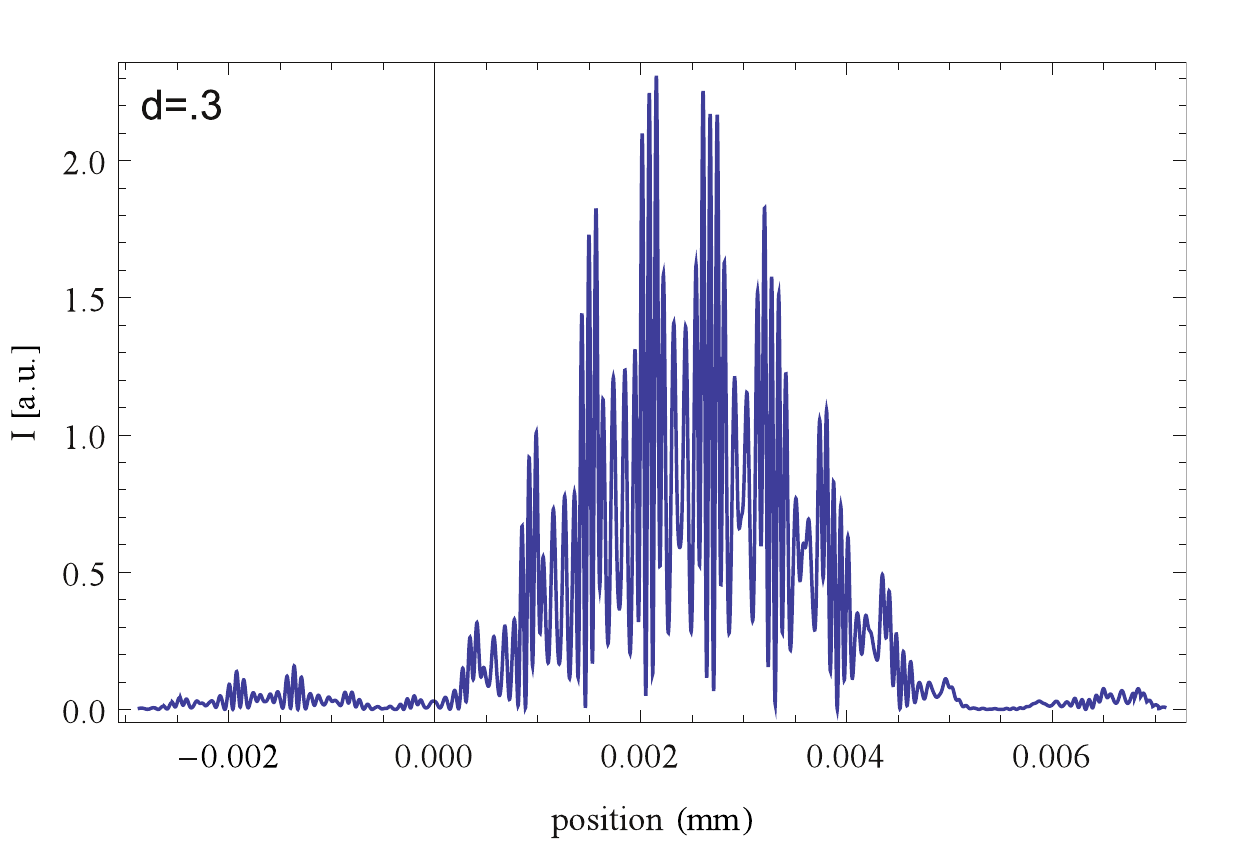}
	\includegraphics[width=8cm]{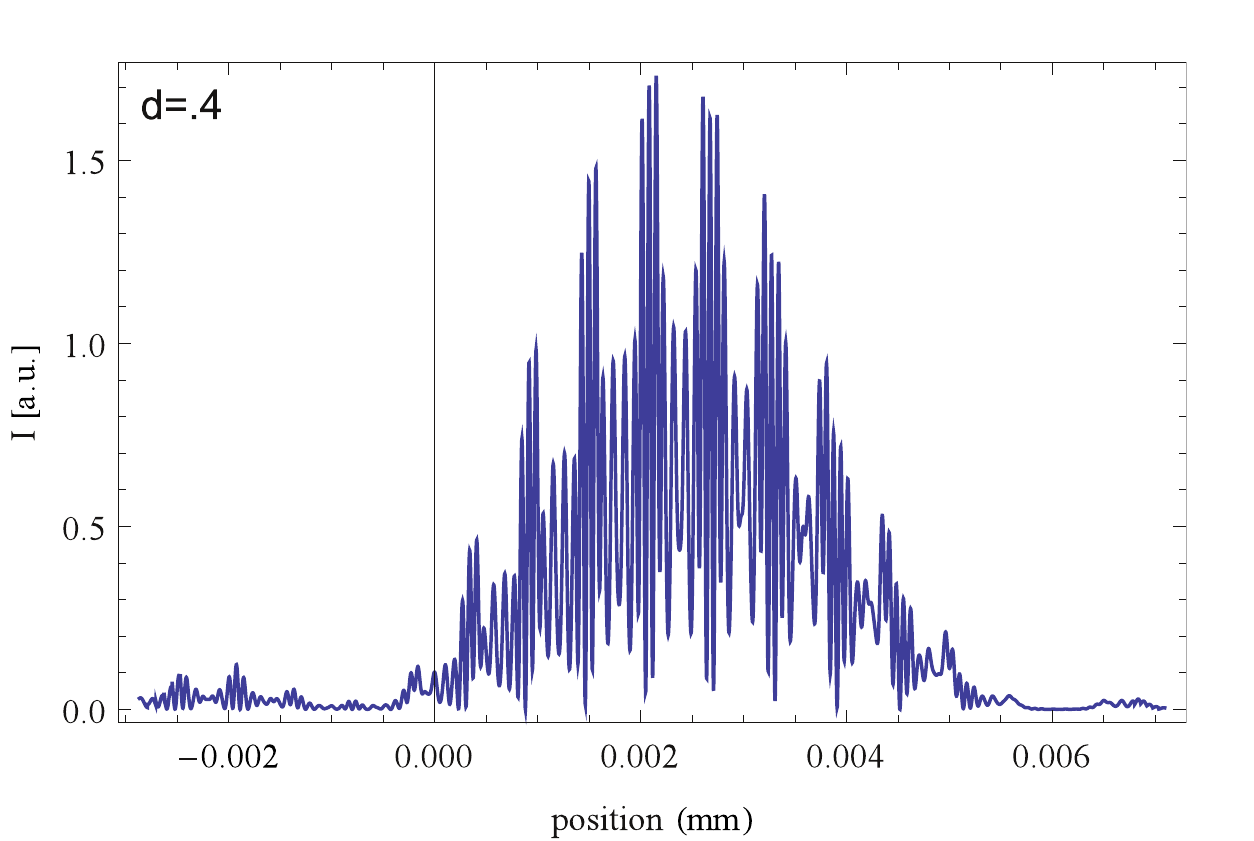}
	\includegraphics[width=8cm]{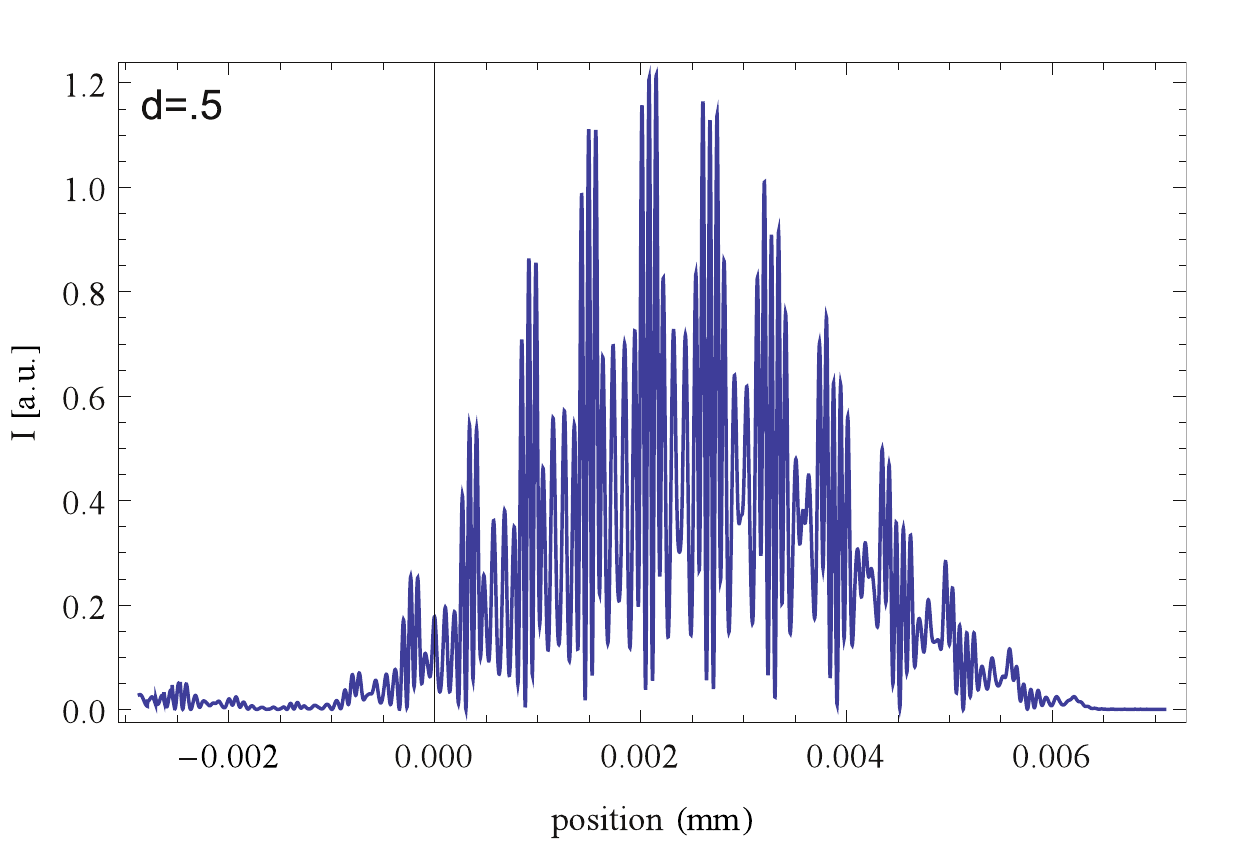}
	\caption{Scattering pattern in the first order. Left to right, top to bottom: $d=0.05$, $0.1$, $0.2$, $0.3$, $0.4$, $0.5$, $0.6$.}
	\label{fig:pattern-deg}
\end{figure}

Fig.\,\ref{fig:pattern-deg} shows the results of the numerical calculations of the scattering pattern in the $\pm45^{\circ}$-direction for different degrees of degradation $d$.
We note that the intensity has been re-scaled in the plots according to the respective intensity maxima, that decrease by a factor of $\approx 4$ from $d=0.05$ to $d=0.6$.
At the same time, the FWHM of the scattering feature (measured on the envelope) increases linearly by $\approx 150\percent$ from $d=5\percent$ to $d=50\percent$.
As the distance of the substructure peaks is maintained, more peaks are observed on the degraded structure.

The most dominating changes  by degradation in the scattering pattern, however, appear due to the loss of the scattering cross section at the Riblet flanks.
Here, the increase of the degree of degradation results in a considerable decrease of the effective area of the Riblet flank for the scattering of the incoming light wave.
Thus, the overall scattering intensity in $\pm45^{\circ}$ direction is affected.
Thales' theorem states that the intensity reflected on these surfaces decreases proportional to $d$.
Indeed, the integrated intensity of the respective scattering pattern decreases proportional to the degree of degradation, i.e. with the factor $(1-d)$.

Apart from this, no other characteristic changes in the features are recognizable.
This is not surprising, because the structure that generates the pattern - i.e., the planes and the remaining part of the flanks - do not change fundamentally.

\subsubsection{Influence of non-ideal degradation}

In practical situation, perfectly smooth spheres as reflected by the spherical tips are unlikely.
More probable is a type of degradation causing general roundings and more or less rough surfaces.
From the current understanding of the relation between scattering pattern and Riblet shape we can conclude that in case of real degradations (and in comparison to the considerations of spherical tips presented above):
\begin{itemize}
\item the principle scattering feature and its fine structure in $\pm 45^{\circ}$-direction will not be altered
\item the dependence of the loss of overall scattering intensity in $\pm 45^{\circ}$-direction on the degree of degradation will remain and may even be amplified
\item a more uniform distribution of the noise background and its overall intensity will appear.
\end{itemize}
These considerations are valid for a huge variety of different types of real degradation with different shapes, surfaces and regularities. We note that ideally flat and smooth surfaces are excluded, here, i.e., degradations that can be understood as significant changes of the principle Riblet geometry.

\begin{infobox}[]
The most dominating effect of idealized degradation is is a decrease of intensity in the $\pm45^\circ$ features coupled with an increase of the overall background intensity. The $0^\circ$ feature is ideally unchanged. In the case of non-ideal degradation -- which is more probable in practical situations -- this basic behavior is maintained and may even be ampified.
\end{infobox}

%% file: experiment.tex
\section{Experimental setup and data}

% \subsection{Tasks of related WPs}
% 
% 
% The task of WP2 for the first six months of the projects is to optically study the interaction of coherent light waves, this means laser light, with the structured, and especially non-degraded riblet surfaces. For this purpose an optical setup was installed to perform the measurements needed to achieve milestone no. 2.1, the comparison of experimental and theoretical pattern of non-degraded riblets. Therefore measurements of intensity vs. apex angle as a function of probe wavelengths as well as beam diameter and sample geometry have been performed.
% 
% The task of WP2 for the second six months of the projects is to optically study the interaction of laser light with the structured and in this case especially degraded riblet surfaces. For this purpose changes to the optical setup were made to perform the measurements needed to achieve milestone no. 2.2: the comparison of experimental and theoretical pattern of degraded riblets. For this purpose, measurements of scattered laser light intensity vs. position have been performed.
% 
% The task of WP2 for the last six months of the projects is to optically study the interaction of laser light with the structured and in this case degraded riblet surfaces with different degrees of degradation. The measurements are performed with the optimized laser system. All this leads to the installment of a preliminary configuration of an optical setup in an optical lab at UOS, which is the final milestone no. 2.3at the end of the project.

\subsection{Studied samples}

Measurement of degraded and undegraded Riblet structures were performed on three sets of samples. Each set was specially chosen for the respective measurements needed in the projects time frame. All samples were provided by IFAM.

\paragraph{Set 1}

Set 1 consists of four samples to first study the undegraded riblet structure and then the difference to a degraded structure. These samples were also used to perform all needed fundamental characterization of both future samples as well as the experimental setups.
An overview is given in Table \ref{tab:samples}. Samples 1 and 2 are unstructured; sample 1 is unstructured riblet paint on TOPCOAT, sample 2 on metal. Samples 3 and for 4 have the riblet paint structured, again sample 3 is on TOPCOAT, sample 4 on metal. The geometry of the riblet structure itself is provided by IFAM. For the studied samples the riblet height is $50\,\mu$m, the period length is $100\,\mu$m and the opening angle is $\alpha = 45^\circ$. On all samples, spatial variations of riblet paint can be observed over the coated area of the sample of $\approx 150$\,cm$^2$, for details see the photographs in the appendix. These variations make it difficult to compare absolute values of of reflected intensity measured at different spots on the sample surface. However, a qualitative comparison is possible.

 \begin{table}[h]
	\centering
\begin{tabular}[h]{|c|c|c|c|}
\hline

\multirow{2}{1cm}{sample no.} & \multirow{2}{2.5cm}{surface} & \multirow{2}{2cm}{base material} & \multirow{2}{2.5cm}{sample size coting size} \\
& & &  \\
\hline
\multirow{2}{1cm}{1} & \multirow{2}{2.5cm}{unstructured} & \multirow{2}{2cm}{TOPCOAT}
& \multirow{2}{2.5cm}{(10.5\,$\times$\,19)\,cm\\(8\,$\times$\,14)\,cm} \\
& & &  \\
\hline
\multirow{2}{1cm}{2} & \multirow{2}{2.5cm}{unstructured} & \multirow{2}{2cm}{metal}
& \multirow{2}{2.5cm}{(10.5\,$\times$\,19)\,cm\\(8\,$\times$\,15)\,cm} \\
& & &  \\
\hline
\multirow{2}{1cm}{3} & \multirow{2}{2.5cm}{riblet} & \multirow{2}{2cm}{TOPCOAT}
& \multirow{2}{2.5cm}{(10.5\,$\times$\,19)\,cm\\(7.8\,$\times$\,16.2)\,cm} \\
& & &  \\
\hline
\multirow{2}{1cm}{4} & \multirow{2}{2.5cm}{riblet} & \multirow{2}{2cm}{metal}
& \multirow{2}{2.5cm}{(10.5\,$\times$\,19)\,cm\\(7.5\,$\times$\,16.5)\,cm} \\
& & &  \\
\hline
\end{tabular}
\caption{Overview of properties like surface structure, base material and sizes of non-degraded samples provides by IFAM.}
\label{tab:samples}
\end{table}

\begin{figure}[ht]
	\centering
	\includegraphics[height=8cm]{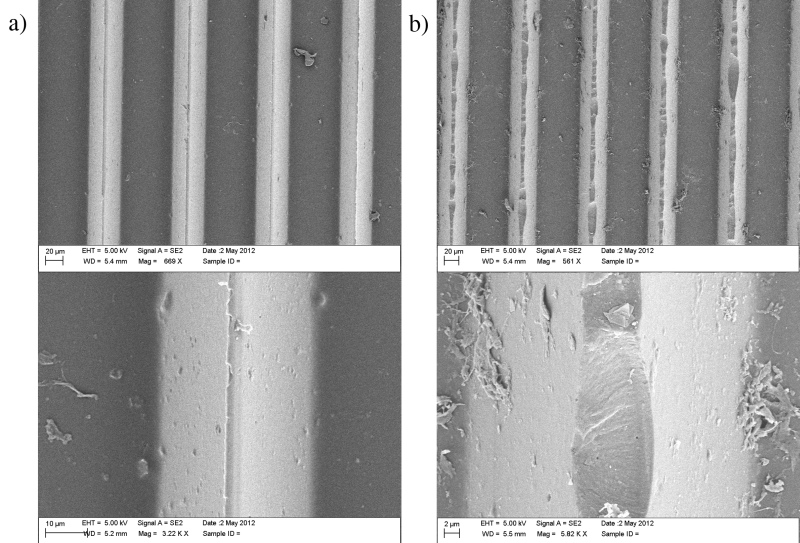}
	\caption{SEM data for a non-degraded (a) and degraded (b) riblet structure. Degradation was achieved by scrubbing with acetone for 14 minutes prior to the SEM measurement. The top images give an overview while the bottom images show a magnification on one tip of the riblet structures.}
	\label{fig:sem}
\end{figure}

Some of the samples of set 1 have been scrubbed with acetone to simulate the degradation process by removing the tip of the riblet. Two samples of this set have been coated with a thin (approx. 3-5$\,$nm) platinum-iridium layer to perform SEM measurements in order to estimate the degree of degradation from a geometrical inspection. Figure~\ref{fig:sem} shows two SEM results from an untreated sample (a) and a sample scrubbed with acetone for 14 minutes (b).

In the image of the untreated, non-degraded sample one can see the nearly perfect structure of the Riblets. The small asymmetry at the top of the Riblet structure, which can be seen in the lower part of Figure~\ref{fig:sem}(a), can be attributed to a shadowing artifact of the measurement. Therefore the riblet can be treated as having a nearly perfect triangular structure.

The image of the treated sample, Figure~\ref{fig:sem}(b), shows the type of degradation which can be seen as typical according to the information of the kick-off meeting. The tip of the riblet is degraded, reducing the height of the riblet while leaving the rest of the structure mostly intact. This is especially correct for the sides/flanks of the riblets and the space between them. From the structural parameters of the Riblets and the information of the SEM image the width of the plateau created by the acetone treatment can be estimated to be $s = (10\pm2)\,\mu$m. This results in a height reduction of $(24\pm5)\,\%$ meaning $d\approx0.25$ leaving the Riblet with a height of $h=(38\pm2)\,\mu$m.

\paragraph{Set 2}

 The second set are imprints of riblet structures mounted on an airplane for test purposes. It consists of three samples taken from different parts of the airplane at different times. An overview picture of this set of samples can be found in the appendix.

\paragraph{Set 3}

The third set of samples consists of 10 samples especially manufactured to study different degrees of degradation as part of the tasks for WP2 for the last six months. All sample are structured riblet paint of TOPCOAT, sample dimensions are $(15 \times 8)\,\rm{cm}^2$ with a riblet coated area of $(10 \times 8)\,\rm{cm}^2$. A picture of one of these samples can be found in the appendix section \ref{sec:samples}.

For measurements of different degrees of degradation and different reflectivities the sample has partly been treated with the acetone treatment mentioned above. In order to get an exact picture of the degraded riblet surface again some samples have been coated with a thin (approx. 3-5$\,$nm) platinum-iridium layer to perform SEM measurements. For this purpose the sample have to be reduced in size as well as protected  during the cutting process to smaller sizes. 
These samples were studied before and after the acetone treatment with the final experimental setup, see section \ref{sec:setup3} representing the preliminary configuration in the lab of UOS as mentioned in milestone 2.3.

\paragraph{Set 4}

The fourth set of sample consists of two sample with differently colored backgrounds, gray and black, and one sample after a QUV test. Photographs of the sample and comparing results of optical measurements can be found in the appendix, see section \ref{sec:colored}.

\subsection{Preliminary configuration of optical setup}
\label{sec:setup3}
Central part of WP2 is the development, construction and test of an optical setup to perform measurements of scattered laser light intensity as a function of position. This setup as a preliminary configuration in the lab of UOS is milestone 2.3 at the end of the project. The setup is constantly improving during the time frame of the project with a focus being on size reduction and reduction of the number of optical components. This is important for the development of a more prototype-like device especially concerning a hand held sensor. Details of the evolution of the optical setup can be found in the respective six months short reports.

\begin{figure}[htb]
	\centering
	\includegraphics{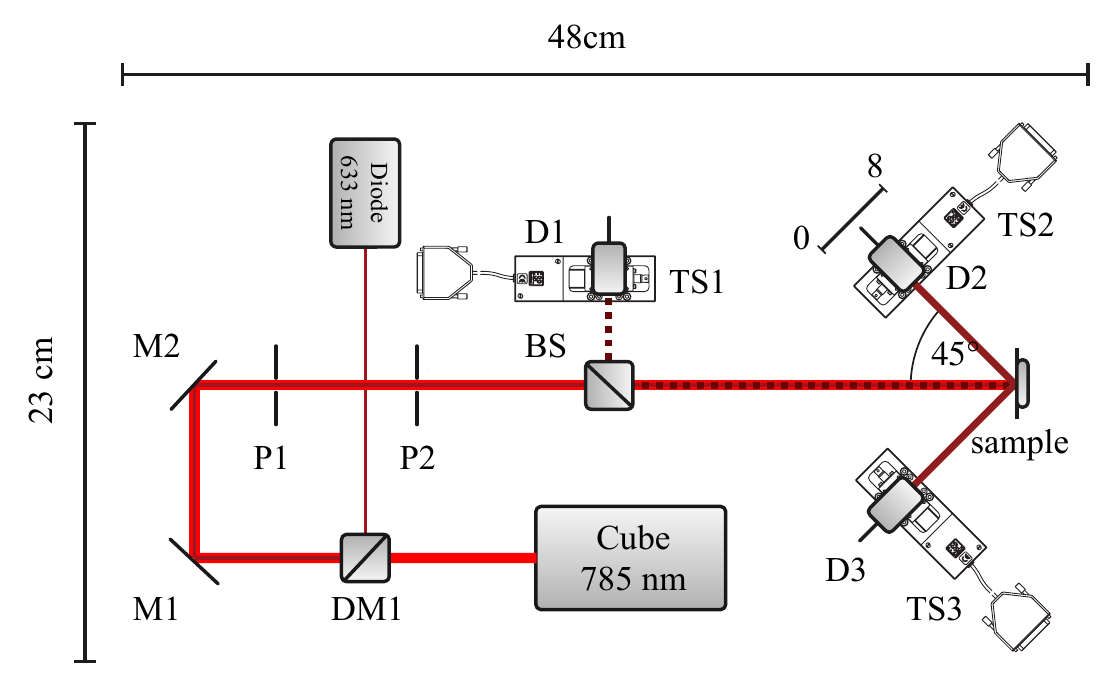}
	\caption{Scheme of the experimental setup allowing for the determination of the angular intensity distribution of waves scattered from a riblet sample. D1-3 SI-PIN photodiodes, BS beamsplitter, M1-2 mirrors, P1-2 pinholes, DM1 dichroic mirror, D1-3 mounted of linear translation stages TS1-3, 0\,mm and 8\,mm positions are marked exemplarily for TS2.}
	\label{fig:setup12}
\end{figure}

Figure~\ref{fig:setup12} sketches the experimental setup used for all measurements performed in the last six months of the project and is the key part of milestone 2.3.

Compared to the former setup used for detection of scattered laser light intensity as a function of position, see Figure~\ref{fig:setup_null} and Figure~\ref{fig:setup_sechs} the setup has been altered:
\begin{itemize}
	\item Now two different types of laser systems are used. A low power laser diode in the red spectral range,$\lambda = 633\,$nm serves as an alignment laser for adjustment of the optical setup. The laser is shut down when the actual measurements are performed due to its higher optical noise. The second laser is a Coherent Cube laser system in the near-infrared spectral range, $\lambda = 785\,$nm. This laser is nearly completely invisible to the naked eye but at the same time its optical noise is very low resulting in an better signal-to-noise ratio. 
	\item For the collinear alignment of the two laser beams a dichroic mirror (DM1) is used. This mirror is highly reflective for light in the visible spectral range. Therefor reflecting the low power adjustment beam. At the same time it is highly transparent for light in the infrared spectrum. Thus, the intensity of the Cube laser is only slightly affected.
	\item To ensure a collinear adjustment of the two laser beams, the pinholes (P1-2) have been incorporated in the optical setup. 
	\item Due to the high overall intensity of the Cube laser system a non-polarizing beamplitter (BS) can be used for the separation of the directly reflected beam  and for the measurement of its intensity by photodiode D1. This results in a reduction of optical components. 
	\item The whole setup is now housed in close box to prevent any stray light from the outside environment from entering the detection system. For details see Figure~\ref{fig:setup_null} and Figure~\ref{fig:setup_sechs} in the appendix.
	\item All photodiodes are equipped with pinholes with an open aperture of 1 mm.  In case of photodiode D1 that is applied for the detection of the directly reflected scattering pattern, a 1mm wide slit is placed in front of the detector to compensate for the increased distance between the detector and the sample. Details of the influence of the pinhole diameter are discussed in the experimental results.
\end{itemize}

The other components of the setup remain unchanged. The Si-PIN photodiodes used to detect the signal are mounted on motorized linear translation stages (TS1-3, \textit{Newport, MFN08CC}) in order to achieve high spatial resolution.

\subsection{Measurement process}
\label{sec:process}

\begin{figure}[htb]
	\centering
	\includegraphics[width=1.00\textwidth]{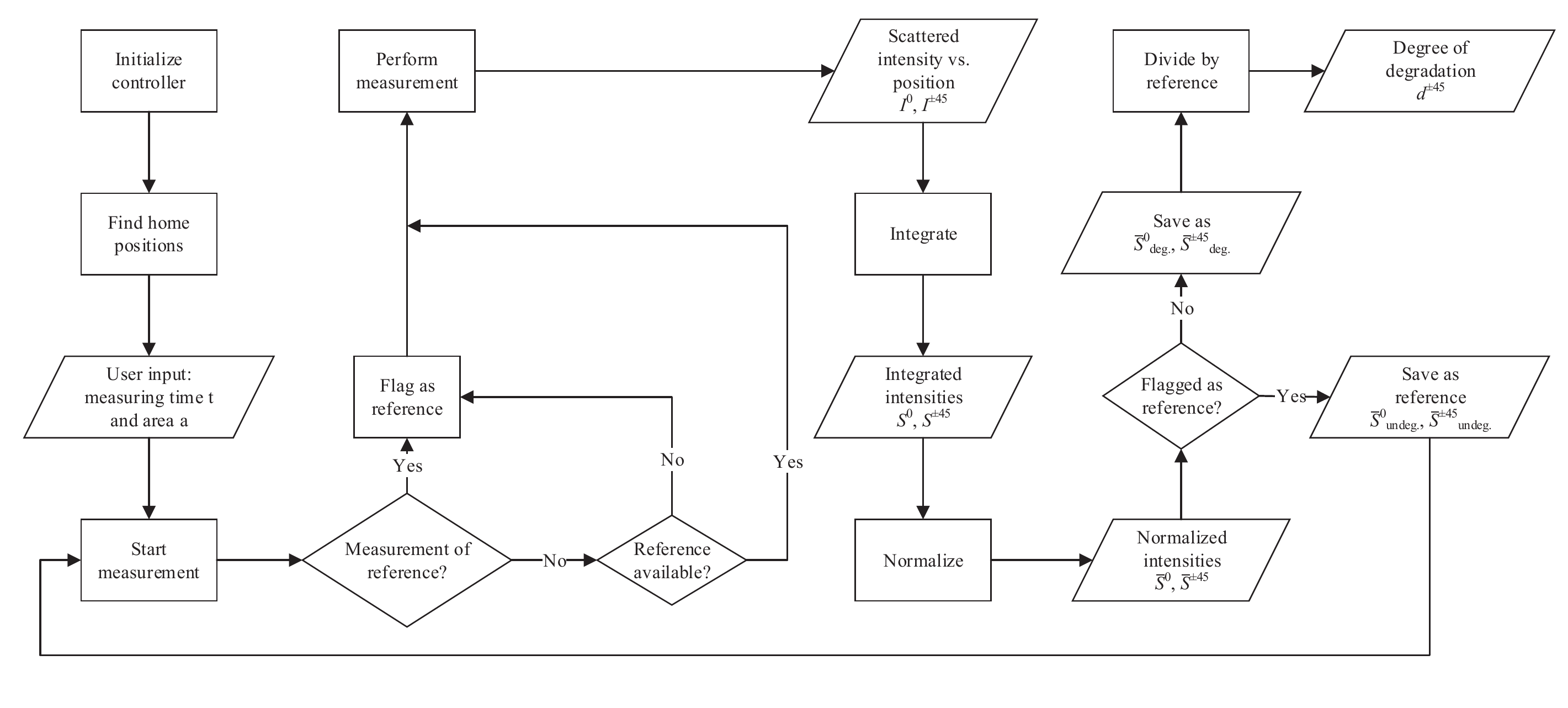}
	\caption{Flowchart of the measuring process to calculated the degree of degradation of a structured riblet surface.}
	\label{fig:flow}
\end{figure}

The complete measuring process is computer-controlled with a single program written in \textit{LabView}. This program controls all technical aspects like positions of the translation stages or amplification of the photodiodes. Furthermore all data analysis needed for the calculation of the degradation parameter $d$ is performed with this software. A flowchart of the measuring process can be found in Figure~\ref{fig:flow}. This flowchart should enable the transfer of the measuring process to any other experimental setup using different components then the used prototype.

At first the software initializes both the motion controller for the translation stages as well as the photometer unit detecting the signals from the photodiodes. The second step is to find the home positions of all translation stages to ensure reproducible starting positions. Next input from the user is needed concerning the measuring time of a scan $t$ and its measured area $a$. At the start of each measurement it has to be stated if the measurement should be used as reference. In this case all data is internally flagged as reference. The software notifies the user if he tries to perform a measurement without a valid reference, in this case the actual measurement will be flagged to be the new reference.

After the actual measurement, the collected data $I^{\pm45}, I^{0}$ is integrated to generate the required values $S^{\pm45}, S^{0}$. 
These quantities correspond to the scalar intensity values from large aperture photodiodes without spatial resolution, which may be used in later productive systems.

\begin{figure}[ht]
	\centering
	\includegraphics[height=6cm]{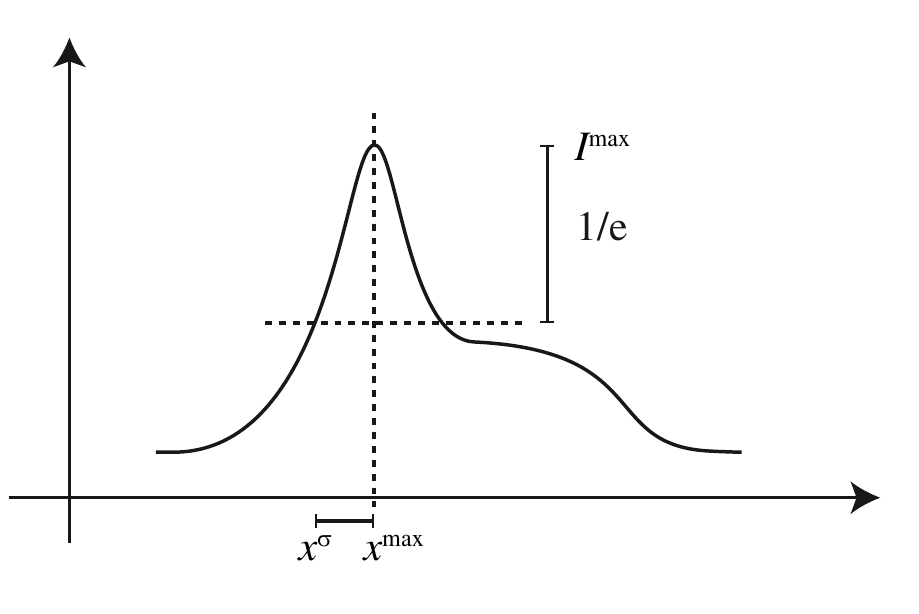}
	\caption{ General shape of the $\pm45^\circ$-pattern. Peak maximum $I^{\pm{\rm max}}$ at $x^{\pm{\rm max}}$ and left-hand peak-width $x^{\sigma\pm}$. }
	\label{fig:result_integrate}
\end{figure}

As shown in fig. \ref{fig:results_set3} the pattern exhibits an additional feature in the angular range of about $50^\circ$.
Initially, this effect has been attributed to diffraction on the riblets tips. 
However, since it is maintained during the degeneration process, it presumably has to be explained with actual deviations of the real structure from the idealized model, probably in the area where flanks and plains meet.
As this signal is not connected to degradation, we do not include it in $I^{\pm45}$.
For this purpose, we introduce a peak derived from the maximum $I^{\pm{\rm max}}$ at $x^{\pm{\rm max}}$ and the left-hand peak-width $x^{\sigma\pm}$ as shown in fig. \ref{fig:result_integrate}.
Possible fitting curves are for example the normal distribution ("`Gauss peak"') or the Cauchy distribution ("`Lorentz peak"').

Indeed, the choice of the exact peak type does not matter due to the fact that the integrals of these and similar well-behaved peak forms is proportional to $I^{\pm{\rm max}} x^{\sigma\pm}$ and only differ in a constant coefficient that is lost during the following normalization process:

The values for the $\pm45^\circ$ reflex are normalized by the directly reflected beam to $\bar{S}^{\pm45}=S^{\pm45}/(S^{0})^2$. In the case that the measurement is flagged as being the reference measurement all data will be save as reference and the values will be renamed to $\bar{S}^{\pm45}_{\rm{undeg.}}=S_{\rm{undeg.}}^{\pm45}/(S_{\rm{undeg.}}^{0})^2$, otherwise the values will be stored as $\bar{S}^{\pm45}_{\rm{deg.}}=S_{\rm{deg.}}^{\pm45}/(S_{\rm{deg.}}^{0})^2$.

In the last step the integrated and normalized data from the measurement, $\bar{S}^{\pm45}_{\rm{deg.}}$ will be divided by the reference values $\bar{S}^{\pm45}_{\rm{undeg.}}$ to calculate the result of the measurement, the degree of degradation:
	\[d^{\pm45}=1-\frac{\bar{S}^{\pm45}_{\rm{deg.}}}{\bar{S}^{\pm45}_{\rm{undeg.}}}
\]
 A screen-shot of the user interface showing the exemplary implementation of this software for the preliminary setup in the optical lab at UOS can be found in the appendix, see Fig.~\ref{fig:software_user} and Fig.~\ref{fig:software_data}. 

\subsection{Experimental results}

\subsubsection{Sample preparation and SEM measurements}

For measurements of different degrees of degradation the samples of Set 3 have partly be treated with the acetone treatment mentioned above. To achieve different degrees of degradation the time of the treatment has been varied in time. 
%Table \ref{tab:set3_sem} gives an overview of the treatment parameters as well as the SEM results.

In order to get an exact picture of the degraded riblet surface again some samples have been coated with a thin (approx. 3-5$\,$nm) platinum-iridium layer to perform SEM measurements. For this purpose the samples have to be reduced in size as well as to be protected during the cutting process to smaller pieces. Figure~\ref{fig:sample_cutting} in the appendix shows a sketch indicating the points of the sample where the measurements are taken as well as the areas needed to cover the samples during the cutting process. 

%SEM pictures of exemplarily chosen samples can be found in Fig.~\ref{fig:set3_sem}. From these picture the degree of degradation can be calculated by measurements of the width of the visible plateau resulting in a hight reduction due to the acetone treatment. The results can be found in Table \ref{tab:set3_sem}. 

\subsubsection{Results of optical measurements}

To perform measurements of different  degrees of degradation using the experimental setup presented in section \ref{sec:setup3} several methods have been tried on the samples of Set 3 to achieve a sufficient degree of degradation. Some of these methods lead to different effects of degradation than the removal of the riblet tip, which is modeled in WP1 (theory)

\paragraph{Acetone treatment}

\begin{figure}[ht]
	\centering
	\includegraphics[height=10cm]{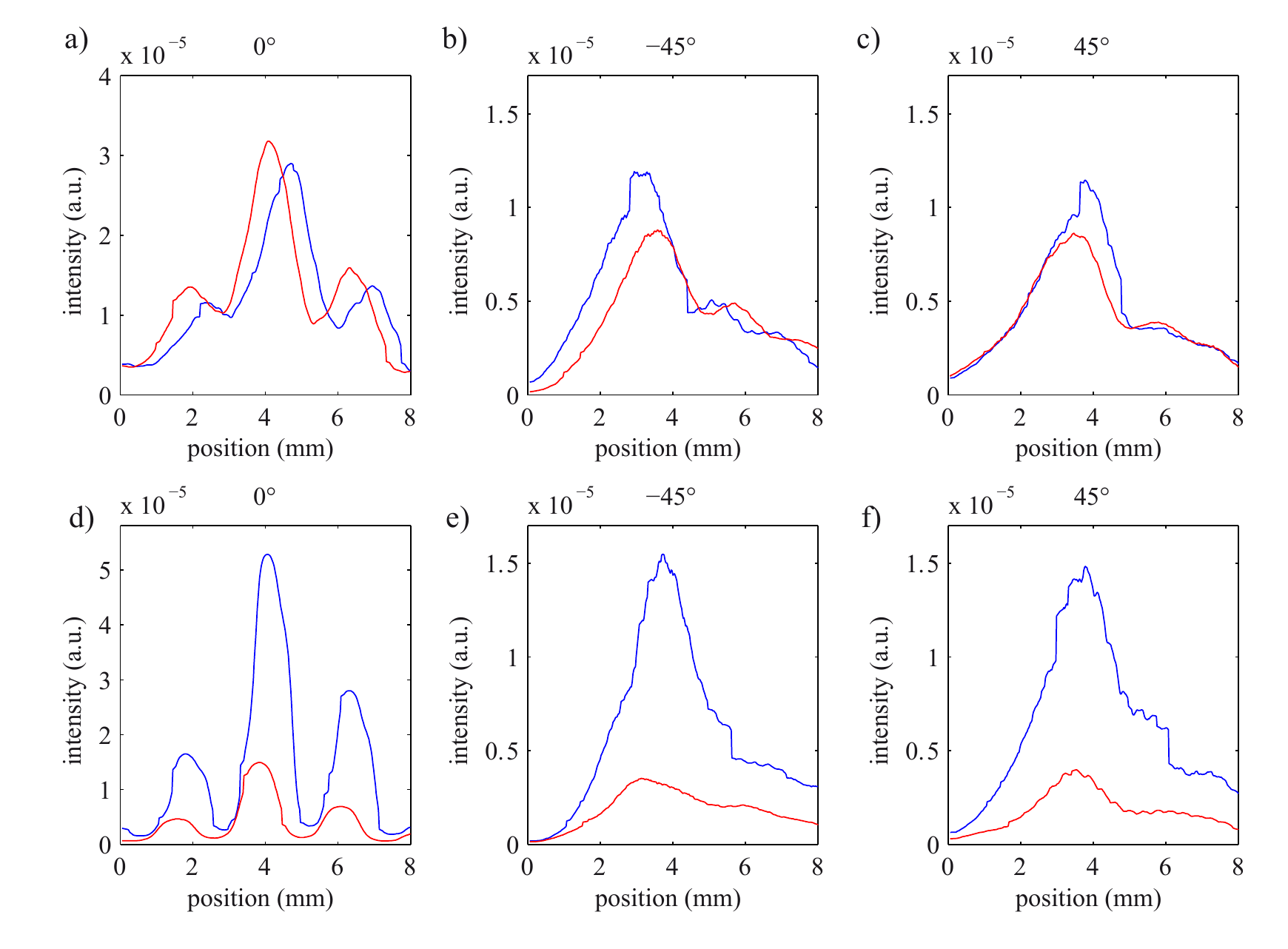}
	\caption{Scattered intensity for the 0$^\circ$ and $\pm45^\circ$-patterns as a function of position for a degraded riblet structure.  In the case of a) to c) the sample was treated with acetone for 4 minutes, in the case of d) to f) the sample was treated for 12 minutes. }
	\label{fig:results_reflec_set3}
\end{figure}

Due to some changes of the coating used for the riblet structure the acetone treatment had different effects on the samples of Set 3 than on the samples of Set 1. It changed the reflectivity of the riblet surface.
Reference measurements are needed
to perform measurements of riblet surfaces with different changes of reflectivity using the experimental setup presented in section \ref{sec:setup3}. These measurements have been performed prior to the acetone treatment. A comparison of the reference measurements with the measurements after the acetone treatment can be found in Fig.~\ref{fig:results_reflec_set3}.

%\todo{Vergleich Bild Messung Aceton}

The blue curve shows the signal prior to the acetone treatment, while the signal after the treatment is indicated by the red curve. One can see a similar decrease of the signal in all patterns, $0^\circ$ and $\pm45^\circ$, of a specific sample. One observe a stronger reduction for the signal  of the reflected intensity of the sample treated longer with the acetone, compare Fig.~\ref{fig:results_reflec_set3} a) to c) with Fig.~\ref{fig:results_reflec_set3} d) to f). 

This behavior corresponds to a degradation of the sample which changes just the reflectivity while leaving the structural parameters of the riblet undisturbed. 
%To verify this, SEM measurement have been performed which can be found in Fig.~\ref{fig:rem_acetone}.
%\todo{REM Bild Aceton}

\begin{infobox}[]
Measurements can be performed to detect a changed overall reflectivity of the riblet structures. In this case the structural geometry of the riblets remains unchanged, i.e. the function for reduction of air resistivity is conserved.
\end{infobox}

\paragraph{Applying pressure using a micrometer screw gauge}

\begin{figure}[ht]
	\centering
	\includegraphics[height=10cm]{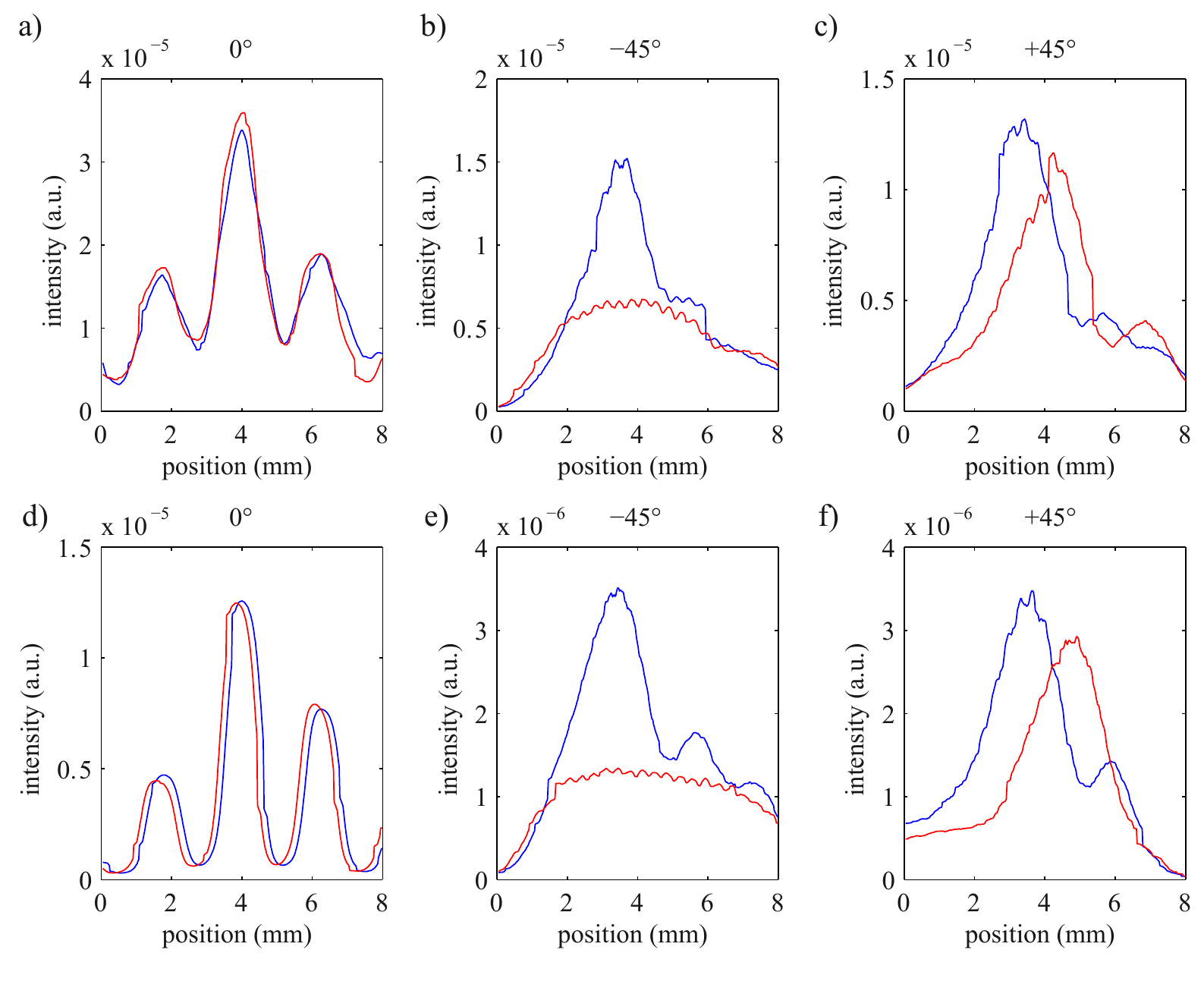}
	\caption{Scattered intensity for the 0$^\circ$ and $\pm45^\circ$-patterns as a function of position for a degraded riblet structure.  In the case of a) to c) the micrometer screw gauge was fastened $5\,\mu$m, in the case of   d) to f) the micrometer screw gauge was fastened $15\,\mu$m. }
	\label{fig:results_micrometer}
\end{figure}

Another approach to degrade just the tip of the riblet structure was to apply pressure on the surface by using a micrometer screw gauge. Therefore the micrometer screw gauge was loosely fastened to the surface, up to this no change to the riblets was made. Then the micrometer screw gauge was further fastened to a specific distance of a few micrometers. This could only be made applying some pressure on the riblet surface and therefore the riblet surface is changed. A comparison of the reference measurements with the measurements after the micrometer screw gauge treatment can be found in Fig.~\ref{fig:results_micrometer}.

The blue curve shows the signal prior to the micrometer screw gauge treatment, while the signal after the treatment is indicated by the red curve. One can see that the $0^\circ$-pattern is unchanged by the treatment. Both signal amplitude and shape stay the same, the slight variations are within the range of a single measurement error. 

Strong variation can be observed in both the $\pm45^\circ$-pattern. The distinct maximum of the $-45^\circ$-pattern is vanished leaving only the background noise of the ghost image beyond the $\pm45^\circ$ region. From this behavior it can be deduced that the complete flank of the riblet has to be damaged extensively therefore reflecting nearly no light in the direction of this pattern.

The $+45^\circ$ peak is shifted and slightly reduced in its height with respect to the reference measurement. From this it is concluded that the angle of the respective riblet flank has been changed resulting in a pattern visible at a changed angle. 

%To proof these estimation SEM measurements have been performed on these samples, the resulting picture can be found in Fig.~\ref{fig:rem_mikrometer}.
%\todo{Rem Bild Mikrometer}
%One can see \todo{Beschreibung REM-Bilder}

\begin{infobox}[]
The method can be used to detect an asymmetric degradation of the riblet structures resulting in a asymmetric change of the signal of the $\pm45^\circ$ patterns.
\end{infobox}

\paragraph{Mechanical tip removal using a lathe}

\begin{figure}[ht]
	\centering
	\includegraphics[height=5cm]{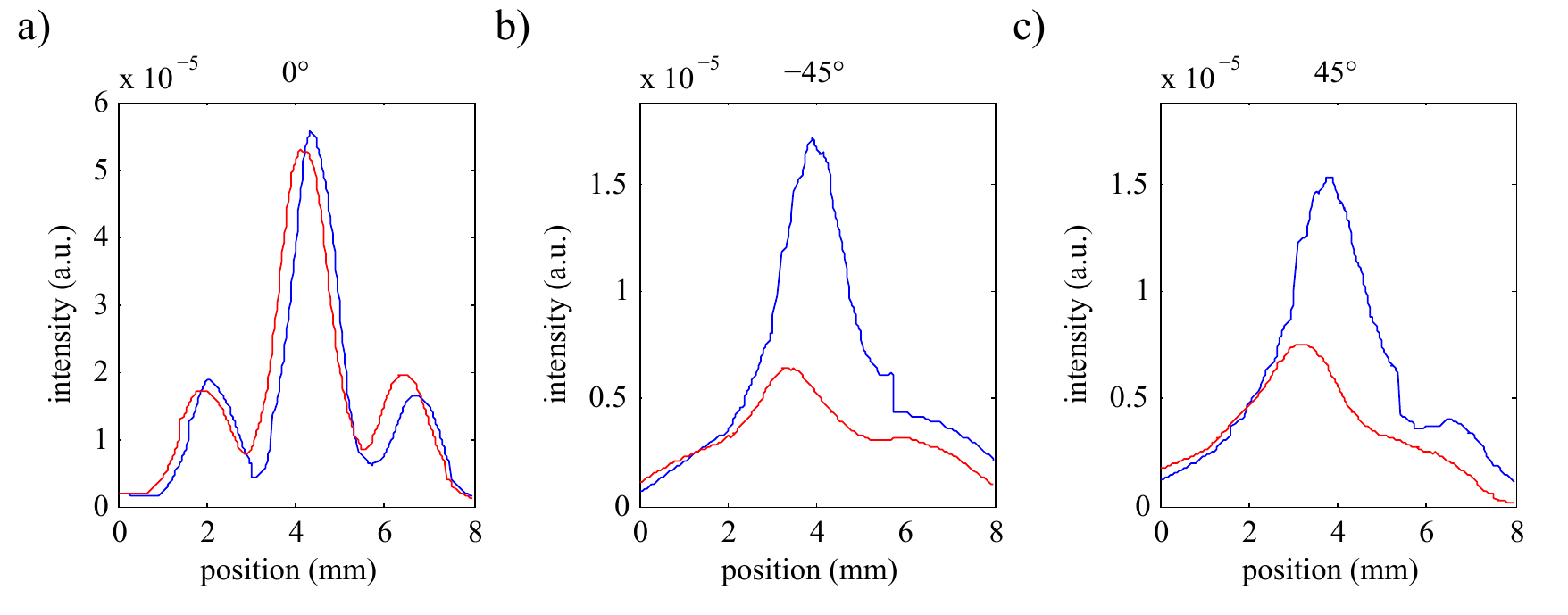}
	\caption{Scattered intensity for the 0$^\circ$ and $\pm45^\circ$-patterns as a function of position for a degraded riblet structure.  In the case of a) to c) a lathe was used to remove about $20\,\mu$m of the riblet tip.}
	\label{fig:results_set3}
\end{figure}

To remove just the tip of the riblet without damaging the flat areas between a lathe is used to remove about $20\,\mu$m of the tip, see Fig. \ref{fig:results_set3}.

The blue curve shows the signal prior to the treatment, while the signal after the treatment is indicated by the red curve. One can see that the $0^\circ$-pattern is unchanged. Both signal high and shape stay the same, the slight variations are within the range of a single measurement error. 

\begin{figure}[ht]
	\centering
	\includegraphics[height=5cm]{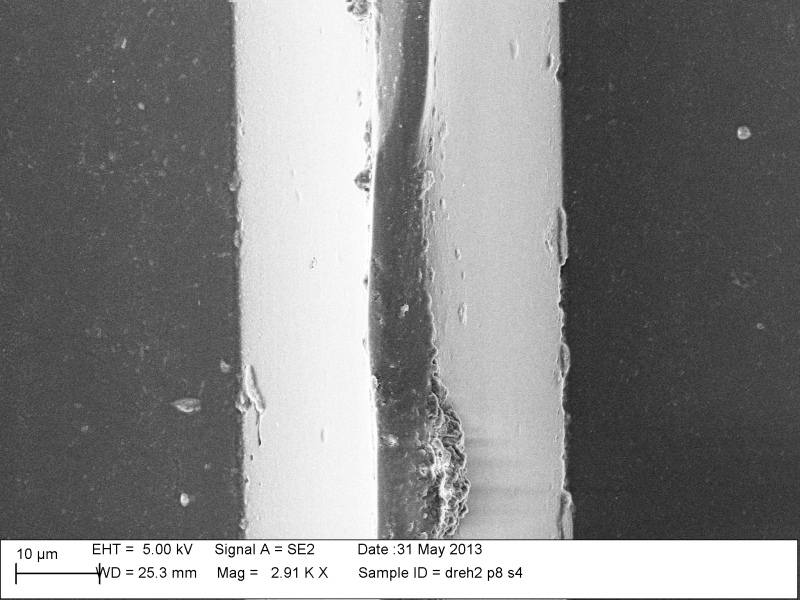}
	\caption{SEM measurement of the sample studied in Fig. \ref{fig:results_set3}, the tip of the riblet was removed using a lathe. }
	\label{fig:rem_set3}
\end{figure}

Variation can be observed in both the $\pm45^\circ$-pattern. In both cases, the signal is reduced and it is 
evaluated by the procedure discussed in section \ref{sec:process}. As a result one gets $d^{+45}= 0,42$ and $d^{-45}= 0,40$.
To proof these SEM measurements have been performed on these samples, the resulting picture can be found in Fig.~\ref{fig:rem_set3}.

From the SEM measurements a plateau width of $s=(3.5\pm0.6)\,\mu$m can be deduced resulting in a height reduction of $\Delta h = (8.45 \pm 1.45)\,\mu$m and therefore yields a degradation value of $d=(0.175\pm 0.25)$. This indicates that the SEM measurements performed from above the riblet structure are not suited to perform precise measurements of the degree of degradation. For this it would be better to perform SEM measurements of a tilted sample or even better on a cut of the samples. This would lead to better results of the topography of the riblet structure. 
%\todo{unterschiede erkl�ren optische messung und rem}

\begin{infobox}[]
The optical method allows to measure even slight reduction of the riblet height due to a removed riblet tip. This  kind of degradation is assumed to be the most common one as discussed in the kick-off meeting. 
\end{infobox}

\subsection{Alternative experimental solutions}

During the time frame of the project the setup used to study the riblets was constantly altered and improved concerning both technical implementation as well as measurement methods. Hereby, the key task is to find an experimental way to measure the not only the scattering patterns at $\pm 45^\circ$ but the directly reflected pattern at $0^\circ$ as well. Therefor in this section other experimental configuration and exemplarily chosen data is shown for comparison. Details of the experimental setup can be found in the respective six months reports.

\subsubsection{Titlted angle incidence}
\label{sec:setup1}

%\textbf{\emph{Erstes Setup bei uns mit Ergebnissen, Winkelabhaengigkeit und Problembenennung}}

\begin{figure}[ht]
	\centering
	\includegraphics[height=6cm]{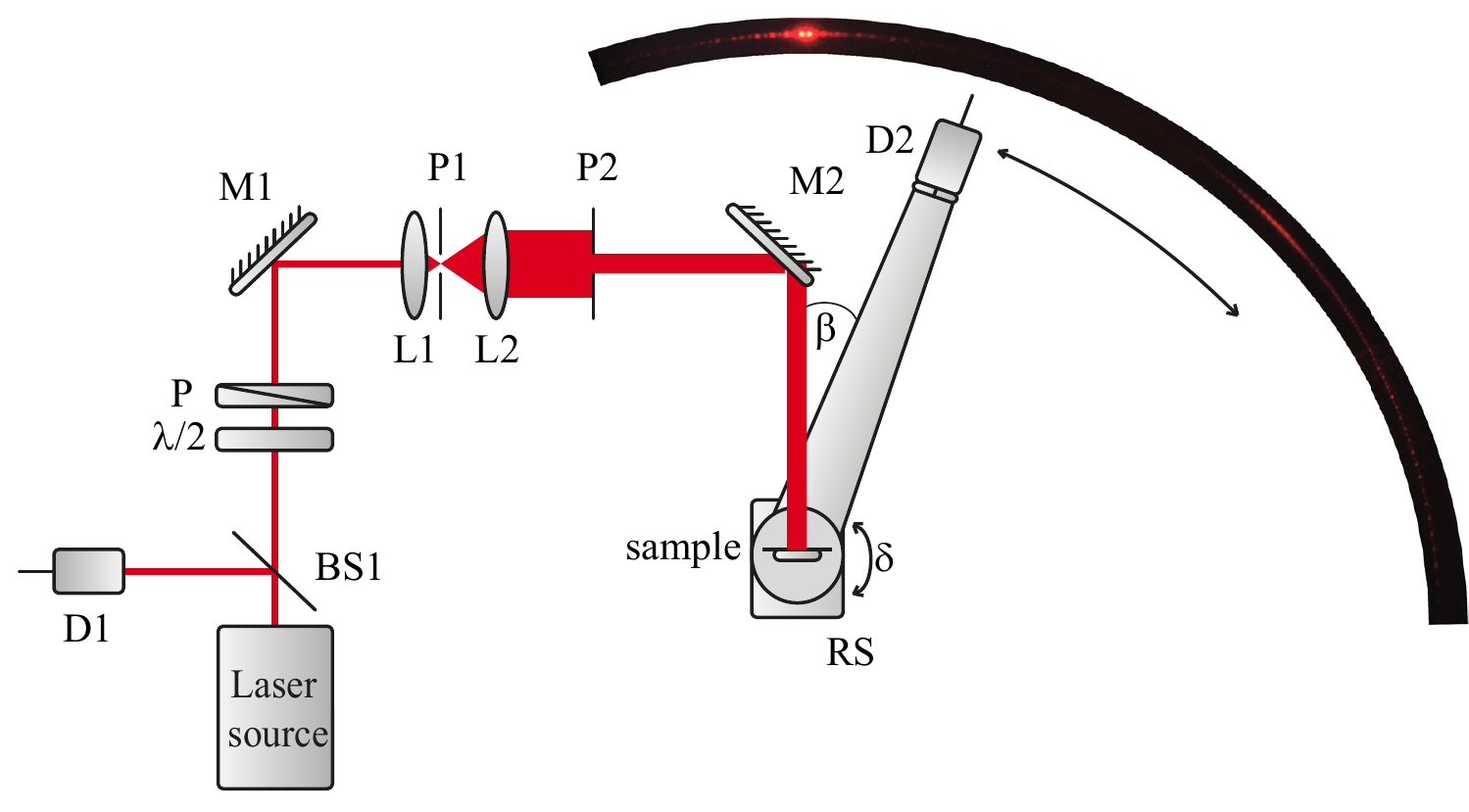}
	\caption{Scheme of the experimental setup allowing for the determination of the angular intensity distribution of waves scattered from a riblet sample. D1-2 SI-PIN photodiodes, BS1 beamsplitter, $\lambda/2$ half-wave plate, P polarizer, M1-2 mirrors, L1-2 lenses, P1-2 pinholes, RS computer-controlled rotationary stages, scattering angle $\beta$, sample angle $\delta$.}
	\label{fig:setup_null}
\end{figure}

One approach to measure but the patterns at $\pm 45^\circ$ as well as at $0^\circ$ was an experimental setup using a a slighly tilted angle between the incident laser beam and the sample. This setup was mainly used during the first six months. An optical setup was developed which allow for exposure of the riblets with laser light of various wavelengths (spectroscopic analysis) and adjustable incidence angles with respect to the riblet surface (analysis of scattering geometry). All these requirements were met with the realized setup (selection and ordering of optical components, setup and testing included) that is schematically sketched in Fig.~\ref{fig:setup_null}.

The complete setup was installed on a vibration isolated, optical table. Different types of lasers were used for the investigations: an optically parametrical amplifier (OPA) pumped by a regeneratively amplified fs-oscillator served for wavelengths in the near-infrared spectral range from $\lambda = 1100\,$nm up to $\lambda = 1600\,$nm. A HeNe gas laser at $\lambda = 633\,$nm and a frequency-doubled solid state laser at $\lambda= 488\,$nm were applied for investigations in the visible spectrum. A combination of a half wave plate ($\lambda/2$) and polarizer (P) was used for precise control of the laser intensity which was monitored by a SI-PIN diode (D1). An optical system consisting of a microscope lense (L1, magnification 20x), a plano-convex lense (L2, f=100\,mm) and a pinhole (P1, $30\,\mu$m) served as spatial frequency filter and beam expander enabling measurements as a function of beam diameter $0.5$\,mm$< d <10$\,mm. These diameters were adjusted with another pinhole (P2). A D-shaped pickoff mirror (M2) assured nearly perpendicular incident of the probe light on the riblet surface. Orientation of riblet surface was perpendicular to the optical table and angle of incidence of the probe beam. This measurement geometry was chosen due to preliminary considerations from WP1 (theory).
For further analysos, the angle of the surface with respect to the incident light $\delta$ can be controlled by a motorized rotation stage (RS, \textit{Newport, UMR 80CC}) and motion controller (\textit{Newport, MM4006}) with an angular resolution of 0.001$^\circ$. The scattering intensity distribution was measured with a SI-PIN photo diode (D2) equipped with a pinhole of $50\,\mu$m. This ensured a high angular resolution. The angle $\beta$ of the photo diode D2 was controlled with a second computer-controlled rotation stage. Alternatively, a 2D-CCD-array was applied to determine the scattering pattern within a large apex angle \textit{Canon EOS 10D}.

\begin{figure}[ht]
	\centering
	\includegraphics[height=10cm]{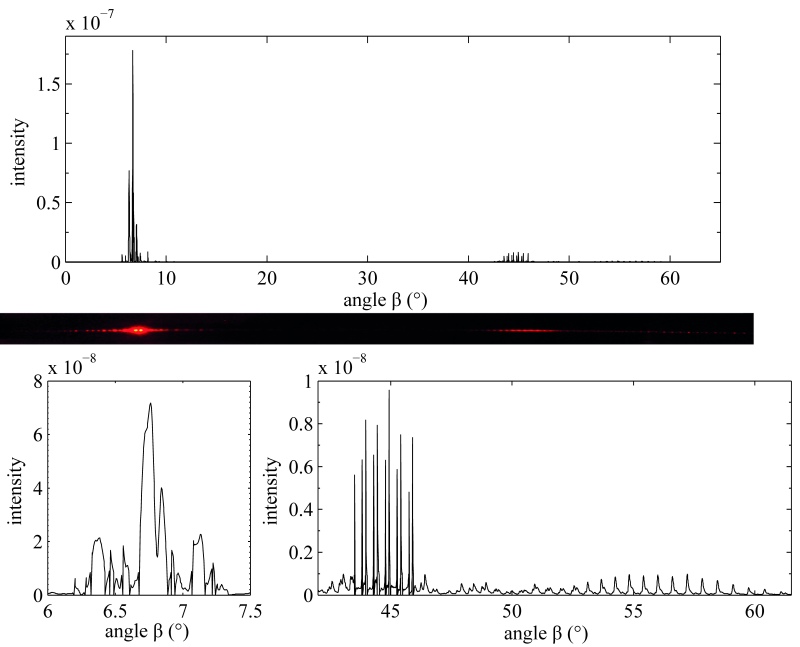}
	\caption{Top: reflected intensity as a function of detector angle $\beta$ for probe wavelength $\lambda = 633\,$nm, beam diameter $d=1\,$mm, $\delta = 3.5^\circ$ on Set 1, sample 4.\newline
	Middle: picture of the reflection pattern.\newline
	Bottom: intrinsic structure of the two peaks of the directly reflected beam (left) and the reflected beam at $\beta = 45^\circ$. }
	\label{fig:results_overview}
\end{figure}

A typical plot for the dependency of the reflected light as a function of angle $\beta$ on on Set 1, sample 4 is shown on Fig.~\ref{fig:results_overview}. Note that the sample was slightly rotated by $\alpha = 3.5^\circ$ (slanted incidence) in order to determine the scattering intensity profile of the first order, as well. The first peak, centered at $\beta = 7^\circ$ because of the slant, is rather narrow with a full width at half maximum about $3^\circ$, and a distinct structure inside. In the angular range from about $15^\circ$ to $40^\circ$ no scattering intensity can be observed within the electronic noise of the detector. A second, broader peak is centered around $45^\circ$ with a broad shoulder extending up to $70^\circ$. Like in the first peak, a substructure with equidistant peaks is observed. In the middle part of Fig.~\ref{fig:results_overview} a picture of the reflection pattern is shown. Both peaks and even their intrinsic structure can be clearly seen by the naked eye. The lower part of Fig. \ref{fig:results_overview} gives a closer view on the intrinsic structure of the peaks of the directly reflected beam (left) and the reflected beam at $\beta = 45^\circ$.

The same measurements were performed to study different dependencies. The results can be found in the appendix. Several measurements were performed at different probe wavelengths $\lambda$. The angular dependence of the reflected intensity exhibit a qualitatively comparable behavior over a broad spectral range. Measurements of reflectivity in the IR spectral range from $\lambda = 1120\,$nm to $\lambda = 1600\,$nm verify a broad spectral ability for peak detection and its substructure at $\beta=45^\circ$. Therefor, from the optical view point there are no restrictions on the selection of the probe wavelength for the measurement of the angular reflectivity within the sensor device.

The second series of measurements studied the probe beam diameter $d$. It defines the area which can be studied within a single surface scan, but also, the spatial resolution of the sensor. A strong dependence of the intrinsic structure on $d$ is found. It is particularly noteworthy, that the peaks of this structure become inseparable for $d>2\,$mm. The overall shape of the pattern, i.e. the main scattering peaks at $\beta = 0, \pm45^\circ$ remain the same. Results can be found in the appendix, see section \ref{sec:dia}, Fig.~\ref{d2} and Fig.~\ref{dd} .

A further series dealt with the angle $\delta$ of the riblet surface in relation to the beam direction, see section~\ref{sec:angle}. Up to an angle $\delta \approx 9^\circ$ no significant change of the position of the peaks at $\beta = 45^\circ$ and only slight changes in intensity are observed. The peak of the directly reflected beam is observed as expected at $\beta = 2\cdot\delta$. Results can be found in the appendix, see section \ref{sec:angle} Fig.~\ref{1grad} and Fig.~\ref{20grad}. This result is important for the determination of the alignment precision to inspect plane areas as well as for the maximum curvature of non-plane area to be inspected.

Lastly, the riblet structure on metal instead of TOPCOAT was studied. The peak of the directly reflected beam is smaller than on TOPCOAT, compare to Fig.~\ref{fig:results_overview}, due to the lower reflectivity of the base material. But the peak is observed at the same angular position. Intensity and position of the peak at $\beta = 45^\circ$ are comparable to the TOPCOAT sample. This shows the applicability of this method to riblet structured coatings on different base materials.

\begin{infobox}[]
Measurements can be performed independently from material of the background material as the riblet paint itself is transparent. This is especially true for painted and unpainted metal. The signal depends solely on the reflectivity of the riblet structure.
\end{infobox}

\subsubsection{Polarized beam splitting}
\label{sec:setup2}

%\emph{\textbf{Setup 2. Halbjahr, Problembenennung mit Entwicklung zu final}}

Another approach to measure the pattern at $0^\circ$ was the use of a polarized beam splitter to extract the directly reflected beam from the incident laser beam. This setup was used mainly during the second six months.
Compared to the setup presented in the first six months report, the optical setup for determination of the scattering pattern has been updated taking into account the deliverables of the accompanying workpackages:
\begin{enumerate}
	\item Theoretical considerations made within the first six months and discussed in the first six months report strongly prefer a non-tilted, this means perpendicular, incidence of the incoming light on the riblet surface. As a consequence, the experimental approach of a d-shaped mirror together with a slightly tilted sample for the detection of the scattering pattern in 0$^\circ$-direction must be reconsidered. This is realized by a polarzing beam splitter as introduced in WP3 (see below).
\item Waves scattered at non-degraded and degraded Riblets propagate mainly in $0, \pm45^{\circ}$-direction. The detection of the full scattering sphere is not required. Thus, the motorized rotational stage is replaced by three  motorized linear stages. 
	\item Some improvements to the design of the setup have been made due to the results and requirements of WP3. This includes the dimensions of the optical baseplate as well as a magnetic sample mount that allows for an easy fixation and replacement of the samples under study.
\end{enumerate}

\begin{figure}[ht]
	\centering
	\includegraphics[height=5cm]{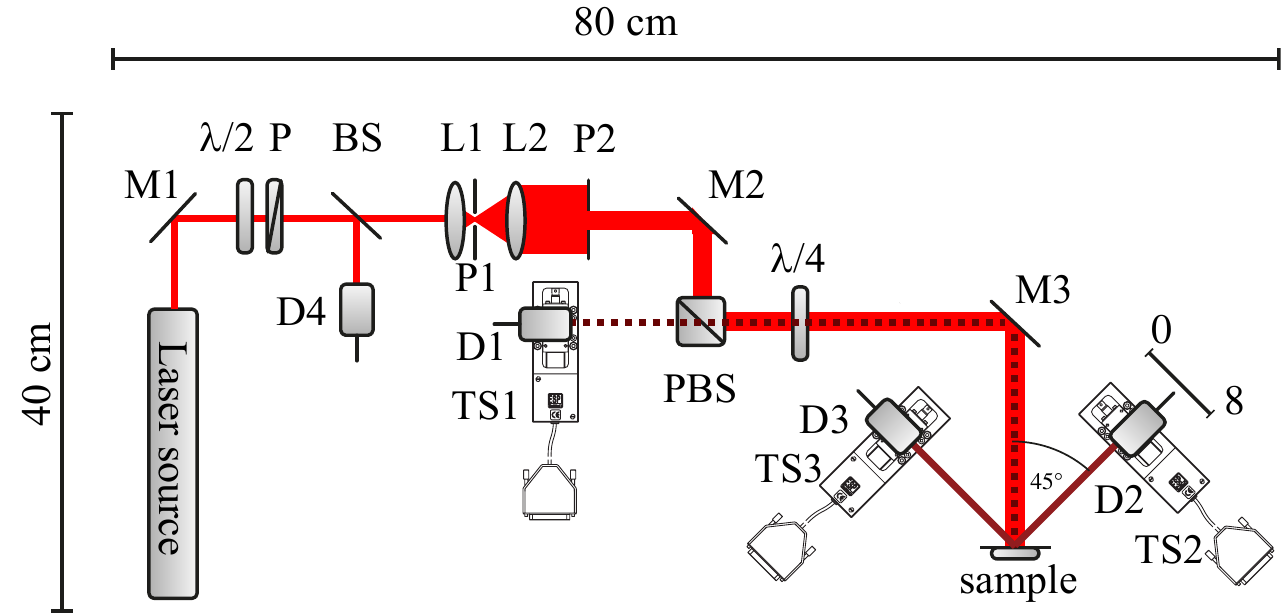}
	\caption{Scheme of the experimental setup allowing for the determination of the angular intensity distribution of waves scattered from a riblet sample. D1-4 SI-PIN photodiodes, D1-3 mounted of linear translation stages TS1-3, BS1 beamsplitter, $\lambda/2$ half-wave plate, P polarizer, M1-3 mirrors, L1-2 lenses, P1-2 pinholes,  PBS polarizing beamsplitter cube, $\lambda/4$ quarter-wave plate. 0\,mm and 8\,mm positions are marked exemplarily for TS2.}
	\label{fig:setup_sechs}
\end{figure}

These considerations and the results of the first six months report from all three work packages, theoretical and experimental measurements as well as the device catalog lead to the development of the optical setup that is sketched in Figure~\ref{fig:setup_sechs}.

This setup meets all state-of-the-art requirements for performing appropriate measurements of scattered laser light intensity as a function of position. The main part of optical components has been described in detail in the report on the first six months. Novel aspects and components are summarized in the following list:

\begin{enumerate}
\item The updated setup allows for scanning both the +45$^\circ$ and -45$^\circ$-pattern at the same time.
\item The number of SI-PIN photodiodes has been increased to simultaneously detect the 0$^\circ$-pattern (D1), +45$^\circ$ and -45$^\circ$-patterns (D2,D3) as well as to monitor the laser intensity (D4).
	\item The motorized rotationary stage used to scan the scattering pattern has been exchanged by motorized linear translation stages (TS1-3, \textit{Newport, MFN08CC}) allowing for the detection of the main scattering features with high resolution. We note that the intensity is thus no longer measured as a function of angle but as function of spatial position in the vicinity of the main scattering angles, see  Figure~\ref{fig:setup_sechs} for details.
	\item For measurements of the 0$^\circ$-pattern pattern a polarizing beamsplitter cube (PBS) and a quarter-wave plate ($\lambda/4$) have been incorporated in the setup. Perpendicular incidence of the laser light is realized. The directly reflected light, the 0$^\circ$-pattern, is captured by TS1 and the following adjustments of light polarizations: The incoming light is s-polarized and therefore reflected by the PBS. After passing the quarter-wave plate the light is circularly-polarized. When the reflected light passes the quarter-wave plate the second time a total rotation of 90$^\circ$ of the polarization with respect to the incoming beam results. The light is now p-polarized being transmitted by the PBS to be detected by photodiode D1 mounted on TS1.
	\item A new magnetic sample holder has been developed allowing for an exchange of the samples and higher mechanical stability.
\end{enumerate}
This setup is an intermediate step towards a much more smaller, prototype-like setup. During the measurements continuous updates to the setup have been made and further improvements can be made, like reduction of the number of mirrors. It was used to study the samples from Set 1 and Set 2. Some exemplary results can be found below.

Principle optical inspection: Set 2 samples show a high degree of degradation which is clearly visible even by inspection with the naked eye. In the systematic studies, the reflectivity of samples without Riblet structure have been studied for the determination of both reflectivity and surface quality. In comparison with the samples of Set 1, the reflectivity is considerably reduced from $\approx 4\%$ to $\approx 2\%$. At the same time, an increase of the background scattering noise is noticed. This can be explained either by a different index of refraction as by a comparably rough surface that results in pronounced light scattering.

\begin{figure}[ht]
	\centering
	\includegraphics[height=10cm]{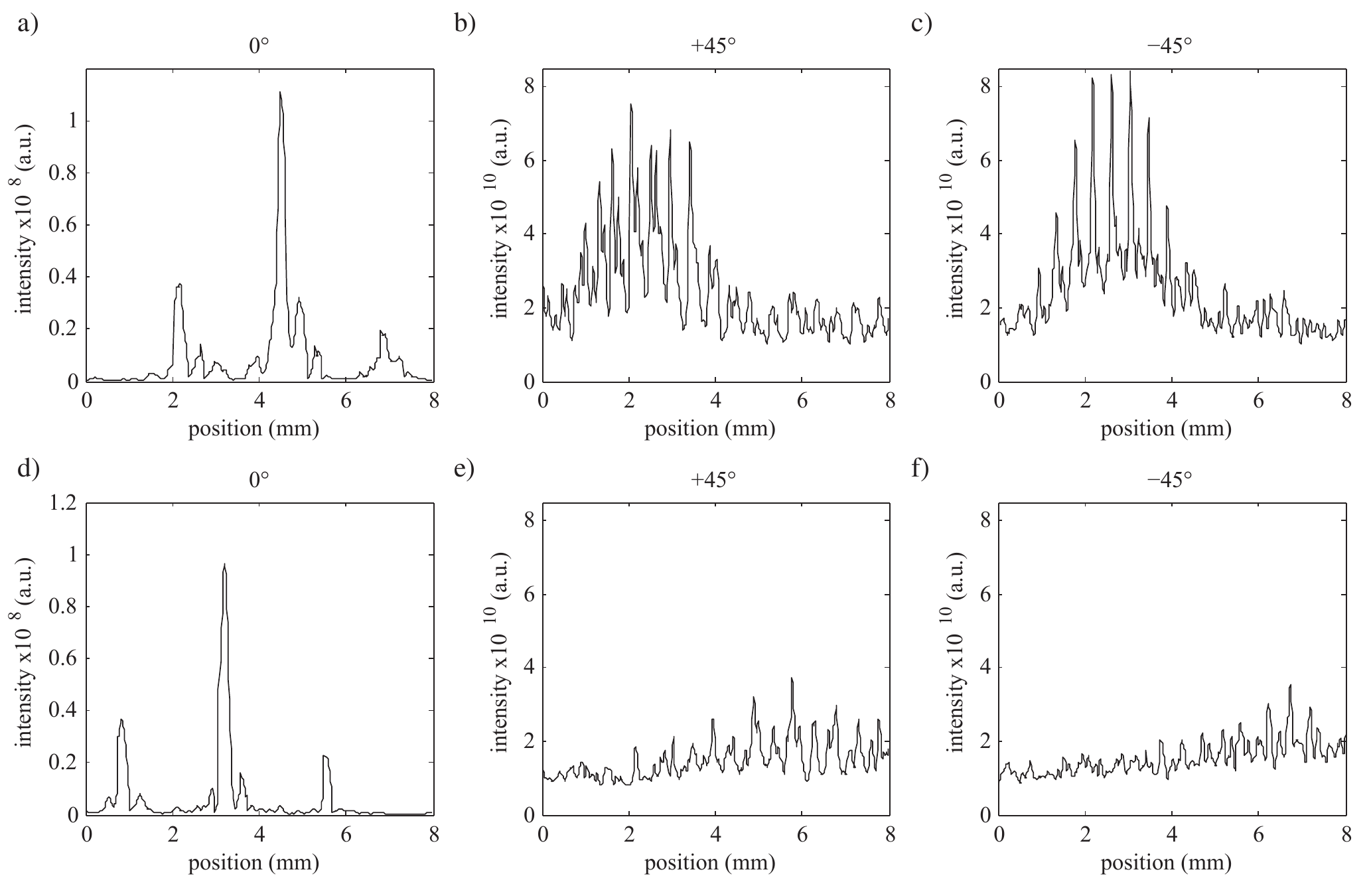}
	\caption{Scattered intensity for the 0$^\circ$ and $\pm45^\circ$-patterns as a function of position for a non-degraded a) to c) and a degraded riblet structure d) to f). }
	\label{fig:degraded}
\end{figure}

Measurements of scattered laser intensity as a function of position were performed with the experimental setup depicted in Figure~\ref{fig:setup_sechs}. The resulting scattering features for Set 1 samples are plotted in Figure~\ref{fig:degraded}: (a,b,c) for a non-degraded riblet structures and(d, e, f) for degraded Riblet structures.

As expected no significant change is found for the 0$^\circ$-pattern between the non-degraded a) and degraded d) riblet structure. This include, both, peak height as well as overall peak shape. The slight difference of absolute position is caused by the measuring process.

Comparing the $\pm45^\circ$-patterns first a good symmetry between the $+45^\circ$ and $-45^\circ$-pattern is found for the non-degraded, b) and c), as well as for the degraded, e) and f), riblet structure. The degradation leads to a significant loss of intensity in these pattern, although the structure is still visible. Obviously the intensity loss is higher than expected from the ideal assumptions of riblet height reduction. This can be attributed to an increased overall roughness of the riblet flanks, making the signal change due to degradation even higher.

Compared to the data from the first six months report background scattering is higher due to the smaller distance between sample and detection system. However, this does not influence the ability of the setup to detect the scattering pattern with sufficient signal to noise ratio but should be kept in mind when comparing the two reports.

The scattering patterns of the structured samples of Set 2 are characterized by a low overall scattering intensity. In direct comparison with the signals obtained with Set 1 samples, the overall scattering intensity in $\pm45^{\circ}$-direction is smaller by a factor of $\approx 10$. Furthermore, the widths of the scattering features has increased significantly and is accompanied with a nearly complete loss of the fine structure. These characteristics are clear indicators that the riblet structure is damaged significantly. Particularly, spatial inhomogeneous degradation of the Riblets must be present that prevents the formation of interference patterns by the superposition of coherent scattered waves. This indicates that the degree of degradation is by far more pronounced as it can be expected within a regular airplane maintenance cycle.

%% file: comparison.tex
\subsection{Comparison for undegraded riblet surfaces}

A comparison of the results of WP1 (Theory) and WP2 (Experiment) is given. All experimental results for nondegraded riblets have been gained using the setup from the first six months of the project, see Figure \ref{fig:setup_null}, as undegraded samples were the focus of the project at that time. First, a comparison of the total scattering  intensity distribution in the angular range from $\beta = 0^\circ$ to  $\beta = 60^\circ$ is performed. Afterwards, the two maxima at  $\beta = 0^\circ$ and  $\beta = 45^\circ$ are inspected in more detail. All data in this context have been normalized to the respective maximum value. 

\begin{figure}[ht]
	\centering
	\includegraphics[width=1.00\textwidth]{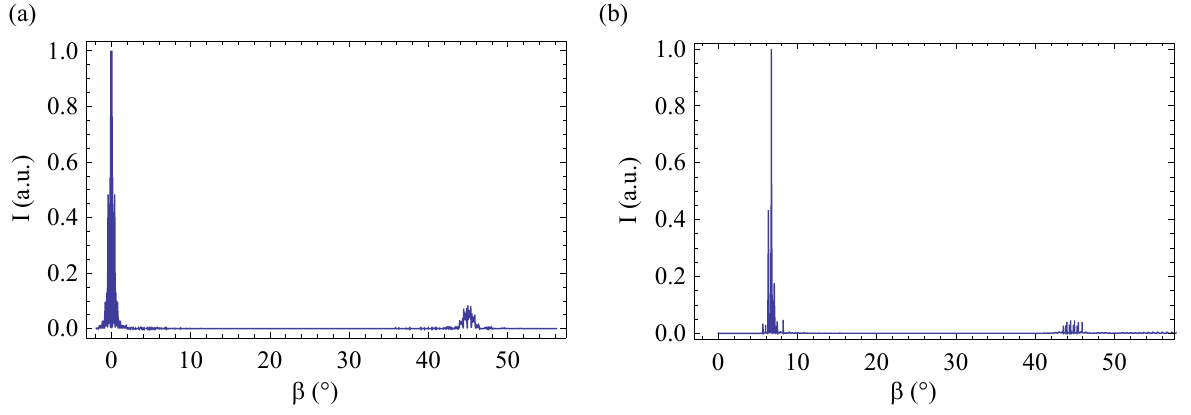}
	\caption{(a) Theoretical and (b) experimental results of the reflected intensity as a function of detector angle $\beta$}
	\label{fig:vergleich-ges}
\end{figure}

Figure \ref{fig:vergleich-ges} shows the total scattering intensity distribution from $\beta = 0^\circ$ up to  $\beta = 60^\circ$ for both the results from WP1 (Theory) (a) and WP2 (Experiment) (b). Due to the experimental setup the peak of the directly reflected beam is shifted by $2\delta \approx 7^\circ$ in Figure 1 (b). 

All calculations and measurements are performed with probe wavelength $\lambda = 633\,$nm and  beam diameter $d = 1$\, mm. 
The intensity distributions are in good agreement concerning peak position and peak width. The experimental data verifies the overall structure of the model derived via the ray-tracing methods.

\begin{figure}[ht]
	\centering
	\includegraphics[width=1.00\textwidth]{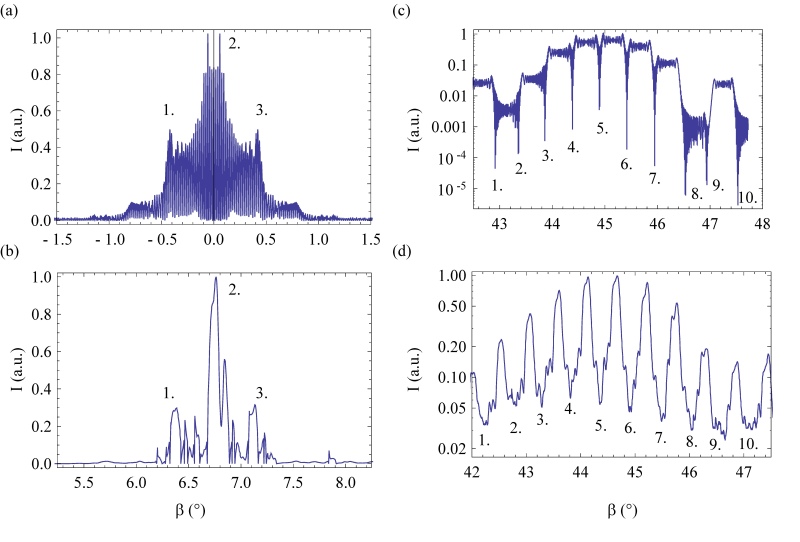}
	\caption{(a) Theoretical and (b) experimental results for the directly reflected beam ($\beta = 0^\circ$). Corresponding maxima are marked from 1 to 3. (c) Theoretical and (d) experimental results for the maximum of the beam reflected at $\beta = 45^\circ$ on a logarithmic intensity scale.  Corresponding minima are marked from 1 to 10.}
	\label{fig:results_overview_2}
\end{figure}

Figure \ref{fig:results_overview_2} shows a detailed comparison for the two maxima at $\beta = 0^\circ$ and  $\beta = 45^\circ$.

Remarkable agreement is found in all data for both peak distance and number of peaks. In Figure 2 (a) three distinct extrema appear that are verified experimentally in Figure 2 (b). In Figure 2 (c) the numerical results of the scattering pattern at $\beta = 45^\circ$ is depicted, that highlights 10 extrema. All 10 extrema are verified in Figure 2 (d) in full accordance to the expectations. This verifies the Huygens-Fresnel approach of the theoretical model for the intrinsic structure of the maxima; i.e. the fine structure can be understood to result from light diffraction at the riblet structure.

\subsection{Comparison for degraded riblet surfaces}

The theoretical plots given in Fig.~\ref{fig:pattern-deg} and experimental data of Fig.~\ref{fig:degraded} are compared with respect to the features of the scattering pattern that are decisive for the determination of the degree of degradation. Nevertheless, we like to point to the striking accordance even in the fine structure of the scattering patterns in $0$- and $\pm 45^{\circ}$-direction that was primarily studied in the first interim report.

The same is valid for the overall scattering intensity distributions, that are characterized by three main scattering features directed in $0$- and $\pm45^{\circ}$-direction in both, theoretical and experimental analysis. As predicted by the theoretical analysis, the scattering intensity in $\pm45^{\circ}$-direction of the degraded sample (Fig.~\ref{fig:degraded}(e,f)) is considerably reduced in comparison with the non-degraded sample (Fig.~\ref{fig:degraded}(b,c)). Here, the degree of degradation $d=0.25$ can be best compared with the theoretical calculations for $d=0.2$ and $d=0.3$ (see Fig.~\ref{fig:pattern-deg}(c,d)). Related to the peak intensities, the signal is reduced by a factor of two in the theoretical analysis, while a factor of 4 is obvious in the experimental data. Comparable pronounced values are obtained in the experimental data from the comparison to theory with respect to the full width at half maximum and integral scattering intensity. At the same time, the scattering feature in $0^{\circ}$-direction Fig.~\ref{fig:degraded}(a,d) remains its characteristics (overall shape, peak intensity, fine structure) in full accordance with the theoretical predictions.

The pronounced sensitivity of the optical sensor to Riblet degradations points to additional sources of intensity losses that are not considered by the theoretical approach, so far. Some indications can be obtained from the magnified SEM pictures of the non-degraded and degraded Riblets in Figs.~\ref{fig:sem}(c,d). Beside the loss of the Riblet tip, the picture  Figs.~\ref{fig:sem}(d) shows a larger amount of deposited material distributed on the Riblet surface. It is likely, that the deposited material is part of the material removed from the Riblet tip and/or is generated during the process to decrease the Riblet height by means of sample scrubbing using acetone. The deposited material will act as additional scattering centers and, thus, will foster losses of the intensity that is scattered into the $\pm45^{\circ}$-direction. At the same time, an increase of the background noise in the entire sphere must result. The small increase can not be deduced from the experimental data in the directions of the main scattering features, but may be detectable with an appropriate diode at a scattering angle in-between. Further sources of additional losses are a change in the index of refraction, that affects the surface reflectivity. This may be due to the chemical procedure using acetone that results in a chemical modification of the Riblet surface. Although such modification is very unlikely for Riblet inspection in the frame of airplane maintenance, a change of the surface reflectivity, e.g. by deposition of dust on the surface or by alterations of the index of refraction due to aging, can be taken into account. An example for such a change of reflectivity and its successful detection by the experimental setup can be found in the results of the QUV-sample, see section \ref{sec:colored}.

%% file: snr.tex
\subsection{Definition of signal and noise and a degradation measure}

The general definition of the signal-to-noise-ratio (SNR) accounts for a definition of both, the signal and noise for the present optical sensor. According to the deliverables of WP1 and WP2, the signal for the determination of Riblet degradation is derived from the ratio between the peak or overall scattering intensity in $\pm45^{\circ}$-direction $I^{\pm45}_{\rm deg}(z)$ and the intensity in $0^{\circ}$-direction $I^{0}(z)$. Due to noise effects, all these quantities deviate from the ideal value by $\Delta I^{0,\pm45}_{\rm deg}(z)$. This quantity can be measured by closing the shutters and measuring only the noise. 

The noise must consider the electronic noise of the detection system as well as the background optical noise. The latter may be determined in two ways: (i) by detection of the scattered light in the angular regimes in-between the main scattering features. However, this also contains important information on the Riblet structure, particularly its surface roughness and contamination with deposited extrinsic material, as it is discussed in WP1 and WP2. According to the corresponding deliverables, the additionally detection of this type of background intensity is suggested, and will, hence, not be considered for noise detection. (ii) By determination of the scattered light that is due to light scattering at the optical components of the sensor setup. This type of noise can be determined by replacing the riblet sample with a black scatterer (e.g. a mechnical shutter) and is applied for the SNR in the following.

The noise impact on the degradation parameter is derived as follows. In a first step, the coordinate-dependent signal is integrated to an scalar intensity value, if the signal is measured with spacial resolution: 
\begin{align}
S^{0,\pm45}&= \left. \int_{0\,\rm{mm}}^{8\,\rm{mm}}I^{0,\pm45}_{\rm deg}(z)dz \right|_{\rm{shutter~open}}, \\
N^{0,\pm45}&= \left. \int_{0\,\rm{mm}}^{8\,\rm{mm}}I^{0,\pm45}_{\rm deg}(z)dz \right|_{\rm{shutter~closed}}.
\end{align}
When measuring with static photodiodes without spacial resolution, this is the initally measured value. 
Consecutively, the normalized quantities
\begin{align}
\bar{S}^{\pm45}&= S^{\pm45}/(S^0)^2\text{,} \\
\bar{N}^{\pm45}&= \frac{S^0 N^{\pm45} + S^{\pm45} N^{\pm45}}{(S^0)^3} 
\end{align}
are derived to account for the reflectivity of the surfaces and scattering on small particles.
From these, the degradation degree $d$ may be calculated via
\begin{equation}
d^{\pm45} = 1-\frac{\bar{S}^{\pm45}_{\rm deg}}{\bar{S}^{\pm45}_{\rm
undeg}}=1-\frac{S^{\pm45}_{\rm deg} S^0_{\rm undeg}}{S^{\pm45}_{\rm
undeg} S^0_{\rm deg}}\text{,}
\end{equation}
where ``deg'' denotes the signal from the sample under study and ``undeg'' denotes a reference signal.

\subsection{Derivation of the signal-to-noise-ratio}

The overall effect of noise, i.\,e. the final SNR, can be developed via standard propagation-of-error methods, resulting in
\begin{align}
\Delta d^{\pm45}&:=\frac{\bar{N}^{\pm45}_{\rm deg} \bar{S}^{\pm45}_{\rm
undeg} +
\bar{S}^{\pm45}_{\rm deg} \bar{N}^{\pm45}_{\rm undeg}}{(\bar{S}^{\pm45}_{\rm
undeg})^2}\\
&= \frac{S^{0}_{\rm undeg} (N^{\pm45}_{\rm deg} S^{0}_{\rm undeg} S^{\pm45}_{\rm
undeg} (S^{0}_{\rm deg}+S^{\pm45}_{\rm deg})+N^{\pm45}_{\rm undeg} S^{0}_{\rm
deg} S^{\pm45}_{\rm deg} (S^{0}_{\rm undeg}+S^{\pm45}_{\rm undeg}))}{(S^{0}_{\rm
deg})^3 (S^{\pm45}_{\rm undeg})^2}
\end{align}

In the expected cases of degradation $d^{45}=d^{-45}=d$. However, we accounted
for possible unsymmetrical degradation here.

The requirements for the detection limit of the optical sensor for $d$ is 10$\percent$ (see results of the projects first report meeting and presentation). Considering the definition of SNR, the requirements of the background noise and signal intensity can be deduced as follows: The degree of degradation results from a Riblet measurement in comparison to a reference sample, both characterized by a SNR, e.g. SNR(sample) and SNR(reference). It is assumed, that both measurements are performed with the same optical sensor, i.e. there is no difference in the sensors' noise. The difference signal must exceed the optical noise, that is at least 0.1. Thus, the limit for the ratio of signal and noise is determined by SNR$>10$. This requirement is well met by the optical sensor presented in this workpackage and the systematic study on $d$ in WP2.

%% file: device.tex
%\subsection{Report of WP3 (Device development)}
\subsection{Device constraints}
\label{sec:req_device}

%Considerations and specifications for device development\\

The task of WP3 for the first six months of the project was to summarize measurement conditions and device pre-requisites taking into account all information on relevant environment and technical aspects of maintenance. For this purpose a device catalog has been assembled, that meets the demands for milestone 3.1, i.e. a device specification list is available. A laser system for the preliminary configuration of an optical sensor is selected and proposed on the basis of the device catalog.

%\subsubsection{Results of WP3}

Relevant information on maintenance as well as of manufacturing processes of aircrafts is taking into account \footnote {Information collected via electronic publications from European Aviation Safety Agency (EASA), EADS, Federal Aviation Administration (FAA)}. The following specifications need to be met by an optical sensor device for the inspection of riblet structures (maintenance ($\ast$); manufacturing ($\ast\ast$)):\\

{\bf Device Catalog, Part A:} List of General Specifications\\
\begin{center}
\begin{tabular}{c}
%\hline
$\bullet$ conditions of operation \\
\hline
\hline
humidity: 5-95\,$\%$\\
temperature: 0-60\,$\deg$C\\
particle contamination: ISO 9 [DIN EN ISO 14644-1]\\
working place exposure to light: 300 lux \footnote{European Agency for Safety and Health at Work}\\
\hline
\\
\\
%\hline
$\bullet$ safety aspects\\
\hline
\hline
laser classes: 1$^{\ast, \ast\ast}$, 1M$^{\ast, \ast\ast}$, 2$^{\ast, \ast\ast}$, 2M$^{\ast, \ast\ast}$, 3R$^{\ast\ast}$, 3B$^{\ast\ast}$ [BGV B2 (VBG 93)]\\
contact-proof/dust-proof/water-proof: IP 51$^{\ast}$, IP54$^{\ast\ast}$ [DIN EN 60529]\\
\hline
\\
\\
%\hline
$\bullet$ energy consumption\\
\hline
\hline
battery powered$^{\ast}$ (rechargeable)\\
mains operation$^{\ast\ast}$ (110-240 V): 3.000 Watts max. \\
\hline
\\
\\
%\hline
$\bullet$ form factor\\
\hline
\hline
weight: 1.500 g max.$^{\ast}$,4.000 g max.$^{\ast\ast}$\\
dimension: handheld sensor, i.e., 0.002\,m$^3$ max. $^{\ast}$, 0.125\,m$^3$ max.$^{\ast\ast}$\\
\hline
\\
\\
%\hline
$\bullet$ costs\\
\hline
\hline
$<$ 500 EUR $^{\ast}$, $<$ 100.000 EUR $^{\ast\ast}$\\
\hline
\\
\\
%\hline
$\bullet$optical pre-requisites\\
\hline
\hline
operation at near-infrared wavelengths preferred (700-1600\,nm)\\
laser power: $<$ 0.5\,mW $^{\ast}$, $<$ 5\,mW $^{\ast\ast}$\\
\hline
\end{tabular}
\end{center}

\clearpage 

{\bf Device Catalog, Part B:} Process Criteria\\
The optical sensor needs to meet the following process criteria:
\begin{center}
\begin{tabular}{c}
$\bullet$ testing time\\
\hline
\hline
$<$ 100\,ms\\
\hline
\\
\\
$\bullet$ scan velocity\\
\hline
\hline
0.01\,m/s min.\\
\hline
\\
\\
$\bullet$ spatial scan resolution = single point scan area\\
\hline
\hline
1\,mm$^2$\\
\hline
\\
\\
$\bullet$ optical sensitivity\\
\hline
\hline
0.5\,A/W (non-amplified,T=25$^{\circ}$, $\lambda=780$\,nm)\\
\hline
\\
\\
Measurement of riblet degradation\\
\hline
\hline
calibration mode by applying reference non-degraded riblet sample\\
sensitivity to riblet degradation: $<5\%$ \\
visual representation of point of measurement\\
\hline
\\
\\
$\bullet$ representation of riblet degradation\\
\hline
\hline
visual via display ($> 2''$, color preferred)\\
audio via sound output\\
\hline
\\
\\
$\bullet$ device mount\\
\hline
\hline
compatible with DIN/ISO 9409-1-A31,5\\
4$\times$M5-screws, concentric d= 40\,mm\\
\hline
\\
\\
\end{tabular}
\end{center}

\clearpage 

Taking into account the device catalog and the deliverables of WP1 and WP2, the following selections rules for laser systems of the device are deduced:
\begin{itemize}
\item laser power: $>$\,1\,mW
\item peak wavelength: 600 - 1600\,nm; visible spectrum $< 800$\,nm preferred
\item spectral width: $<10$\,nm
\item beam divergence: $\theta < 30^{\circ}$
\item power stability: $< 1\%$\, RMS
\item slope efficiency: $>0.4$\,W/A
\item operating temperature: 0 - 70 $^{\circ}$
\item costs $< 100$\, EUR
\item dimensions $<2\cdot 10^{-5}$\,m$^{3}$
\item power consumption $< 1$\,W (voltage $\leq 24$\,VDC preferred)
\end{itemize}
Concluding these selection rules, either a telecommunication laser system operating in the third optical window ($\approx 1500$\,nm) or a laser diode in the visible/near-infrared spectral range (e.g. 785\,nm) are proposed.

%\subsubsection{Deliverables from WP3 to the other WP}
%Deliverables of WP3 are
%\begin{itemize}
%\item device catalog, part A: list of general specifications
%\item device catalog, Part B: process criteria
%\item selection rules for laser system
%\item proposal for laser systems
%\end{itemize} 

\subsection{Selection of laser system}

Following the device catalog presented in the first six months report and the results from WP1 and WP2 the selection rules and therefor the recommendation for a laser system made in the first six months report are still valid as of the end of the second six months of the project. 

According to the definition of the SNR and its limit of SNR$<10$, and considering the set of experimental data presented in WP2, it is concluded that the minimum laser power to fulfill the SNR limit is 10\,mW. 

We note, that according to the experienced work with the optical setup in the lab, a laser system emitting light at the upper end of the visible spectrum (600-700\,nm) is advantageous with respect to the optical adjustment. Then, however, the laser power, required for a sufficient SNR, may exceed the device requirements for laser safety. In order to fulfill both, a sufficient SNR and a visibility of the probing light, a superposition of two laser beams, one emitting light in the near-infrared spectrum with large power and one emitting light in the visible with low power, may be considered in the development of a demonstration platform or prototype. It is emphasized that the optical adjustment with visible laser light is not a necessary requirement for the principle function of the detector.

\clearpage

\subsection{Preliminary device configuration and its functionality}
%\todo{Preliminary device configuration and its functionality}

\begin{figure}[ht]
	\centering
	\includegraphics[height=10cm]{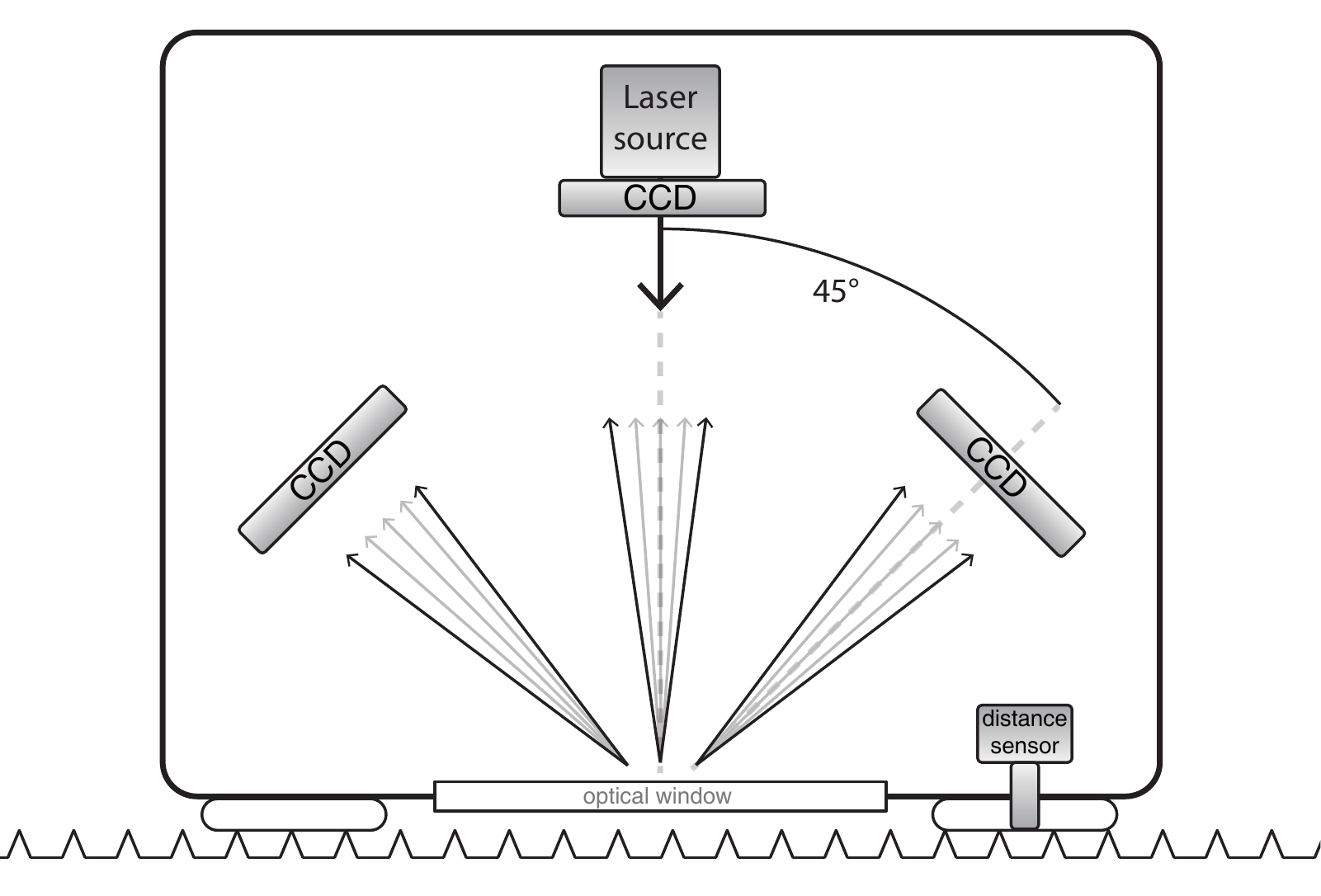}
	\caption{Sketch of a preliminary configuration of the riblet scanning device. Illumination of sample with a laser source, detection of signals at $0^\circ, \pm45^\circ$ with CCD-cameras, optical window shielding optics, distances sensor ensuring appropriate distance to sample. Signals of all three CCDs are analyzed using micro-controllers}
	\label{fig:device}
\end{figure}

One task of WP3 is a sketch of preliminary device configuration and its functionality. Taking in to account the measuring process, section \ref{sec:process}, and the preliminary configuration of the optical setup on an optical table at UOS, section \ref{sec:setup3}, a sketch giving an overview of the arrangement of all important components of a scanning device can be found in Fig. \ref{fig:device}. 

\subsubsection{Basic functionality}

To illuminate the sample a non-visible laser beam from a laser source meeting all the requirements mentioned above is used. Illumination is performed perpendicular to the riblet surface. The resulting scattering patterns at $0^\circ$ and $\pm45^\circ$ are detected  with three CCD cameras instead of photodiodes on translation stages. This allows for the device to be constructed without any moving parts. 

All optics of the device are located inside a light-weight housing which meets all requirements for the sealing of the device mentioned above in section \ref{sec:req_device}. An optical window is incorporated in housing. The device needs some kind of distance sensor to restrict measurements to a distinct distance range allowing for best use of the detection areas of the CCDs.

Signal processing can be performed using micro controllers as all mathematical operations needed to calculate the degree of degradation can be implemented as low-level operations being performed directly on the micro controllers. Optical or acoustical feedback can be given depending on a preset threshold level.

% \subsubsection{Device design}
% 
% The next step to design the final device is to realize the experimental setup as one monolithic block with all optical components glued together. Instead of distinct laser systems only semiconductor laser diodes (HLD) and photodiodes (PD) are used. This would results once again in a great reduction in the setup size. Fig. \ref{fig:device_block} shows a comparison of the implementation of the preliminary setup on an optical table in the lab of UOS and a sketch of the experimental setup realized as one monolithic block.  A reduction of the sensor dimension by about a factor of 10 is likely. Thia allows for the construction of a lightweight  handheld sensor.
% 
% \begin{figure}[htb]
% 	\centering
% 	\includegraphics{images/device_block2.pdf}
% 	\caption{a) Sketch of the preliminary experimental setup realized on an optical table in the lab of UOS, b) Sketch of the experimental setup realized in on monolithic block. }
% 	\label{fig:device_block}
% \end{figure}
% 

%% file: socio.tex
\section{Socio-economic impacts}

As described in the abstract, the the sensor whose preliminary configuration is presented in this report is of crucial importance for the application of structured coatings on airplane surfaces.
These coatings may have significant impact on the fuel consumption in commercial air traffic, which can not only reduce cost but also considerably decreases exhaust of greenhouse gases.

Apart from these ecological aspects, the sensor itself and the connected coating carry economic potential by creating jobs in the European Union and contributing to the European exports in the high-tech field.
Having presented the preliminary configuration, as a next step the device has to be developed towards a commercial device.
Consequently, it has to be produced in bulk quantities and distributed in and outside the EU.
Regarding the coating instruments for the application have to be manufactured and it has to be applied on parts of the airplane.
These requirements make room for several start-up companies and a considerable amount of jobs.
Additionally, existing European airplane manufacturers gain market advantages by applying the techniques developed in this project. 

Finally, users in both ground and air crews have to be instructed in the maintenance procedures, creating additional jobs in this sector.

%% file: appendix.tex
\clearpage
\subsection{Milestone list}
\begin{center}
\begin{tabular}{|p{0.7cm}|p{4cm}|p{1.6cm}|p{2.1cm}|p{4cm}|p{1cm}|}
\hline
\multicolumn{2}{|l|}{\bf{Milestone}} & & & & \\\hline
\textbf{No.} & \textbf{Name} & \textbf{WP(s) involved} & \textbf{Expected date} & \textbf{Means of verification} & \textbf{Done} \\ \hline
 1.1 & Theoretical calculations on non-degraded riblet structures & WP1 & 6 & Plot of intensity-apex angle available & $\surd$ \\ \hline
 1.2 & Theoretical calculations on degraded riblet structures & WP1 & 12 & Plot of intensity-apex angle available & $\surd$ \\ \hline
 1.3 & Theoretical estimates for device features & WP1 & 18 & 2 & $\surd$ \\ \hline
 2.1 & Comparison of experimental and theoretical pattern on non-degraded riblets & WP2 & 6 & Scattering features of non-degraded riblets highlighted & $\surd$ \\ \hline
 2.2 & Comparison of experimental and theoretical pattern on degraded riblets & WP2 & 12 & Scattering features of degraded riblets highlighted & $\surd$ \\ \hline
 2.3 & Preliminary configuration of optical setup& WP2 & 18 & Preliminary configuration of optical setup in optical lab of UOS & $\surd$ \\ \hline
 3.1 & Assembly of device catalog & WP3 & 4 & Device specification list available & $\surd$ \\ \hline
 3.2 & Selection of laser systems based on catalog and measurements & WP3 & 6 & Device catalog available & $\surd$ \\ \hline
 3.3 & Checking and verification of the selection of laser systems& WP3 & 12 & Laser system selection available & $\surd$ \\ \hline
 \end{tabular}
\end{center}

\subsection{WP2: photographs of riblet samples}
\label{sec:samples}

\begin{minipage}[htb]{\textwidth}

%\begin{figure}[htb]
	\centering
	\includegraphics[width=9cm]{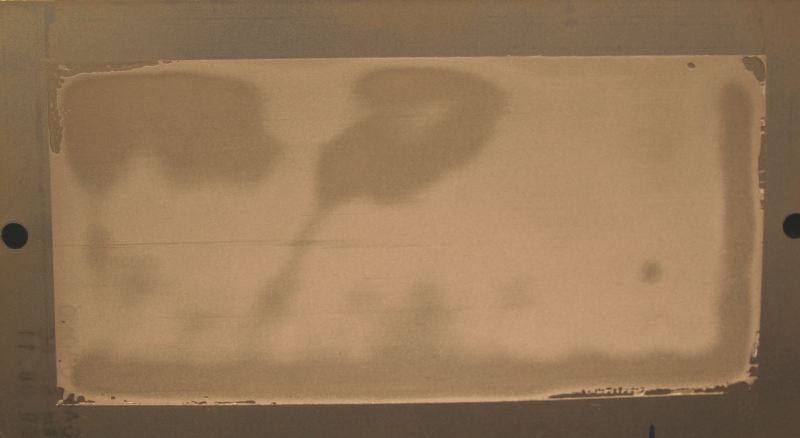}
	\captionof{figure}{The photograph was taken as an example from the first set of riblet samples under study, \textbf{Set 1}. An inhomogeneous distribution of the riblet coating over the sample area is obvious and affects the signal-to-noise ratio of the measurements series. 
}
	\label{fig:sample_0}
%\end{figure}
\vspace{.5cm}
%\begin{figure}[htb]
	%\centering
	\includegraphics[width=12cm]{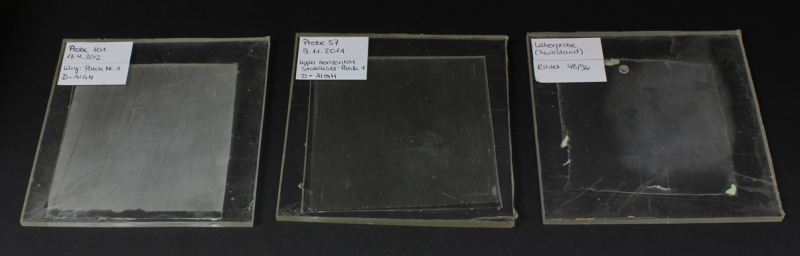}
	\captionof{figure}{This photograph shows the samples of \textbf{Set 2}, imprints of riblet structures mounted on an airplane for test purposes. }
	\label{fig:sample_6}
%\end{figure}
\vspace{.5cm}
%\begin{figure}[bth]
	%\centering
	\includegraphics[height=5cm]{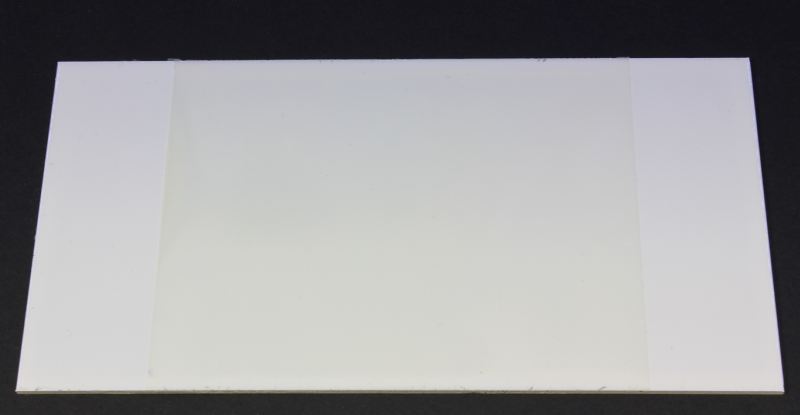}
	\captionof{figure}{The photograph shows an exemplary sample of \textbf{Set 3}. These samples were used to to study different degrees of degradation. }
	\label{fig:sample_12}
%\end{figure}

\end{minipage}

\clearpage

\subsection{WP2: sketch of samples being reduced in size for SEM measurements}

\begin{figure}[ht]
	\centering
	\includegraphics{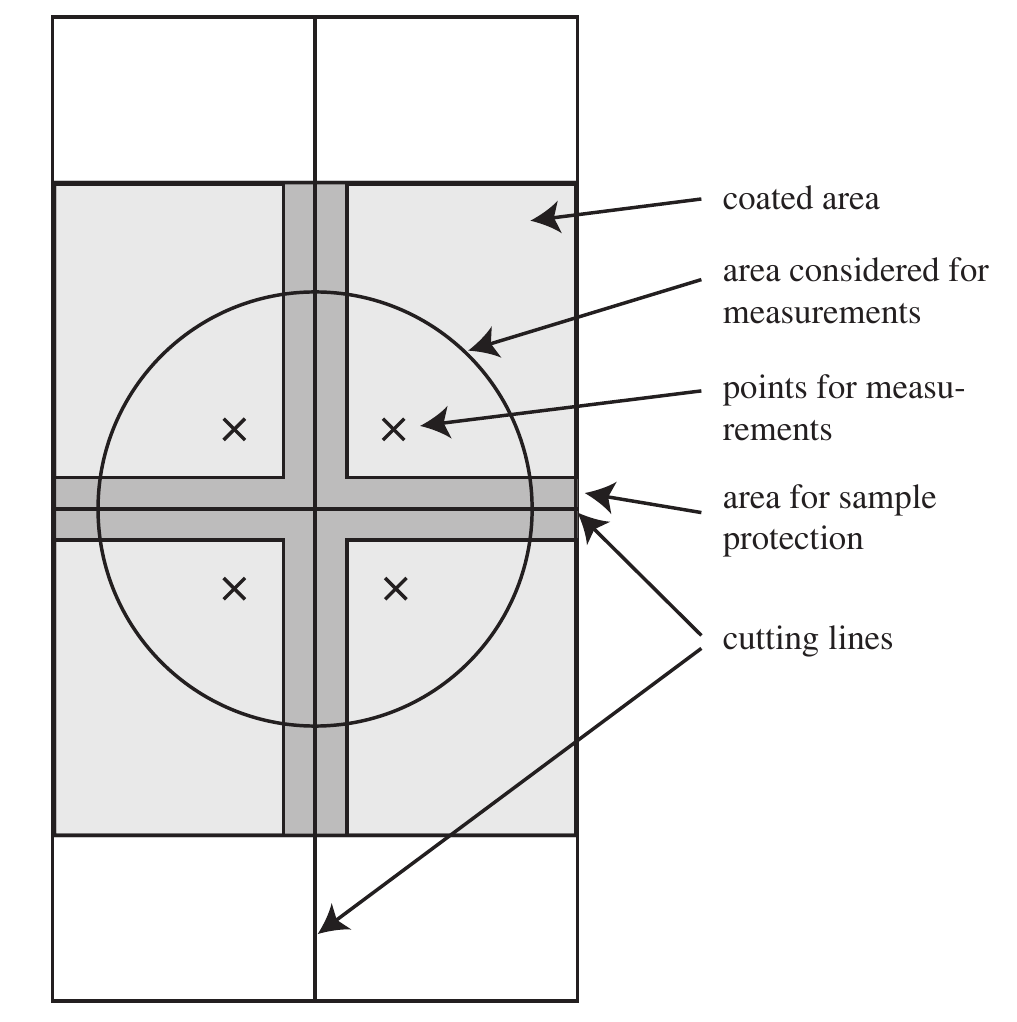}
	\caption{The samples of set 3 are reduced in size to allow SEM measurements. From the whole area coated with the riblet surface only the inner part of the sample where the riblet structure is more homogeneous is considered for measurements. Due to the cutting process some areas of the sample have to be used for protection during this process. }
	\label{fig:sample_cutting}
\end{figure}

\clearpage

\subsection{WP2: photographs of experimental setups}

\begin{figure}[hbt]
	\centering
	\includegraphics[height=7cm]{images/foto_setup_2}
	\caption{Photograph of the experimental setup used during the second six months of the project as described as the polarized beam splitting technique. Details can be found in section \ref{sec:setup2} and in the second six months short report.}
	\label{fig:foto_setup_2}
%\end{figure}

\vspace{.3cm}

%\begin{figure}[hbt]
	\centering
	\includegraphics[height=7cm]{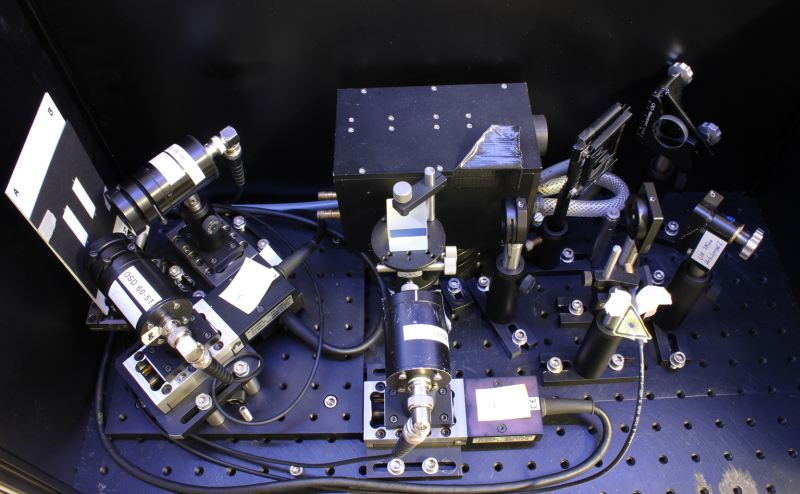}
	\caption{Photograph of the preliminary configuration of the optical setup as described in this report in section \ref{sec:setup3}. Obviously the setup has been greatly reduced in size as well as in number of optical components compared to the setup used in the polarized beam splitting technique, see Fig.~\ref{fig:foto_setup_2} }
	\label{fig:foto_setup_3}
\end{figure}

\clearpage

\subsection{WP2: additional experimental results}

\subsubsection{Influence of background color and QUV-test}
\label{sec:colored}

\begin{figure}[hbt]
	\centering
	\includegraphics[height=7cm]{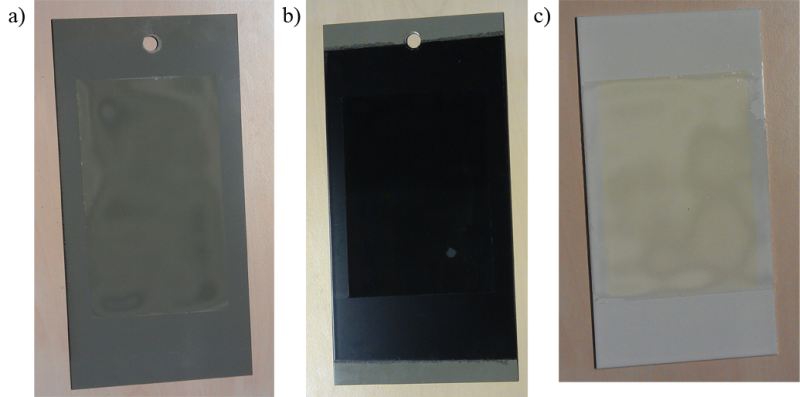}
	\caption{Photograph of the samples of set 4, a) riblet structures on a grey colored background,  b) on a black colored background, c) after a QUV-test.}
	\label{fig:foto_farbige_proben}
\end{figure}

Photographs of the samples of set 4, Fig.~\ref{fig:foto_farbige_proben} a) shows the riblet structures on a grey colored background, Fig.~\ref{fig:foto_farbige_proben} b) on a black colored background and Fig.~\ref{fig:foto_farbige_proben} c) after a QUV-test. 

\begin{figure}[hbt]
	\centering
	\includegraphics{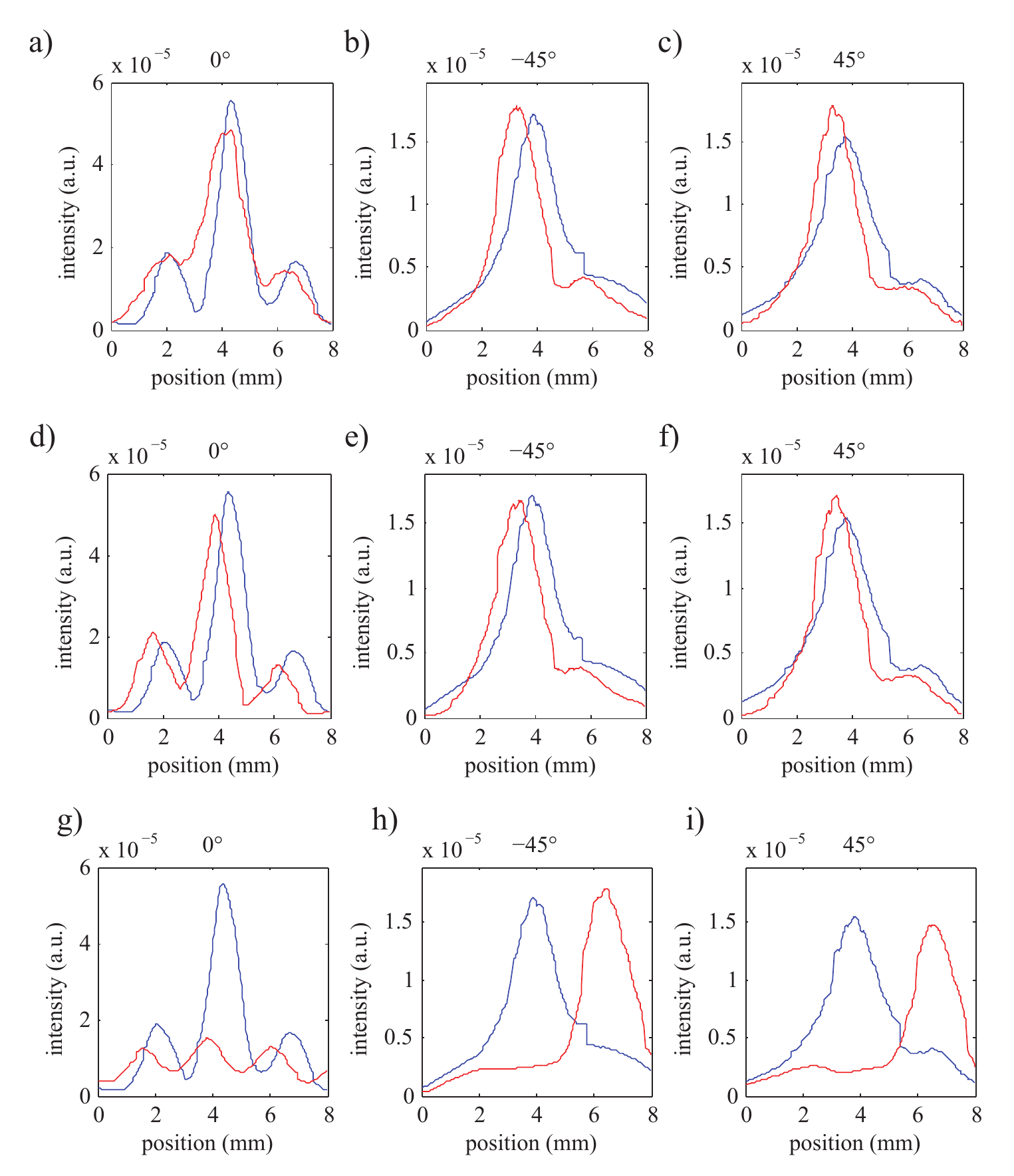}
	\caption{Optical measurements of the samples of set 4 in comparison to undegraded riblet structures on white background. a)-c) grey sample, d)-f) black sample, g)-i) after QUV test.}
	\label{fig:results_farbige_proben}
\end{figure}

Optical measurements have been performed on all there samples and are shown in Fig.~\ref{fig:results_farbige_proben} in comparison to an undegraded riblet structure on a white background.

Fig.~\ref{fig:results_farbige_proben} a) to c) shows the results of the sample with the grey background, d) to f) the results from the black background sample, g) to i) after the QUV test. 

One can see that the results from the grey and black sample are in perfect agreement to an undegraded riblet sample conserning both signal height and shape. It should be noticed that the signal height is comparable without adjustment due to comparable reflectivities. Therefor, the experimental approach works independent from the background coloring.

The QUV sample shows a shifted position of the $\pm45^\circ$ pattern. This results probably from a wider angle of the riblet structure due to UV induced shrinkage of the riblet paint. The width of the peaks remains unchanged indicating that the riblet flanks are still in a very good condition and that a possible shrinkage has affected the whole riblet paint homogeneously.
One should note that the $0^\circ$ pattern is reduce in its signal height. This can probably be attributed to the area between the riblet being damaged by the QUV-test or a overall changed reflectivity.

\begin{infobox}[]
Measurements can be performed independently from color of the background material as the riblet paint itself is transparent. This is especially true for dark colors like gray and black. The signal depends solely on the reflectivity of the riblet structure.
\end{infobox}

\clearpage

\subsubsection{Dependence of scattering pattern on probe beam diameter}
\label{sec:dia}
The following photographs were collected with a 2D-CCD array scanning scattering intensity distribution from the riblet surface for the case of illumination to a laser beam with different beam diameters with the setup in Fig.~\ref{fig:setup_null}. Scattering is depicted for the direction of the 1st order (double reflection on the riblet surface included).
\begin{figure}[h]
\begin{minipage}[h]{7.2cm}
	\centering
	\includegraphics[width=7.2cm]{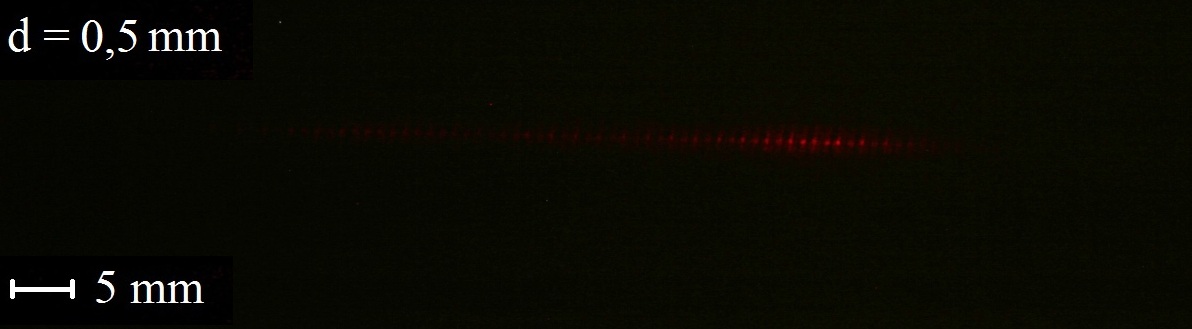}
\end{minipage}
\begin{minipage}[h]{7.2cm}
	\centering
	\includegraphics[width=7.2cm]{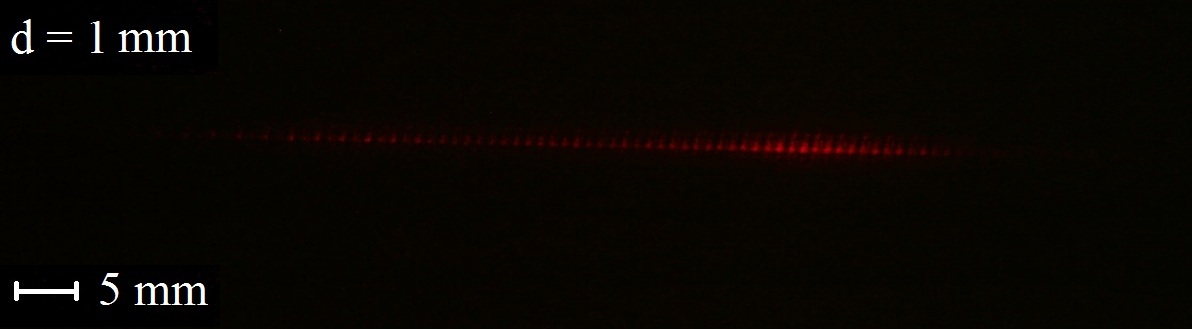}
\end{minipage}

\vspace{0.5mm}

\begin{minipage}[h]{7.2cm}
	\centering
	\includegraphics[width=7.2cm]{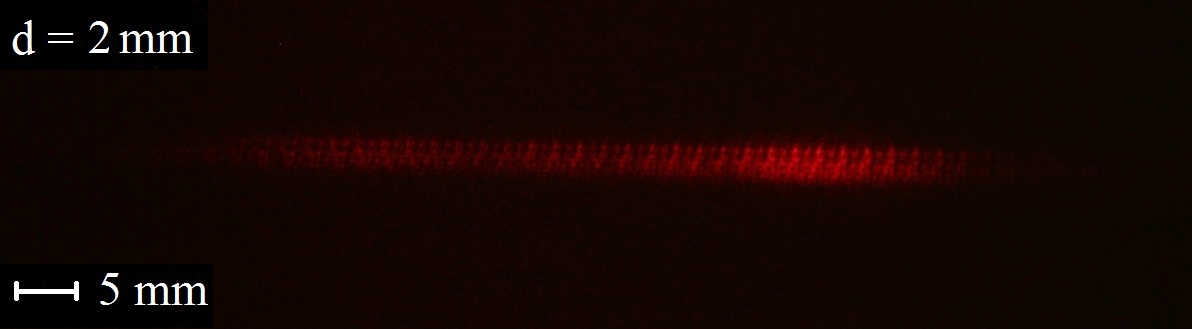}
\end{minipage}
\begin{minipage}[h]{7.2cm}
	\centering
	\includegraphics[width=7.2cm]{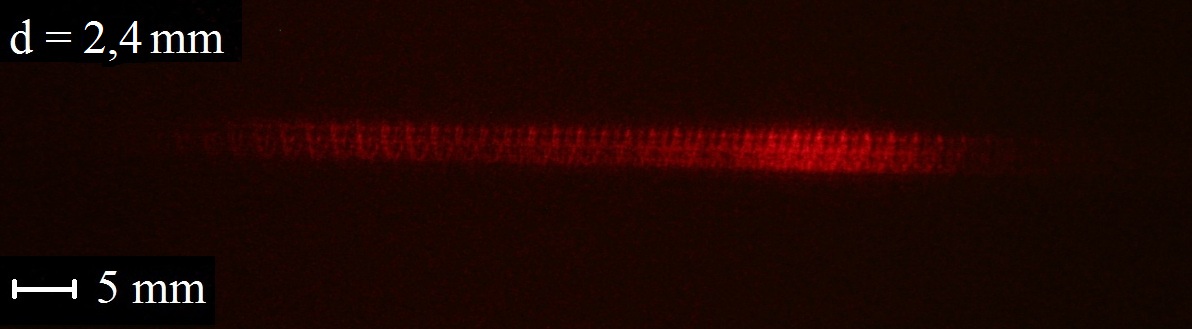}
\end{minipage}

\vspace{0.5mm}

\begin{minipage}[h]{7.2cm}
	\centering
	\includegraphics[width=7.2cm]{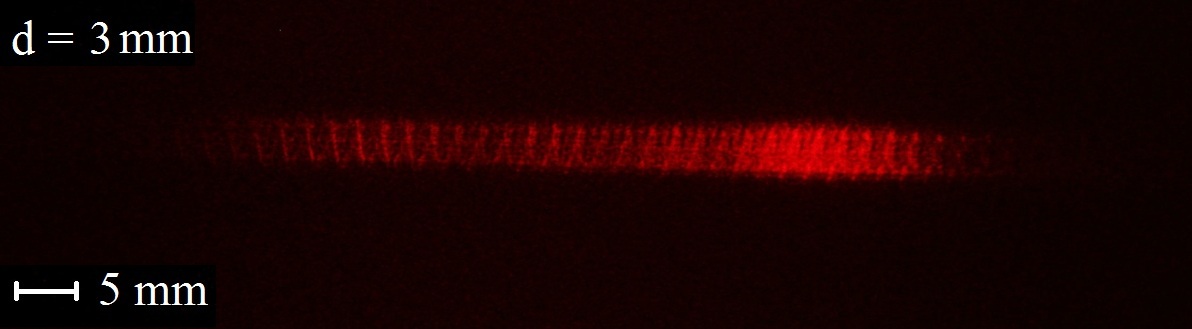}
\end{minipage}
\begin{minipage}[h]{7.2cm}
	\centering
	\includegraphics[width=7.2cm]{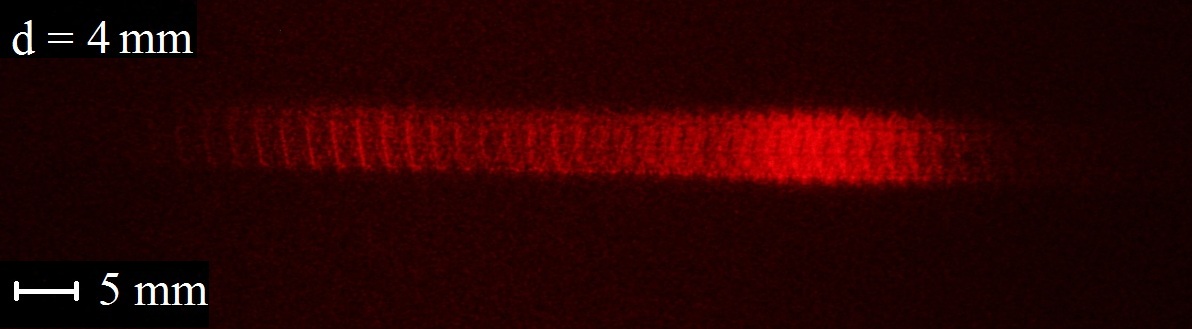}
\end{minipage}

\vspace{0.5mm}

\begin{minipage}[h]{7.2cm}
	\centering
	\includegraphics[width=7.2cm]{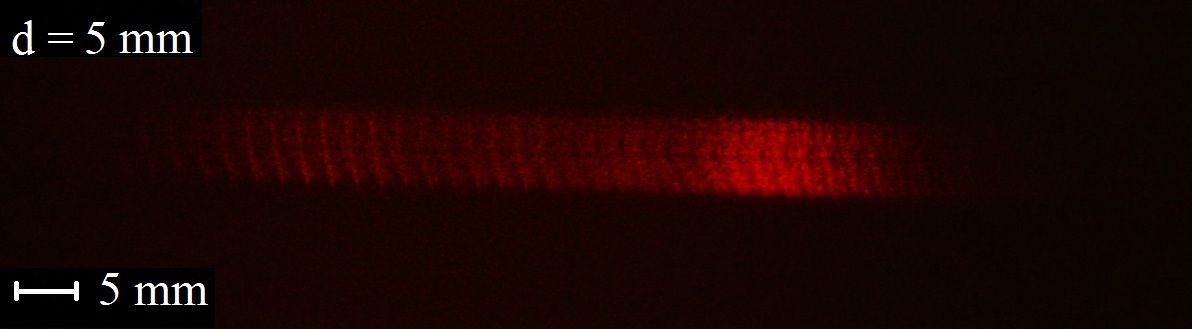}
\end{minipage}
\begin{minipage}[h]{7.2cm}
	\centering
	\includegraphics[width=7.2cm]{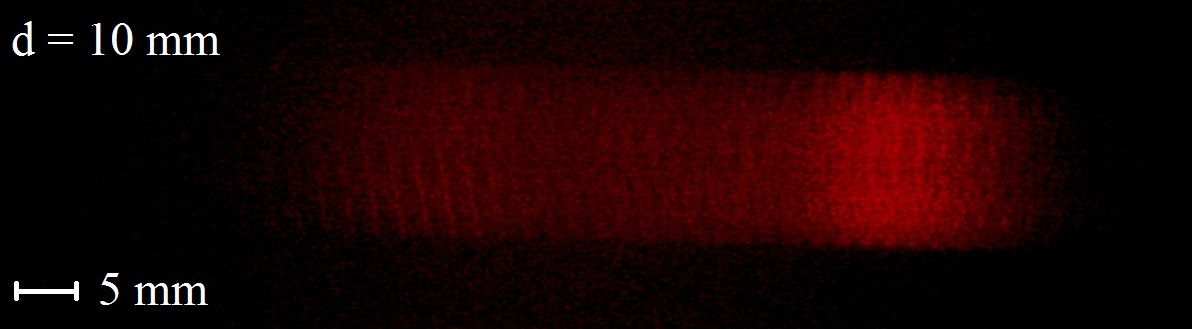}
\end{minipage}

\caption{Screen photos of intensity distribution at $45^{\circ}$ with different beam diameters $d$ at $\lambda_1$\,=\,633\,nm, $\alpha$\,=\,XX Measurement performed with Set 1, sample \,2.}
\label{d2}
\end{figure}

The scattering intensity distribution is additionally scanned with high angular resolution by means of a Si-PIN detector with appropriate pinhole. Exemplarily, the angular scattering distribution is shown for two diameters of the probing beam: 1 and 10\,mm.
\begin{figure}[ht]
	\centering
		\includegraphics[width=1\textwidth]
		{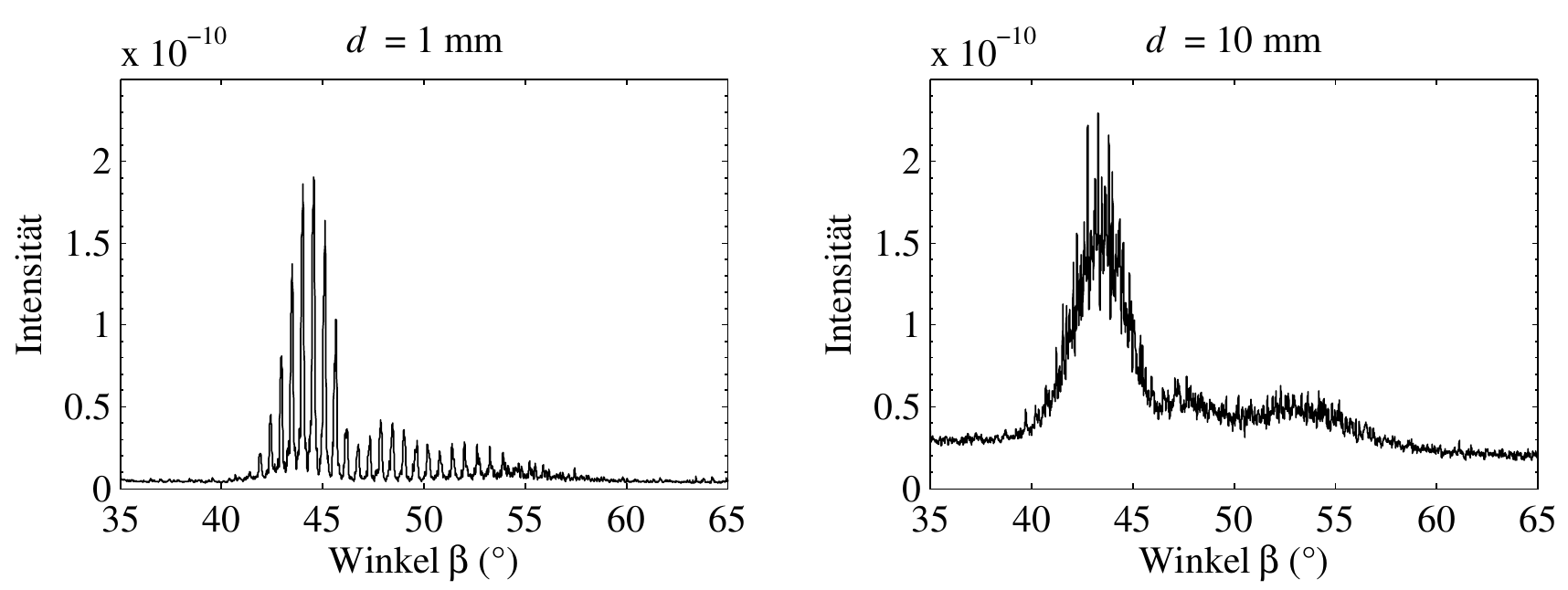}
		\caption{Intensity-apex from $\beta$\,=\,35$^{\circ}$ to $\beta$\,=\,65$^{\circ}$for beam diameter $d$\,=\,1\,mm und $d$\,=\,10\,mm bei $\lambda_1$\,=\,633\,nm. Measurement performed with set 1, sample \,2.}
		\label{dd}
\end{figure}

\clearpage

\subsubsection{Dependence on angle of incidence}
\label{sec:angle}
The following scattering intensity distributions were collected to study the influence of slant angle of the incident probing beam on the scattering pattern. Slight deviations of the angle of incidence with respect to the samples' normal strongly affect the scattering intensity distribution. This may be applied for a more detailed analysis of the type of riblet degradation (content of WP2 within the months 7-12).
\begin{minipage}[htb]{\textwidth}

	\centering
	    \includegraphics[width=.75\textwidth]{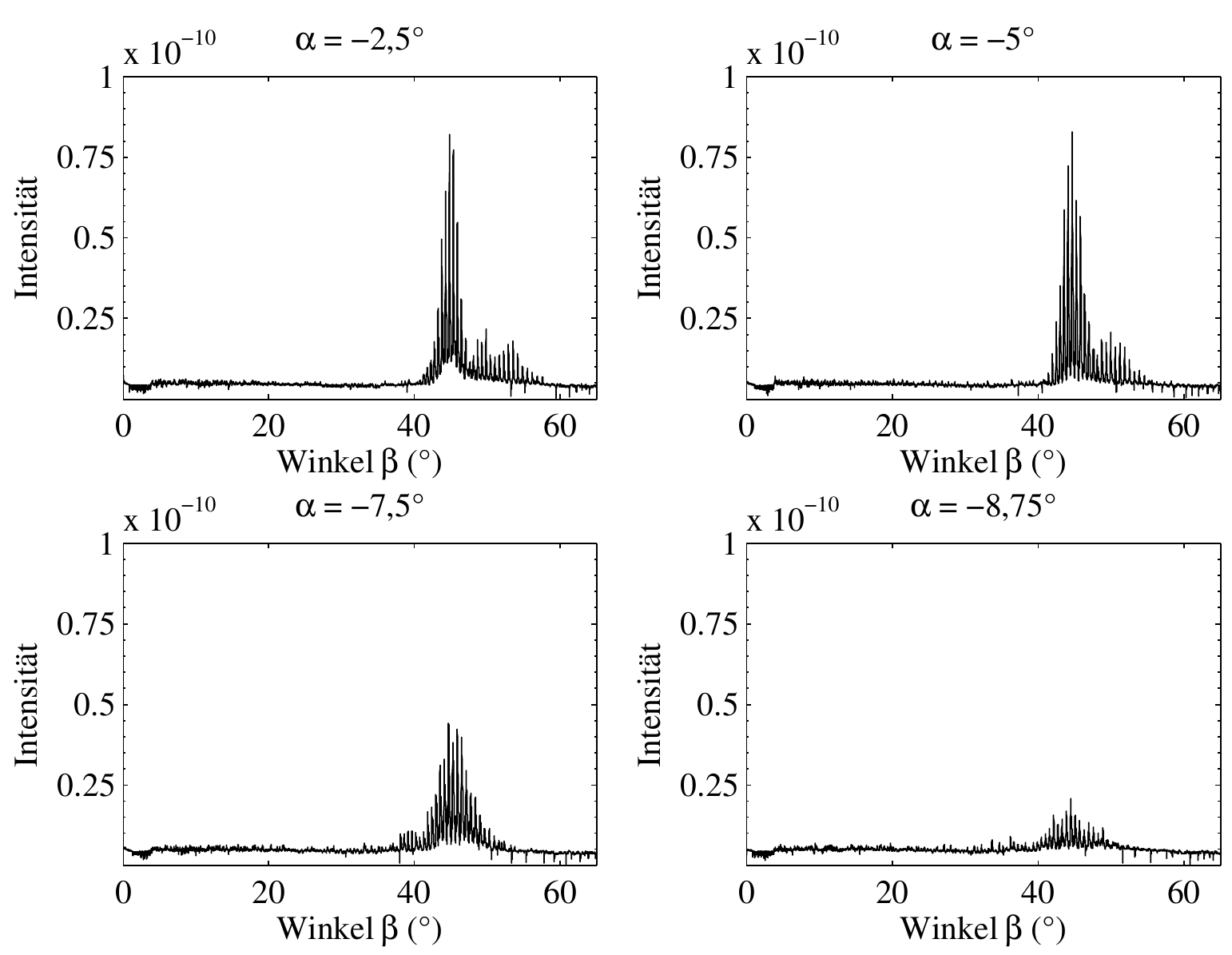}
		 \includegraphics[width=.75\textwidth]{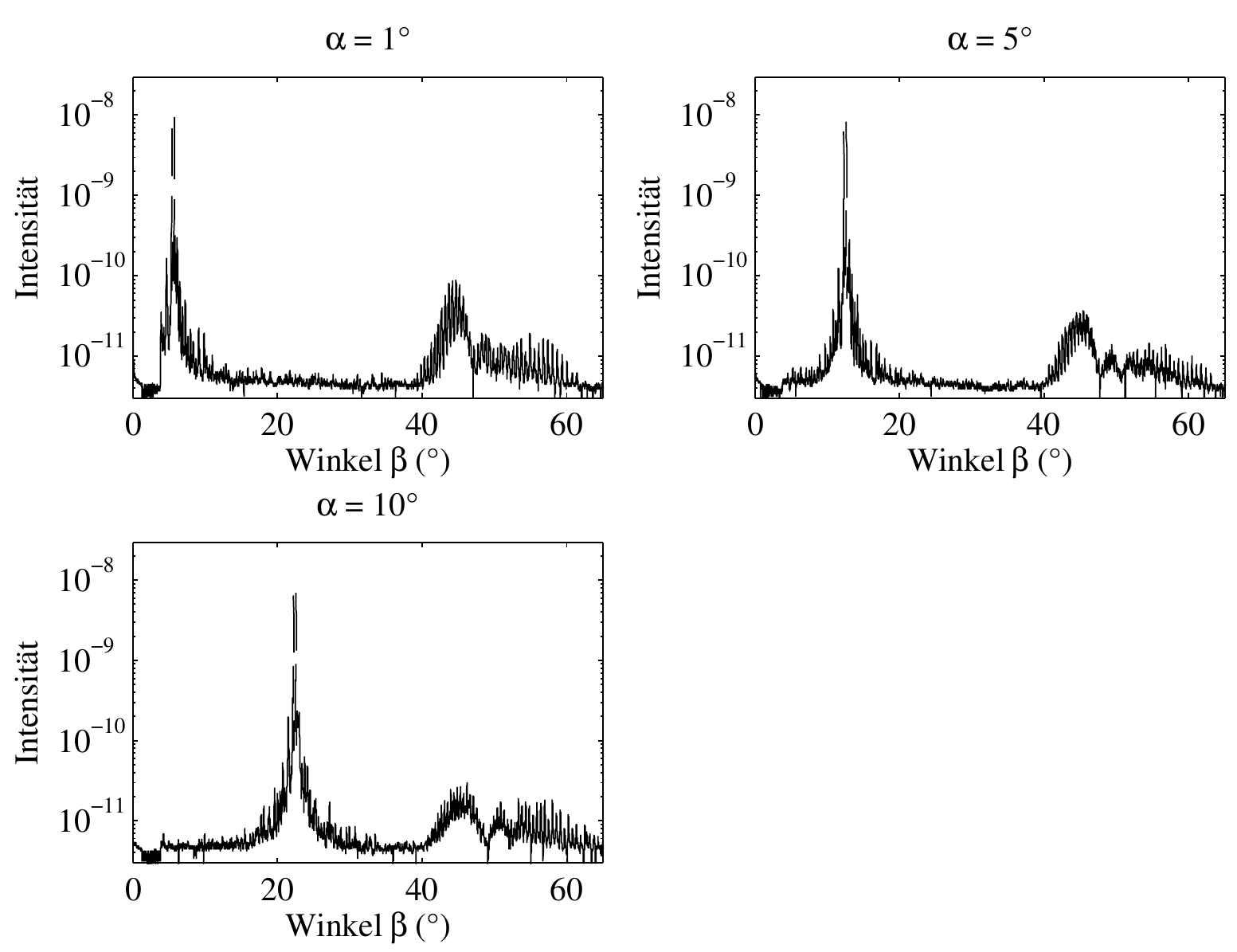}
		\captionof{figure}{Intensity-apex for angle of incidence $\alpha$\,=\,-2,5$^{\circ}$ $\alpha$\,=\,-5$^{\circ}$ $\alpha$\,=\,-7,5$^{\circ}$ $\alpha$\,=\,-8,75$^{\circ}$ $\alpha$\,=\,1$^{\circ}$ $\alpha$\,=\,5$^{\circ}$ and $\alpha$\,=\,10$^{\circ}$ at $\lambda_1$\,=\,633\,nm, $d$\,=\,1\,mm. Measurement performed with Set 1, sample \,2.}
		\label{1grad}
\end{minipage}

\clearpage

\begin{figure}[h]
	\centering
		 \includegraphics[width=.8\textwidth]{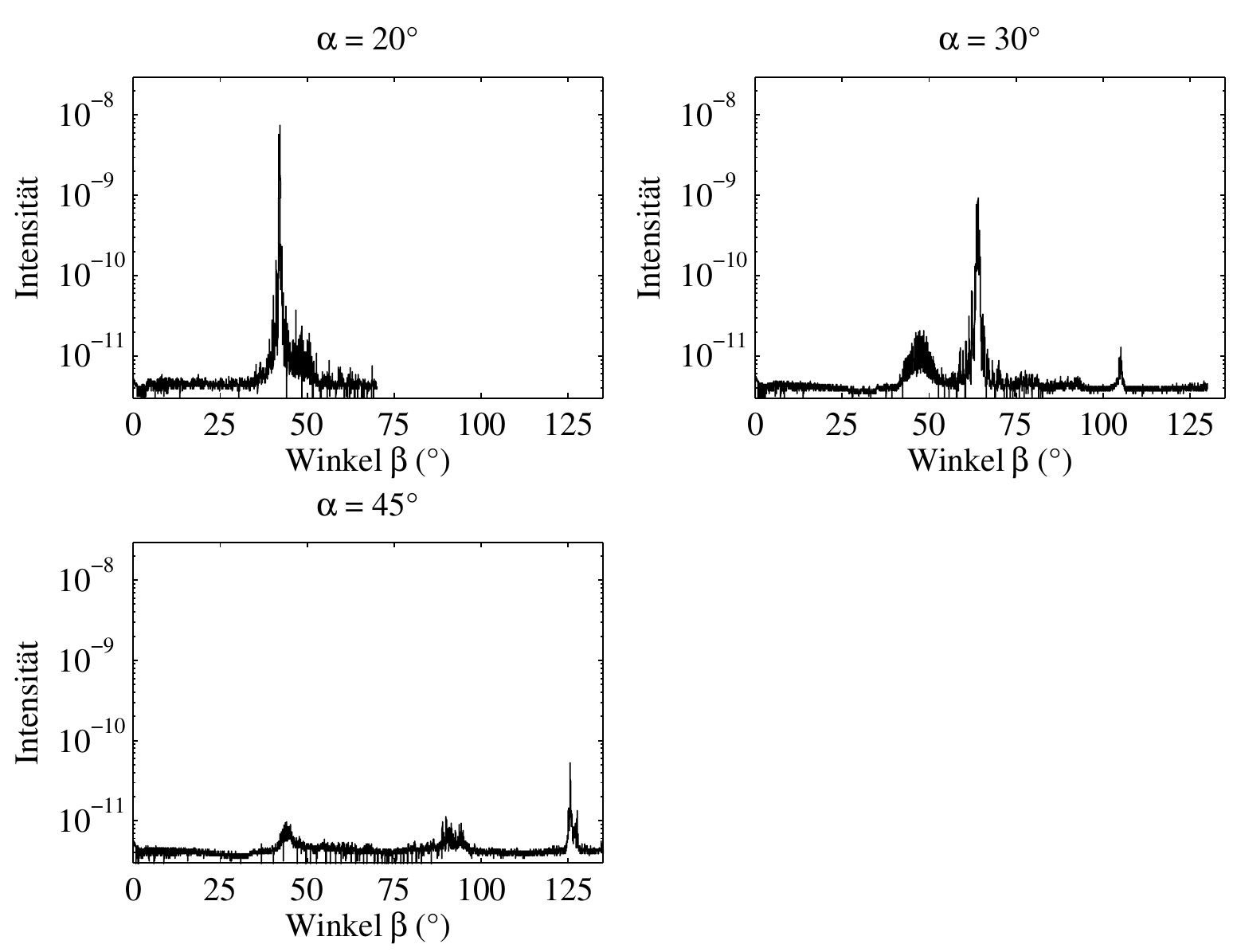}
		\caption{Intensity-apex for angle of incidence $\alpha$\,=\,20$^{\circ}$ $\alpha$\,=\,30$^{\circ}$ and $\alpha$\,=\,45$^{\circ}$ at $\lambda_1$\,=\,633\,nm, $d$\,=\,1\,mm. Measurement performed with Set 1, sample \,2.}
		\label{20grad}
\end{figure}

\clearpage

\subsubsection{Dependence on detection diode diameter}

\begin{figure}[h]
	\centering
		 \includegraphics[width=.8\textwidth]{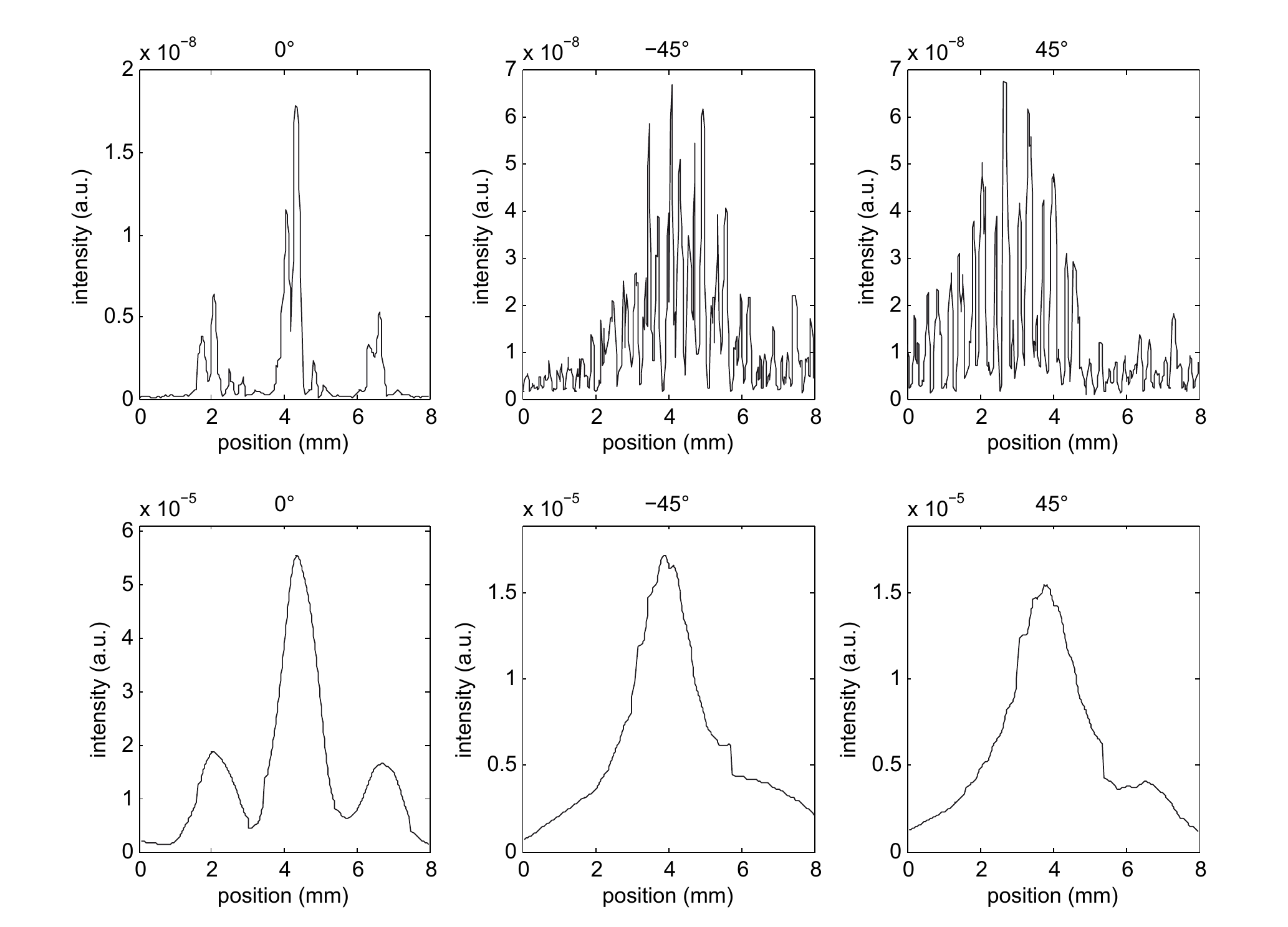}
		\caption{Comparison of experimental results using two different diameters of the detection area of the photo diodes. Top row: diameter of $50\,\mu$m, bottom row: $1\,$mm }
		\label{fig:pinhole_diameter}
\end{figure}

The influence of the size of the detection area was studied using pinholes with different diameters in front of the photo diodes. Fig. \ref{fig:pinhole_diameter} shows the two results for a detection area of $50\,\mu$m, top row, and $1\,$mm, bottom row.

\clearpage

%12062013_vergleich_dioden_abstand.pdf

\subsection{WP2: screenshots of software implementation}

\begin{figure}[h]
	\centering
		 \includegraphics[width=.8\textwidth]{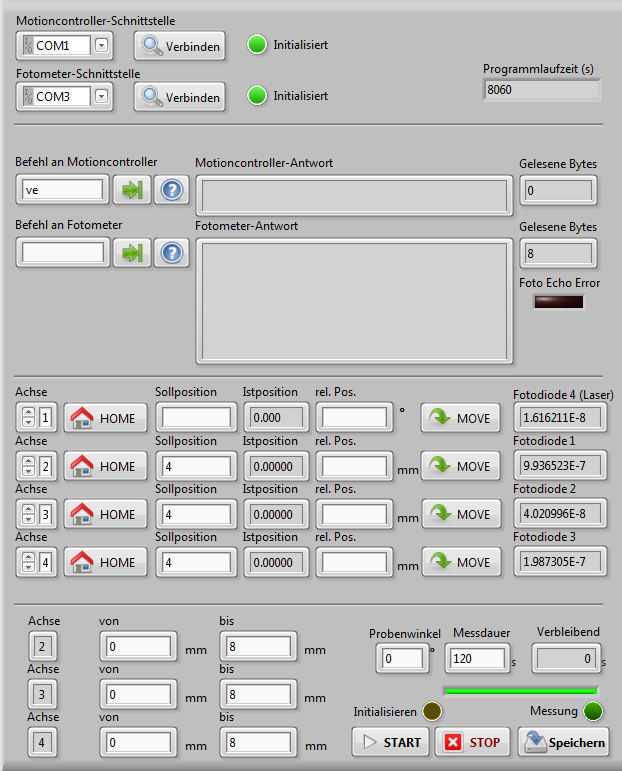}
		\caption{Screen-shot of the user interface of the software implementation using LabView for the preliminary setup in the optical lab at UOS}
		\label{fig:software_user}
\end{figure}

\clearpage

\begin{figure}[h]
	\centering
		 \includegraphics[width=.8\textwidth]{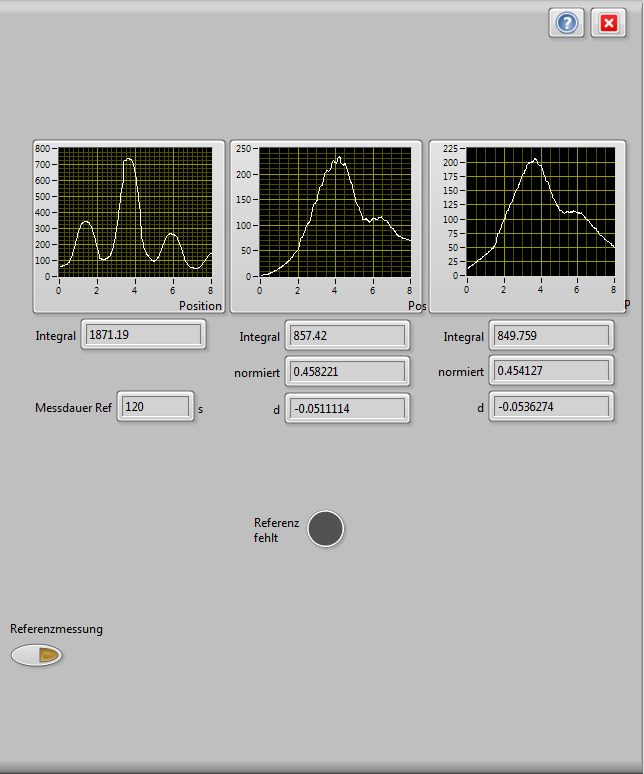}
		\caption{Screen-shot of the data acquisition of the software implementation using LabView for the preliminary setup in the optical lab at UOS}
		\label{fig:software_data}
\end{figure}
\clearpage

\clearpage
%\subsection{Code}
%
%\includepdf[pages=-]{appendix/RayTracing-v3-degraded.pdf}
%\includepdf[pages=-]{appendix/FrenHuy-v7-degraded-chain.pdf}